\documentclass[acmsmall,screen, nonacm]{acmart}
\makeatletter
\let\@authorsaddresses\@empty
\makeatother
\AtBeginDocument{%
  }

\setcopyright{none}

\renewcommand\footnotetextcopyrightpermission[1]{}

\usepackage{xcolor}
\definecolor{RoyalBlue}{RGB}{0,113,188}
\usepackage[utf8]{inputenc}
\usepackage[T1]{fontenc}
\usepackage{microtype}
\usepackage{graphicx}
\usepackage{listings}
\usepackage{multicol}
\usepackage{wrapfig}
\usepackage{float}
\usepackage{mathpartir}
\usepackage{enumitem}
\usepackage{stmaryrd}
\usepackage{hyperref}
\usepackage{subcaption}
\usepackage{xspace}

\newtheorem{theorem}{Theorem}[]

\newtheorem{exam}{Theorem}[]
\newtheorem{appexam}{Theorem}[]
\newtheorem{proposition}[theorem]{Proposition}
\newtheorem{definition}[theorem]{Definition}
\newtheorem{lemma}[theorem]{Lemma}

\newtheorem{corollary}[theorem]{Corollary}
\newtheorem{example}[exam]{Example}
\newtheorem{appexample}[appexam]{Example}

\newtheorem{innercustomgeneric}{\customgenericname}
\providecommand{\customgenericname}{}
\newcommand{\newcustomtheorem}[2]{%
  \newenvironment{#1}[1]
  {%
   \renewcommand\customgenericname{#2}%
   \renewcommand\theinnercustomgeneric{##1}%
   \innercustomgeneric
  }
  {\endinnercustomgeneric}
}

\newcustomtheorem{customlem}{Lemma}
\newcustomtheorem{customthm}{Theorem}
\newcustomtheorem{customcor}{Corollary}
\newcustomtheorem{customdef}{Definition}

\newenvironment{relproof}{\noindent \textit{Relational Proof.}}{\hfill$\square$}
\newenvironment{traproof}{\noindent  \textit{Standard Proof.}}{\hfill$\square$}

\newcommand{\defeq}{\triangleq} 

\newcommand{\tabitem}{~~\llap{\textbullet}~~}

\newcommand{\nrmdeno}[1]{\llbracket#1\rrbracket_{\mathsf{nrm}}}
\newcommand{\errdeno}[1]{\llbracket#1\rrbracket_{\mathsf{err}}}

\newcommand{\nrmfcdeno}[1]{\llbracket#1\rrbracket^{\chi}_{\mathsf{nrm}}}

\newcommand{\relasrt}[1]{\mathcal{#1}}
\newcommand{\reltriple}[3]{\langle\mathcal{#1}\rangle \ {#2}\ \langle\mathcal{#3}\rangle}
\newcommand{\reltriplenofont}[3]{\langle {#1}\rangle \ {#2}\ \langle {#3}\rangle}
\newcommand{\unarytriple}[3]{\{ {#1}\} \ {#2}\ \{{#3}\}}
\newcommand{\quadruple}[4]{\{ \mathbb{#1} \}\ {#2} \precsim {#3} \ \{ \mathbb{#4}\}}

\newcommand{\encasrt}[1]{\llparenthesis {#1}\rrparenthesis}

\newcommand\sepimp{\mathrel{-\mkern-6mu*}}

\newcommand{\absc}{c^{\text{\normalfont H}}}
\newcommand{\conc}{c^{\text{\normalfont L}}}

\newcommand{\conSigma}{\Sigma^{\text{\normalfont L}}}
\newcommand{\absSigma}{\Sigma^{\text{\normalfont H}}}

\newcommand{\const}{\sigma^{\text{\normalfont L}}}
\newcommand{\absst}{\sigma^{\text{\normalfont H}}}
\newcommand{\gst}{\sigma^{\text{\normalfont G}}}

\newcommand{\absStmts}{\text{\normalfont Prog}^{\text{\normalfont H}}}

\newcommand{\conasrt}[1]{{#1}^{\text{\normalfont L}}}
\newcommand{\absasrt}[1]{{#1}^{\text{\normalfont H}}}
\newcommand{\conasrtenc}{\bar{P}^{\normalfont L}}

\newcommand{\pro}{\rho}

\newcommand{\absprog}{\text{Prog}^{\text{\normalfont H}}}

\newcommand{\conprog}{\text{Prog}^{\text{\normalfont L}}}
\newcommand{\choice}[2]{\text{\normalfont\textsf{choice}}({#1},\, {#2})}

\newcommand{\conf}{f^{\text{\normalfont L}}}

\newcommand{\conb}{b^{\text{\normalfont L}}}

\newcommand{\reldelta}{\Gamma}

\newcommand{\onlyExec}{{\color{RoyalBlue}\mathsf{Exec}}}
\newcommand{\needExecX}[3]{{\color{RoyalBlue} \mathsf{Exec}}_{#3}{\color{RoyalBlue}(}{#1}{\color{RoyalBlue},} {\color{purple}#2}{\color{RoyalBlue})}}

\newcommand{\hasrt}[1]{\lceil {#1} \rceil}
\newcommand{\lasrt}[1]{\lfloor {#1} \rfloor}
\newcommand{\hspeccolor}[1]{{\color{purple}[{#1}]}}

\newcommand{\cskip}{\text{\normalfont \textsf{skip}}}

\newcommand{\pvar}[1]{\mathsf{{#1}}}

\newcommand{\convar}[1]{\pvar{#1}^{\text{\normalfont L}}}
\newcommand{\absvar}[1]{\pvar{#1}^{\text{\normalfont H}}}

\newcommand{\conexp}{e^{\text{\normalfont L}}}
\newcommand{\absexp}{e^{\text{\normalfont H}}}

\newcommand{\absbexp}{b^{\text{\normalfont H}}}

\newcommand{\constorepre}{\mathsf{storePre}^{\text{\normalfont L}}}
\newcommand{\absstorepre}{\mathsf{storePre}^{\text{\normalfont H}}}

\newcommand{\constorepost}{\mathsf{storePost}^{\text{\normalfont L}}}
\newcommand{\absstorepost}{\mathsf{storePost}^{\text{\normalfont H}}}

\newcommand{\emp}{\textsf{emp}}

\newcommand{\demovdash}{\vdash_{\forall}}

\newcommand{\angelvdash}{\vdash_{\exists}}

\newcommand{\cassert}[1]{\text{\normalfont \textsf{assert }}({#1})}

\newcommand{\cwhile}[2]{\text{\normalfont \textsf{while }}{#1} \ {\normalfont \textsf{do }} {#2}}

\newcommand{\ccall}[1]{\text{\normalfont  \textsf{call }}{#1}}

\newcommand{\ctest}[1]{\text{\normalfont  \textsf{assume }}{#1}}
\newcommand{\nil}{\text{\normalfont empty}}
\newcommand{\pnull}{\text{\texttt{NULL}}}

\newcommand{\fid}{\text{\normalfont fid}}

\newcommand{\conmerge}{\texttt{merge}^{\text{\normalfont L}}}
\newcommand{\absmerge}{\texttt{merge}^{\text{\normalfont H}}}

\newcommand{\sll}[2]{\textsf{sll}({#1}, {#2})}
\newcommand{\sllbseg}[3]{\textsf{sllbseg}({#1}, {#2}, {#3})}

\newcommand{\len}[1]{\text{\textsf{len}}({#1})}

\newcommand{\ptrue}{\normalfont\textsf{True}}
\newcommand{\pfalse}{\normalfont\textsf{False}}

\newcommand{\abswhile}{\cwhile{\absvar{j}<8}{\ldots}}

\newcommand{\figref}[1]{Fig. \ref{#1}}
\newcommand{\secref}[1]{Sec. \ref{#1}}
\newcommand{\propref}[1]{Prop. \ref{#1}}
\newcommand{\defref}[1]{Def. \ref{#1}}
\newcommand{\theoref}[1]{Theo. \ref{#1}}
\newcommand{\weakestpre}[2]{\textsf{wlp}({#1},{#2}) }
\newcommand{\wpre}[2]{\textsf{wp}({#1},{#2}) }

\newcommand{\wpreghost}[3]{\textsf{wp}^{\text{gst}}_{{#3}}({#1},{#2}) }

\newcommand{\cupdot}{\charfusion[\mathbin]{\cup}{\cdot}}
\newcommand{\bitmaskexample}{\hyperref[fig:toyprog]{bitmask/set example}\xspace}
\newcommand{\inlinecodes}[1]{\lstinline[mathescape=true,language=C, basicstyle=\normalfont]!#1!}

\makeatletter
\def\moverlay{\mathpalette\mov@rlay}
\def\mov@rlay#1#2{\leavevmode\vtop{%
   \baselineskip\z@skip \lineskiplimit-\maxdimen
   \ialign{\hfil$\m@th#1##$\hfil\cr#2\crcr}}}
\newcommand{\charfusion}[3][\mathord]{
    #1{\ifx#1\mathop\vphantom{#2}\fi
        \mathpalette\mov@rlay{#2\cr#3}
      }
    \ifx#1\mathop\expandafter\displaylimits\fi}
\makeatother

\makeatletter
\DeclareFontFamily{OMX}{MnSymbolE}{}
\DeclareSymbolFont{MnLargeSymbols}{OMX}{MnSymbolE}{m}{n}
\SetSymbolFont{MnLargeSymbols}{bold}{OMX}{MnSymbolE}{b}{n}
\DeclareFontShape{OMX}{MnSymbolE}{m}{n}{
    <-6>  MnSymbolE5
   <6-7>  MnSymbolE6
   <7-8>  MnSymbolE7
   <8-9>  MnSymbolE8
   <9-10> MnSymbolE9
  <10-12> MnSymbolE10
  <12->   MnSymbolE12
}{}
\DeclareFontShape{OMX}{MnSymbolE}{b}{n}{
    <-6>  MnSymbolE-Bold5
   <6-7>  MnSymbolE-Bold6
   <7-8>  MnSymbolE-Bold7
   <8-9>  MnSymbolE-Bold8
   <9-10> MnSymbolE-Bold9
  <10-12> MnSymbolE-Bold10
  <12->   MnSymbolE-Bold12
}{}

\let\llangle\@undefined
\let\rrangle\@undefined
\DeclareMathDelimiter{\llangle}{\mathopen}%
                     {MnLargeSymbols}{'164}{MnLargeSymbols}{'164}
\DeclareMathDelimiter{\rrangle}{\mathclose}%
                     {MnLargeSymbols}{'171}{MnLargeSymbols}{'171}
\makeatother

\begin{document}

\title{Encode the \texorpdfstring{$\forall\exists$}{PlainText}  Relational Hoare Logic into Standard Hoare Logic
}
\subtitle{Extended Version}
\author{Shushu Wu}
\orcid{0009-0000-4060-5635}
\affiliation{%
  \institution{Shanghai Jiao Tong University}
  \country{China}
}
\email{Ciel77@sjtu.edu.cn}
\author{Xiwei Wu}
\orcid{0009-0006-2469-3800}
\affiliation{%
  \institution{Shanghai Jiao Tong University}
  \country{China}
}
\email{yashen@sjtu.edu.cn}
\author{Qinxiang Cao}
\orcid{0000-0002-5678-6538}
\affiliation{%
  \institution{Shanghai Jiao Tong University}
  \country{China}
}
\email{caoqinxiang@gmail.com}



\begin{abstract}
 Verifying a real-world program’s functional correctness can be decomposed into (1) a refinement proof showing that the program implements a more abstract high-level program and (2) an algorithm correctness proof at the high level.
Relational Hoare logic serves as a powerful tool to establish refinement but often necessitates formalization beyond standard Hoare logic.
Particularly in the nondeterministic setting, the $\forall\exists$ relational Hoare logic is required.
Existing approaches encode this logic into a Hoare logic with ghost states and invariants, yet these extensions significantly increase formalization complexity and soundness proof overhead.
This paper proposes a generic encoding theory that reduces the $\forall\exists$ relational Hoare logic to standard (unary) Hoare logic. 
Precisely, we propose to redefine the validity of relational Hoare triples while preserving the original proof rules and then encapsulate the $\forall\exists$ pattern within assertions. 
We have proved that the validity of encoded standard Hoare triples is equivalent to the validity of the desired relational Hoare triples. 
Moreover, the encoding theory demonstrates how common relational Hoare logic proof rules are indeed special cases of standard Hoare logic proof rules, and relational proof steps correspond to standard proof steps.
Our theory enables standard Hoare logic to prove $\forall\exists$ relational properties by defining a predicate $\onlyExec$, without requiring modifications to the logic framework or re-verification of soundness.

\end{abstract}




\maketitle

\section{\label{sec:intro}Introduction}
Hoare Logic \cite{Floyd1967Flowcharts,10.1145/363235.363259} is widely used to prove the functional correctness of programs formally.
Nevertheless, when dealing with complex programs, it is often more convenient to decompose the proof into two parts \cite{DBLP:conf/cpp/Appel22,lammich2015framework}: (1) a refinement proof showing that the concrete low-level program refines a more abstract high-level program and (2) an algorithm correctness proof of the high-level program.
This verification mode  \footnote{This mode differs from verification tools like Why3, which rely on ghost code to help verify properties of real programs. 
A detailed discussion on these tools is provided in \secref{subsec:rhlworks}. } enables significant proof reuse. For instance, once the algorithm correctness proof of a high-level program is established, it can be reused across different low-level implementations; likewise, once a refinement proof is established, it can be leveraged for new usage scenarios focusing on different aspects of algorithm correctness.
To support proofs that relate two programs, relational Hoare logic \cite{DBLP:conf/popl/Benton04} has been introduced and is widely used to establish program refinement \cite{DBLP:conf/popl/LiangF16,DBLP:journals/corr/abs-2006-13635,timany2019mechanized,DBLP:journals/pacmpl/GaherSSJDKKD22}.
In practice, one might formalize a program logic in a proof assistant like Rocq for deductive verification. 
However, the  verification mode above requires formalizing both standard Hoare logic (for functional correctness of simple programs) and relational Hoare logic (for refinement proofs), each with its own soundness proof, which can become cumbersome. 
Then, to reason both single- and multi-program properties in a unified logic framework, researchers have explored how to encode relational Hoare logic into standard Hoare logic.

A notable technique is self-composition \cite{1310735,conf/sas/TerauchiA05}, which reduces the $\forall\forall$ relational Hoare logic, also known as \textit{2-safety},  to standard Hoare logic.
The $\forall\forall$ logic requires that for all pairs of program executions, if their initial states satisfy the precondition, then any terminal states must satisfy the postcondition. 
For deterministic programs, this logic is sufficient to capture refinement.
Nevertheless,  when dealing with nondeterministic programs, the $\forall\forall$ logic cannot express program refinement. 
Instead, program refinement is captured through the $\forall\exists$ relational Hoare logic.
This logic requires that for any execution of the low-level program, there exists a corresponding execution of the high-level program such that the relational postcondition holds.
The $\forall\exists$ relational Hoare logic can be encoded into a Hoare logic augmented with ghost states 
and invariants
\cite{DBLP:conf/pldi/LiangF13,DBLP:conf/popl/TuronTABD13,DBLP:conf/esop/TassarottiJ017,DBLP:journals/pacmpl/GaherSSJDKKD22}.
In this logic, the judgment is  defined in a $(\forall\exists)^{\omega}$ pattern \footnote{Detailed explanation for this $(\forall\exists)^{\omega}$ pattern can be found in \secref{sec:relatedwork}.}: for any update of the physical state, there exists a way to update the ghost state such that the invariant holds, and the two states can continue to be updated as described before or terminate at states satisfying the postcondition.
While powerful, this Hoare logic with ghost states and invariants is significantly more complex in terms of both formalization and proof of soundness compared to standard Hoare logic.
To our knowledge, only a few frameworks provide machine-checked frameworks for Hoare logic with ghost states and invariants, most notably Iris \cite{DBLP:journals/jfp/JungKJBBD18} and the Verified Software Toolchain (VST) \cite{conf/nfm/Appel12}. 
Even in the Iris documentation, the presentation begins with a weakest precondition definition \footnote{Iris uses weakest preconditions as the underlying logical primitive; Hoare triples can be defined using weakest preconditions.}  without ghost states and adds ghost states with view shifts only later. 
Similarly, VST initially lacks ghost reasoning and only supports it after extensive extensions to its memory model and a subsequent re-verification of soundness.

Before this paper, it remains unknown how  the $\forall\exists$ relational Hoare logic can be reduced to standard Hoare logic.
Although this question may appear purely theoretical,  such a reduction could enable standard Hoare logic to reason $\forall\exists$ properties and thus  have significant practical implications for simplifying verification frameworks and proofs.
To answer this question, this paper explores the theoretical connections between the $\forall\exists$ relational Hoare logic and standard Hoare logic and presents a generic encoding theory.
This encoding theory shows that a lightweight extension to standard Hoare logic—by defining a $\needExecX{\absasrt{P}}{\absc}{X}$ predicate and its related rules—suffices to achieve relational reasoning without modifying the framework or redoing  soundness proof.


\subsection{Known theoretical connection: Similarities in proof rules}
\begin{figure}[ht]
    \centering
    \vspace{-15pt}
    \begin{tabular}{c|l}
      \begin{minipage}{0.45\textwidth}
    \centering
        \includegraphics[width = 0.64\textwidth]{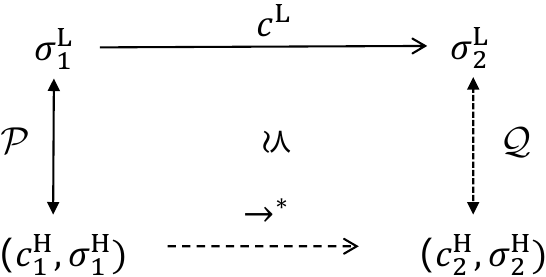}
        \caption{\label{fig:reltriple}Relational Hoare Triples}
          \end{minipage} &
    \begin{minipage}{0.35\textwidth}
    \centering
      \includegraphics[width = 0.6\textwidth]{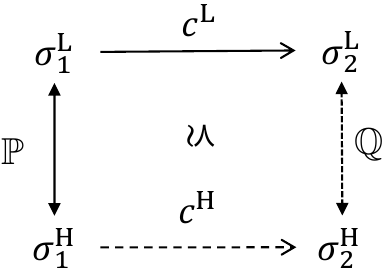}
      \caption{
      \label{fig:refinement}Program Refinement}
    \end{minipage}
    \end{tabular} 
\end{figure}

In relational Hoare logic, various forms of judgments have been proposed in the literature. 
We develop our encoding theory based on relational Hoare triples \footnote{Some frameworks choose to employ different judgments like Hoare quadruples. However, a Hoare quadruple  $\quadruple{P}{\conc}{\absc}{Q}$ can be transformed into a equivalent relational Hoare triple  $\reltriplenofont{\mathbb{P} \land \hspeccolor{\absc}}{\conc}{\mathbb{Q} \land \hspeccolor{\cskip}}$.}, which use \textit{program-as-resource} assertions \cite{DBLP:conf/pldi/LiangF13,DBLP:conf/popl/TuronTABD13,10.1145/2500365.2500600}.
This concept originates from the field of concurrent program verification and can be applied to verify sequential programs.
We choose the relational Hoare triple to explore encoding theory because it shares a similar set of proof rules with standard Hoare logic.
The validity of  relational Hoare triples is defined as follows: 

\begin{definition}[\label{def:rhtintro}Relational Hoare Triples] As shown in \figref{fig:reltriple}, relational Hoare triple $\langle \mathcal{P} \rangle \ \conc \ \langle \mathcal{Q} \rangle$ is valid if given any initial states $(\const_1, \absst_1)$ and a high-level statement $\absc_1$ such that  $(\const_1, \absst_1, \absc_1) \models \mathcal{P}$, for any final state $\const_2$ after execution of the low-level statement $\conc$, 
there exists a multi-step transition from $(\absst_1, \absc_1)$ to $(\absst_2, \absc_2)$ such that  $(\const_2, \absst_2, \absc_2) \models \mathcal{Q}$.
 \end{definition}
\noindent To express the refinement between a low-level command $\conc$ and a high-level command $\absc$, as illustrated in \figref{fig:refinement}, we lift the binary assertions $\mathbb{P}$ and $\mathbb{Q}$  \footnote{For clarity, this paper uses different fonts to distinguish assertions as follows. In appendix, Table~\ref{tab:notation} summarizes all notations.

\begin{tabular}{@{}l@{\hspace*{0.5em}}l@{\hspace*{0.5em}}l@{\hspace*{0.5em}}l@{\hspace*{0.5em}}l@{\hspace*{0.5em}}l@{}}
\toprule
Notation & $P$ & $\mathbb{P}$ & $\mathcal{P}$ & $\conasrt{P}$ & $\absasrt{P}$ \\
Meaning & Unary assertion & Binary assertion & Program-as-resource assertion & Low-level assertion & High-level assertion \\
\bottomrule
\end{tabular}}, as well as the high-level program $\absc$, into program-as-resource assertions over triples $(\const, \absst, \absc)$.
\begin{definition}[Assertion Lifting]Any High-level program $\absc_0$, binary assertion $\mathbb{P}$, unary assertions $\conasrt{P}$ and $\absasrt{P}$ (about low-level and high-level program states respectively) are lifted as follows:
\begin{enumerate}
\item $(\const,\absst,\absc) \models \hspeccolor{\absc_0}$ \text{iff.} $\absc = \absc_0$
\item $(\const,\absst,\absc) \models \mathbb{P}$ \text{iff.} $(\const,\absst) \models \mathbb{P}$
\item $(\const,\absst,\absc) \models \lasrt{\conasrt{P}}$ \text{iff.} $\const \models \conasrt{P}$
\item $(\const,\absst,\absc) \models \hasrt{\absasrt{P}}$ \text{iff.} $\absst \models \absasrt{P}$
\end{enumerate}
\end{definition}
\noindent 
We include unary assertion liftings for completeness, though they become more relevant later when we discuss proof rules.
With these lifted assertions, the refinement between $\conc$ and $\absc$ can be described by the triple  $\reltriplenofont{\mathbb{P} \land \hspeccolor{\absc}}{\conc}{\mathbb{Q} \land \hspeccolor{\cskip}}$ \footnote{While program-as-resource assertions are common in separation logic frameworks, $\hspeccolor{-}$ can adapt to a separating context.}, where  $\cskip$ denotes that the high-level program has terminated.
\begin{figure}[ht]
  \centering
 \begin{tabular}{@{\hspace{1em}}p{0.30\linewidth}|@{\hspace{1em}}p{0.25\linewidth}|@{\hspace{1em}}p{0.35\linewidth}@{}}
\begin{tabular}{@{}l@{}}
  \textcolor{gray}{\texttt{bit\_mask}} \\
        \begin{lstlisting}[aboveskip=0\baselineskip,mathescape=true, language=C, 
        xleftmargin=0.2cm,morekeywords={stack, list}]
$\convar{x}$ $\coloneq$ 0;
$\convar{x}$ $\coloneq$ $\convar{x}$ | (1 << $a_0$);
$\convar{x}$ $\coloneq$ $\convar{x}$ | (1 << $a_1$);
    \end{lstlisting}
  \end{tabular} & 
    \begin{tabular}{@{}l@{}}
  \textcolor{gray}{\texttt{set\_union}} \\
   \begin{lstlisting}[aboveskip=0\baselineskip,mathescape=true,language=C, 
    xleftmargin=0.5cm,morekeywords={stack, list}]
$\absvar{s}$ $\coloneq$ { };
$\absvar{s}$ $\coloneq$ $\absvar{s}$ $\cup$ {$a_0$};
$\absvar{s}$ $\coloneq$ $\absvar{s}$ $\cup$ {$a_1$};
    \end{lstlisting}
  \end{tabular}
  & 
    \begin{tabular}{@{}l@{}}
  \textcolor{gray}{\text{A relational triple}} \\
   \begin{lstlisting}[aboveskip=0\baselineskip,mathescape=true,language=C, 
    xleftmargin=0.5cm,morekeywords={stack, list}]
$\langle \hspeccolor{\texttt{set\_union}}  \rangle$ 
$\quad\texttt{bit\_mask}$
$\langle \convar{x} = \Sigma_{a\in \absvar{s}} 2^a \land \hspeccolor{\cskip} \rangle$ 
    \end{lstlisting}
  \end{tabular}\\
  \end{tabular}
  \caption{The bitmask/set example: a low-level program uses bitmask operations to record constants, a high-level program uses set union to build a set, and a relational triple expresses their refinement.}
  \label{fig:toyprog} 
\end{figure}
Consider the \bitmaskexample in \figref{fig:toyprog}: a low-level bitmask program that records constants $a_0$ and $a_1$, and a high-level set program with similar functionality.
The refinement is then expressed by the triple $\reltriplenofont{\hspeccolor{\texttt{set\_union}}}{\texttt{bit\_mask}}{ \convar{x} = \Sigma_{a\in \absvar{s}} 2^a \land \hspeccolor{\cskip}}$, where the postcondition relates the two program variables $\convar{x}$ and $\absvar{s}$, and the precondition is trivial (i.e., $\ptrue$) and therefore omitted.
This example serves as a running illustration throughout the section.

On the other hand, standard Hoare logic commonly reasons about the safety property of one program, and the validity for standard Hoare triples  is defined as follows:
\begin{definition}[Standard Hoare Triples]Standard Hoare triple $\unarytriple{P}{c}{Q}$ is $\forall$-valid \footnote{Standard Hoare triples are $\forall$-valid for nondeterministic programs and we use "$\forall$-valid" to distinguish them from the angelic  triple in rule \hyperref[intro:absfocus]{\textsc{High-Focus}}.} if, for any initial state $\sigma_1 \models P$ and for any final state $\sigma_2$ after executing program $c$, we have $\sigma_2 \models Q$.
  \label{def:ht}
\end{definition}

\noindent 
Therefore in comparison a standard Hoare judgment presents a $\forall$-structure, while to illustrate program refinement a relational judgment presents a $\forall\exists$ pattern as shown in \defref{def:rhtintro}.



Despite this semantic difference, it is known that $\forall\exists$  relational Hoare logic has similar proof rules to standard Hoare logic's rules, as depicted  in \figref{fig:rules}.
In Transfinite Iris \cite{Transfinite_Iris}
the rule \hyperref[intro:relseq]{\textsc{Rel-Seq}} 
\footnote{In their paper,  the programming language is a higher-order functional language. We depict these rules for an imperative language to explicitly compare with standard Hoare logic.} 
can be derived to evaluate the currently focused statement in a manner similar to the reasoning employed by the rule \hyperref[intro:seq]{\textsc{Seq}}  in standard Hoare logic.
Besides,  relational Hoare logic frameworks support the typical rule \hyperref[intro:rhtex]{\textsc{Rel-ExIntro}} to introduce the existential variables in preconditions, which corresponds to the rule \hyperref[intro:ex]{\textsc{ExIntro}}.
In real program verification,  relational proofs often proceed by focusing on and evaluating one side—either the low-level or high-level program—at a time~\cite{DBLP:journals/pacmpl/GaherSSJDKKD22}.
To support such reasoning, many frameworks represent assertions in a \emph{decomposed form} that separates the components related to the low-level state, high-level state, and high-level program~\cite{DBLP:journals/corr/abs-2006-13635, DBLP:conf/cpp/VindumFB22}:
\[
\exists\, \vec{a}.\, B(\vec{a}) \land \lasrt{\conasrt{P}(\vec{a})} \land \hasrt{\absasrt{P}(\vec{a})} \land \hspeccolor{\absc}
\]
Here, $\vec{a}$ represents a list of existential logical variables, and $B$ is a pure logical assertion used to constrain $\vec{a}$. 
The unary low-level assertion $\conasrt{P}$ and high-level assertion $\absasrt{P}$ are both lifted to program-as-resource assertions.
For \bitmaskexample, the relational triple is then rewritten into
\begin{equation}
    \reltriplenofont{\hspeccolor{\texttt{set\_union}}}{\texttt{bit\_mask}}{\exists\,l.\, \lasrt{\convar{x} = \sum\nolimits_{a\in l}2^a} \land \hasrt{\absvar{s} = l} \land \hspeccolor{\cskip} }. 
    \label{eq:toy_decomposed_reltriple}
\end{equation}
Decomposed assertions enable focusing rules, such as  \hyperref[intro:absfocus]{\textsc{High-Focus}} to angelically evaluate the currently focused high-level statement. Here the triple $\angelvdash \unarytriple{\absasrt{P}}{\absc_1}{\absasrt{R}}$ represents that if an initial state $\absst_1 \models \absasrt{P}$, then there exists a final state $\absst_2$ after executing program $c$ such that $\absst_2 \models \absasrt{R}$.
As shown in \figref{fig:introexample_relproof}, the high-level assignment \inlinecodes{$\absvar{s}\coloneq\,$ \{ \}} can be independently evaluated, reducing the precondition to $\hasrt{\absvar{s} = \emptyset} \land \hspeccolor{\inlinecodes{$\absvar{s}\,$ $\coloneq$ $\,\absvar{s}\,$ $\,\cup\,$ \{$a_0$\};}\cdots}$.
In view of the low-level program, the rule \hyperref[intro:absfocus]{\textsc{High-Focus}} weakens the precondition, similar to the rule \hyperref[intro:conseq]{\textsc{Conseq-pre}} in standard Hoare logic.

\begin{figure*}[t]
    \centering
    \begin{tabular}{c|l}
    \begin{minipage}{0.55\textwidth}
  \begin{mathpar}
  \inferrule[\label{intro:relseq}Rel-Seq]
    { \reltriple{P}{\conc_1}{R} \\ \reltriple{R}{\conc_2}{Q} }
    { \reltriple{P}{\conc_1; \conc_2}{Q} }
  \\
  \inferrule[\label{intro:rhtex}Rel-ExIntro]
    { \forall a. \,\, \reltriplenofont{\mathcal{P}(a)}{\conc}{\mathcal{Q}} }
    {\reltriplenofont{\exists a. \mathcal{P}(a)}{\conc}{\mathcal{Q}}  }
      \\
  \inferrule[\label{intro:absfocus}High-Focus]
    { \angelvdash \unarytriple{\absasrt{P}}{\absc_1}{\absasrt{R}}\\  \reltriplenofont{\lasrt{\conasrt{F}} \land \hasrt{\absasrt{R}} \land \hspeccolor{\absc_2}}{\conc}{\mathcal{Q}} }
    {  \reltriplenofont{\lasrt{\conasrt{F}} \land  \hasrt{\absasrt{P}} \land \hspeccolor{\absc_1;\absc_2}}{\conc}{\mathcal{Q}} }
  \end{mathpar}
    \end{minipage}
    &
    \begin{minipage}{0.38\textwidth}
  \begin{mathpar}
  \inferrule[\label{intro:seq}Seq]{\demovdash \{P\} \ c_1 \ \{R\} \\ \demovdash \{R\} \ c_2 \ \{Q\} }
  {\demovdash \{P\} \ c_1;c_2 \ \{Q\} }
  \\
  \inferrule[\label{intro:ex}ExIntro]{ \forall a. \,\, \demovdash \unarytriple{P(a)}{c}{Q} }
  {\demovdash \unarytriple{\exists a. P(a)}{c}{Q} } 
  \\
  \inferrule[\label{intro:conseq}Conseq-Pre]{P \Rightarrow R \\ \demovdash \{R\} \ c \ \{Q\} }
  {\demovdash \{P\} \ c\ \{Q\} } 
  \end{mathpar}
    \end{minipage}
    \end{tabular}
    \caption{Proof Rules in Relational and Standard Hoare logic} 
    \label{fig:rules}
  \end{figure*} 

\subsection{Preview of Our Encoding Theory: From Relational to Standard Reasoning}
This paper presents an encoding theory that defines assertion encoding $\encasrt{-}_X$ to transform program-as-resource assertions into unary assertions. 
This encoding guarantees that the validity of encoded standard Hoare triples is equivalent to the validity of corresponding relational Hoare triples.
\begin{theorem}[\label{lem:enc}Encoding Relational Triples] For any low-level statement $\conc$ and program-as-resource assertions $\mathcal{P}$ and $\mathcal{Q}$, the relational Hoare triple ${\langle \mathcal{P} \rangle \ \conc \ \langle \mathcal{Q}\rangle}$ is valid if and only if, given any assertion $X$ on high-level states, the standard Hoare triple $\{ \encasrt{\mathcal{P}}_X \}  \ \conc \ \{ \encasrt{\mathcal{Q}}_X \}$ is $\forall$-valid.
 \end{theorem}
\noindent Notably, the variable $X$ serves only as a necessary placeholder. In subsequent discussions, we will see that it does not play any actual role in the encoded relational rules and proofs.

\begin{figure*}[t]
    \centering
        \begin{subfigure}[b]{0.47\textwidth}
        \begin{lstlisting}[aboveskip= 0\baselineskip,mathescape=true,language=C, basicstyle=\footnotesize, escapechar=\&]
// $\langle \hspeccolor{\absvar{s} \coloneq \{\,\}; \absvar{s} \coloneq \absvar{s} \cup \{a_0\};  ;  \ldots} \rangle$ 
// $\color{gray}\text{high-level step}$
// $\langle \hasrt{\absvar{s}= \emptyset} \land \hspeccolor{ \absvar{s} \coloneq \absvar{s} \cup \{a_0\};  \ldots} \rangle$  
$\convar{x}$ $\coloneq$ 0;
// $\color{gray}\text{low-level step}$
// $\langle  \lasrt{\convar{x} = 0} \land \hasrt{\absvar{s}= \emptyset} \land \hspeccolor{ \absvar{s} \coloneq \absvar{s} \cup \{a_0\}; \ldots}  \rangle$ 
$\ldots$
          \end{lstlisting}
          \caption{Part of Proof based on relational Hoare triples}
    \label{fig:introexample_relproof}
   \end{subfigure}
   \hfill
      \begin{subfigure}[b]{0.48\textwidth}
        \begin{lstlisting}[aboveskip= 0\baselineskip,mathescape=true,language=C, basicstyle=\footnotesize, escapechar=\&]
// $\{ \needExecX{\textsf{True}}{\absvar{s} \coloneq \{\,\}; \absvar{s} \coloneq \absvar{s} \cup \{a_0\}; \ldots}{X} \}$ 
// $\color{gray}\text{consequence rule to update the predicate}$ 
// $\{ \needExecX{\absvar{s}= \emptyset}{ \absvar{s} \coloneq \absvar{s} \cup \{a_0\};  \ldots}{X} \}$ 
$\convar{x}$ $\coloneq$ 0;
// $\color{gray}\text{sequencing rule to evaluate low-level assignment}$
// $\{\needExecX{\absvar{s}= \emptyset}{ \absvar{s} \coloneq \absvar{s} \cup \{a_0\};  \ldots}{X} \land {\convar{x} = 0} \}$ 
$\ldots$
          \end{lstlisting}
  \caption{Part of Proof based on standard Hoare triples}
    \label{fig:introexample_stdproof}
    \end{subfigure}
        \caption{Relational and standard proofs for the bitmask/set example.}
\end{figure*}

To apply this encoding in practice, our encoding theory also introduces a predicate $\needExecX{\absasrt{P}}{\absc}{X}$ to transform decomposed program-as-resource assertions into decomposed unary assertions:
\[
\encasrt{ \exists \vec{a}.\, B(\vec{a}) \wedge \lasrt{\conasrt{P}} \wedge \hasrt{\absasrt{P}} \wedge \hspeccolor{\absc} }_X
\iff
\exists \vec{a}.\, B(\vec{a}) \wedge \conasrt{P}(\vec{a}) \wedge \needExecX{\absasrt{P}}{\absc}{X}
\]
\noindent 
This equivalence shows that the execution constraints of the high-level program can be encoded into a pure logical predicate $ \needExecX{\absasrt{P}}{\absc}{X}$, while preserving the low-level assertion unchanged.
As a result, with predicate $\needExecX{\absasrt{P}}{\absc}{X}$, standard Hoare logic can not only express program refinement but also support refinement proofs using standard reasoning rules.
For the \bitmaskexample,
Our encoding transforms relational triple \eqref{eq:toy_decomposed_reltriple}
into a standard triple:
\begin{equation}
    \unarytriple{\needExecX{\ptrue}{\texttt{set\_union}}{X} }{\texttt{bit\_mask}}{\exists\,l.\, \needExecX{\absvar{s}= l}{\cskip}{X}\land {\convar{x} = \sum\nolimits_{a\in l}2^a}}.
    \label{eq:toy_decomposed_triple}
\end{equation}
\figref{fig:introexample_stdproof} shows part of the proof using standard Hoare logic, where the proof proceeds using standard consequence and sequencing rules.
The first step of the proof weakens the precondition by performing an update of the predicate $\needExecX{\absasrt{P}}{\absc}{X}$:
\begin{mathpar}
  \inferrule
      {\angelvdash \unarytriple{\absasrt{P_1}}{\absc_1}{\absasrt{P_2}} }
      { \needExecX{\absasrt{P_1}}{\absc_1;\absc_2}{X} \Rightarrow \needExecX{\absasrt{P_2}}{\absc_2}{X} }
\end{mathpar}
This update reflects a single step of high-level evaluation.
In this case, $\absc_1$ is the assignment \inlinecodes{$\absvar{s} \coloneq\, \{\}$}.
The second step applies the standard rule \hyperref[intro:seq]{Seq}, and evaluates the low-level assignment \inlinecodes{$\convar{x}\coloneq\,$ 0}.
At this point, the precondition includes the updated predicate $\needExecX{\absvar{s}= \emptyset}{ \absvar{s} \coloneq \absvar{s} \cup \{a_0\};  \ldots}{X}$, describing the remaining behavior of the high-level program. Since this predicate does not mention or depend on the low-level state, it can be freely carried across the evaluation of the assignment.
The remaining proof proceeds by alternating updates to the predicate $\needExecX{\absasrt{P}}{\absc}{X}$ and standard reasoning steps on the low-level program, similar to the first two steps.

This example shows how standard proof rules can prove program refinement.
Importantly, our encoding equips standard Hoare logic with comparable proof power to relational Hoare logic.
The theoretical justification lies in the fact that relational proof rules can be encoded into standard Hoare rules.
For instance, independently evaluating the high-level program via rules like \hyperref[intro:absfocus]{High-Focus} corresponds to a consequence step in standard logic, via updating the predicate $\needExecX{\absasrt{P}}{\absc}{X}$:
\begin{mathpar}
\inferrule{\needExecX{\absasrt{P}}{\absc_1}{X} \Rightarrow \needExecX{\absasrt{R}}{\absc_2}{X} \\ \demovdash \{ \needExecX{\absasrt{R}}{\absc_2}{X} \land\conasrt{F}\} \ \conc \ \{ \encasrt{\mathcal{Q}}_X\} }
{\demovdash \{ \needExecX{\absasrt{P}}{\absc_1}{X} \land \conasrt{F}\} \ \conc \ \{ \encasrt{\mathcal{Q}}_X\} } 
\end{mathpar}
Then applying the rule \hyperref[intro:absfocus]{High-Focus} in relational Hoare logic corresponds to applying the consequence rule in standard Hoare logic. 
\secref{sec:encRule} summarizes how core relational rules correspond to derivable rules in standard Hoare logic.
These correspondences ensure that with our encoding, standard Hoare logic is expressive enough to prove relational triples over decomposed assertions.

Moreover, relational Hoare logic frameworks support two types of vertical composition rules: composing a refinement proof with the functional correctness proof of the high-level program, and composing two refinement proofs.
\begin{mathpar}
  \inferrule[\label{rule:vc}\scshape VC-FC]
  {\reltriplenofont{ 
    \mathbb{P}\land \hspeccolor{\absc}
  }{\conc}{\mathbb{Q} \land \hspeccolor{\cskip}} \\ 
  \demovdash \{  \absasrt{P} \} \ \absc \ \{\absasrt{Q}\} 
  }
  { \demovdash \{ \mathbb{P} \odot \absasrt{P}  \} \  \conc \ \{\mathbb{Q} \odot \absasrt{Q} \} } \\
  \inferrule[\label{rule:vcr}\scshape VC-Refine]
  {
  \reltriplenofont{ 
    \mathbb{P}_1 \land \hspeccolor{c_2}
  }{c_1}{\mathbb{Q}_1 \land \hspeccolor{\cskip}}  \\ 
  \reltriplenofont{ 
      \mathbb{P}_2 \land \hspeccolor{c_3}
    }{c_2}{\mathbb{Q}_2 \land \hspeccolor{\cskip}}
  }
  {  \reltriplenofont{ 
    \mathbb{P}_1 \circ \mathbb{P}_2  \land \hspeccolor{c_3}
  }{c_1}{\mathbb{Q}_1 \circ \mathbb{Q}_2 \land \hspeccolor{\cskip}} }   
\end{mathpar}
\noindent Here, the operator $\odot$ denotes linking a binary assertion with a unary assertion, and the operator $\circ$ denotes linking two binary assertions. They are formally defined in \secref{sec:encRule}.
Under our encoding,  both rules can be derived using the consequence rule and rule \hyperref[intro:ex]{\textsc{ExIntro}} in standard Hoare logic.
The central idea of our encoding theory is simple yet powerful: leveraging logical variables to embed $\forall\exists$ patterns in preconditions and postconditions.
To clarify, our goal is not to develop a new program logic.
Rather,  by introducing a predicate $\needExecX{\absasrt{P}}{\absc}{X}$, we enable standard Hoare logic to provide comparable reasoning power for $\forall\exists$ relational proofs. 
We believe that our encoding theory will allow provers to employ the extensive body of work and tools developed for standard Hoare logic to verify relational properties.

Our main contributions are as follows:
\begin{itemize}
  \item We introduce the first encoding theory (\theoref{lem:enc}) to reduce the $\forall\exists$ relational Hoare logic to standard Hoare logic. 
  Our encoding theory relies on a relaxed validity of relational Hoare triples based on configuration refinement (\secref{sec:relax}).
  Importantly, we only encode the $\forall\exists$ pattern in assertions (\secref{sec:enc}).
  Since a relational Hoare triple is encoded into a standard Hoare triple of the exact low-level program involved, this triple can be used in compositional reasoning with other standard Hoare logic judgments.
  \item We present a syntactic encoding of decomposed program-as-resource assertions into unary low-level assertions (\secref{sec:encdecompose}).
  This encoding introduces the execution predicate $\needExecX{\absasrt{P}}{\absc}{X}$ to capture high-level behavior as a purely logical condition.
  The result enables direct application of our encoding for relational reasoning in standard Hoare logic.
  \item Our encoding fully preserves the original reasoning capabilities of relational Hoare logic.
  We illustrate how core inference rules in relational Hoare logic can be encoded into standard Hoare rules and present proof rules for the execution predicate $\needExecX{\absasrt{P}}{\absc}{X}$ (\secref{sec:encRule}).
  \item We demonstrate the expressiveness and practicality of our approach through case studies, including merge sort, binary search trees, depth-first search, and the Knuth–Morris–Pratt (KMP) algorithm.
In particular, \secref{sec:casestudy} presents a detailed proof of the merging algorithm.
  \item Our results are formalized and machine-checked in Rocq, including all meta theorems and case studies.
  Our encoding theory can be extended to support additional language features, including function calls, undefined behavior, and separation logic.
  These extensions are presented in Appendix~\ref{app:ext} and have been formalized in Rocq.
\end{itemize}


\section{\label{sec:background} Background: Relational Hoare Logic with Programs as Resources} 
The relational framework proposed by Turon et al. \cite{DBLP:conf/popl/TuronTABD13} and other works \cite{DBLP:conf/popl/LiangF16,DBLP:conf/pldi/LiangF13,DBLP:conf/esop/TassarottiJ017,DBLP:journals/corr/abs-2006-13635,Transfinite_Iris,10.1145/2500365.2500600} share a notable innovation: treating high-level programs as resources.
In this section, we provide a brief overview of this approach and illustrate how to prove program refinement based on it.
To focus on the core ideas of both relational and standard Hoare logic, program assertions are defined semantically as subsets of states before \secref{sec:syntactic}.

\subsection{High-level Programs as Resources}
Treating high-level programs as resources involves extending a binary assertion $\mathbb{P} \subseteq \conSigma \times \absSigma$ with a high-level program $\absc \in  \text{Prog}^{\text{H}}$.
Here, $\conSigma$ and $\absSigma$ represent the set of low-level program states and the set of high-level program states, respectively.
Then the main judgment based on this approach is the relational Hoare triple $\reltriple{P}{\conc}{Q}$, whose definition is restated as follows:

\begin{customdef}{1}[Relational Hoare Triples] The relational Hoare triple $\langle \mathcal{P} \rangle \ \conc \ \langle \mathcal{Q} \rangle$ is valid if for any $(\const_1, \absst_1, \absc_1) \models \mathcal{P}$ and any $\const_2$ such that $(\const_1, \const_2) \in \nrmdeno{\conc}$, there exist $\absst_2$ and $\absc_2$ such that $(\absst_1, \absc_1) \rightarrow^* (\absst_2, \absc_2)$ and $(\const_2, \absst_2, \absc_2) \models \mathcal{Q}$.
\label{def:reltriple_small}
\end{customdef}
\noindent Here $\nrmdeno{-}$ represents the normal evaluation of denotational semantics.
For any program statement $c \in \text{Prog}^{\text{L}}$,  $(\const_1,\const_2) \in \nrmdeno{c}$ if the execution of $c$ from initial state $\const_1$ may terminate at state $\const_2$.
Besides,  $\rightarrow^*$ represents multi-step transition based on small-step semantics and  $(\absc_1, \absst_1) \rightarrow^* (\absc_2, \absst_2)$
indicates that $(\absc_1, \absst_1)$ reduces to $(\absc_2, \absst_2)$ in zero or more steps.
If the small-step semantics and denotational semantics both characterize the high-level programming language, it is expected that they would be consistent with each other.
Specifically, the multi-step transitions $\rightarrow^*$ and the 
denotational semantics function $\nrmdeno{-}$ both describe a large reduction:
\begin{proposition}\label{pro:sd}  $(\absst_1, \absc) \rightarrow^* (\absst_2, \cskip)$ is equivalent to $(\absst_1, \absst_2) \in \nrmdeno{\absc}$.
\end{proposition}
\noindent Then the relational Hoare triple $ \langle  \mathbb{P} \land \hspeccolor{\absc} \rangle \ \conc \ \langle \mathbb{Q} \land \hspeccolor{\cskip} \rangle $ represents that the low-level program $\conc$ refines the high-level program $\absc$, provided that the assertions $\mathbb{P}$ and $\mathbb{Q}$ meaningfully relate their respective initial and final states. 
If $\mathbb{P}$ and $\mathbb{Q}$ are both false, the triple is trivially valid but vacuous, and thus does not constitute a real refinement.

\subsection{\label{sec:rhtproof}Proofs Based on Relational Hoare triples}
\figref{fig:relrules} shows core proof rules of relational Hoare logic for sequential imperative programs.
The first two rules are focusing rules over decomposed assertions, used to evaluate either the low-level program with standard Hoare triples or the high-level program with angelic Hoare triples. The third rule is a loop rule for low-level programs.
These rules support the following three key examples: the refinement proof for two nondeterministic programs, the refinement proof for two simple sequential programs in \bitmaskexample, and the refinement proof for programs involving loops. 
To distinguish between program variables and logical variables,  program variables are written in \textsf{sans-serif} font and logical variables are written in the standard $italic$ font.

\begin{figure}[t]
    \centering
    \small
    \begin{mathpar}
      \inferrule[\label{confocus}\textsc{Low-Focus}]
  { \demovdash \{\conasrt{P}\} \ \conc_1 \ \{\conasrt{R}\}  \\ \reltriplenofont{\lasrt{\conasrt{R}} \land \hasrt{\absasrt{P}} \land \hspeccolor{\absc}}{\conc_2}{\mathcal{Q}}}
  { \reltriplenofont{\lasrt{\conasrt{P}} \land \hasrt{\absasrt{P}} \land \hspeccolor{\absc}}{\conc_1;\conc_2}{\mathcal{Q}} } \hfill
    \inferrule[\label{absfocus}High-Focus]
    { \angelvdash \unarytriple{\absasrt{P}}{\absc_1}{\absasrt{R}}\\  \reltriplenofont{\lasrt{\conasrt{F}} \land \hasrt{\absasrt{R}} \land \hspeccolor{\absc_2}}{\conc}{\mathcal{Q}} }
    {  \reltriplenofont{\lasrt{\conasrt{F}} \land  \hasrt{\absasrt{P}} \land \hspeccolor{\absc_1;\absc_2}}{\conc}{\mathcal{Q}} } \\
\inferrule[\label{relwh}\textsc{Rel-Wh}]{
    \reltriplenofont{\mathcal{P}\land \lasrt{\conb}}{\conc}{\mathcal{P}}
    }{ \reltriplenofont{\mathcal{P}}{\text{\textsf{ while} } \conb \text{\textsf{ do }} \conc}{\mathcal{P} \land \lasrt{\neg \conb}}}
    \end{mathpar}
    \caption{Core Relational Proof Rules}
    \label{fig:relrules}
\end{figure}

\paragraph{\bfseries Nondeterministic Programs} Consider the following two programs both using the statement $\textsf{nondet}(n, m)$ to nondeterministically select an integer within the range $[n, m]$.

\begin{tabular}{@{\hspace{1em}}p{0.45\linewidth}|@{\hspace{1em}}p{0.45\linewidth}}
\begin{tabular}{@{}l@{}}
        \begin{lstlisting}[aboveskip=0\baselineskip,mathescape=true, language=C, 
        xleftmargin=0.5cm,morekeywords={stack, list}]
  $\convar{x}$ $\coloneq$ $\textsf{nondet}$(0,1); 
    \end{lstlisting}
  \end{tabular} & 
    \begin{tabular}{@{}l@{}}
   \begin{lstlisting}[aboveskip=0\baselineskip,mathescape=true,language=C, 
    xleftmargin=0.5cm,morekeywords={stack, list}]
  $\absvar{y}$ $\coloneq$ $\textsf{nondet}$(0,2); 
    \end{lstlisting}
  \end{tabular}\\
  \end{tabular}
\begin{example}\label{exam:nondet}
The low-level program  $\convar{x} \coloneq \textsf{nondet}(0, 1)$ assigns either $0$ or $1$ to $\convar{x}$. In contrast, the high-level program (on the right) allows $\absvar{y}$ to nondeterministically take on any value of $0$, $1$, or $2$.
We aim to prove the relational Hoare triple $\reltriplenofont{\hspeccolor{\inlinecodes{$\absvar{y}\coloneq\,$ $\textsf{ nondet}$(0,2)}}}{{\inlinecodes{$\convar{x}\coloneq\,$ $\textsf{ nondet}$(0,1)}}}{ \convar{x} = \absvar{y} \land \hspeccolor{\cskip}}$.
\end{example}

\noindent 
The relational proof for this example is presented as follows, with annotations in the low-level program.
We first focus on the low-level program and demonically evaluate the assignment $\convar{x} \coloneq \textsf{nondet}(0,1)$, where the result value is nondeterministic $0$ or $1$.
Then, we can apply the \hyperref[absfocus]{\textsc{High-Focus}} rule to angelically assign the value of $\convar{x}$  to $\absvar{y}$ in the high-level program.
This example demonstrates that, for any demonic execution of the low-level program, we can leverage the \hyperref[absfocus]{\textsc{High-Focus}} rule to ensure that the high-level program captures its behavior.

  \begin{lstlisting}[aboveskip= 0\baselineskip,mathescape=true,language=C,basicstyle=\normalfont\footnotesize, escapechar=\&]
// $\langle \hspeccolor{\absvar{y} \coloneq \textsf{nondet}(0,2)} \rangle$ 
$\convar{x}\coloneq$ $\textsf{nondet}$(0,1);
// $\color{gray}\text{low-level step}$ $\langle  \exists n. n \in \{0, 1\} \land \lasrt{\convar{x} = n} \land  \hspeccolor{\absvar{y} \coloneq \textsf{nondet}(0,2)} \rangle$ 
// $\color{gray}\text{high-level step}$ $\langle n \in \{0, 1\} \land \lasrt{\convar{x} = n} \land  \hasrt{\absvar{y} = n} \land \hspeccolor{\cskip} \rangle$ 
  \end{lstlisting}

\paragraph{\bfseries Simple Sequential Programs}
Consider the  \bitmaskexample introduced in \secref{sec:intro}, where the low-level program uses bitwise OR operations to set specific bits in program variable $\convar{x}$ while the high-level program directly builds a set $\absvar{s}$ using set union.
We need  to prove the triple $\reltriplenofont{\hspeccolor{\texttt{set\_union}}}{\texttt{bit\_mask}}{ \convar{x} = \Sigma_{a\in \absvar{s}} 2^a \land \hspeccolor{\cskip}}$.
Part of the relational proof for this example is shown in \figref{fig:introexample_relproof}. 
The proof begins with a high-level step, applying the rule \hyperref[absfocus]{\textsc{High-Focus}} to independently focus on and evaluate the high-level assignment $\absvar{s}$ = $\{$ $\}$. 
Similarly, the proof proceeds with a low-level step, applying the rule \hyperref[confocus]{\textsc{Low-Focus}} to focus on and evaluate the low-level assignment $\convar{x}$ = 0.
Next, we observe that the remaining statements in the low-level and high-level programs correspond to each other, and each corresponding pair can be evaluated 
by applying both the rule \hyperref[absfocus]{\textsc{High-Focus}} and the rule \hyperref[confocus]{\textsc{Low-Focus}}.
Finally, the resulting postcondition ${ \lasrt{\convar{x} = 2^{a_0} + 2^{a_1}} \land \hasrt{\absvar{s}= \{a_0,a_1\}} \land \hspeccolor{\cskip}}$ implies the desired relational postcondition.
This example illustrates that singleton statements, such as an assignment statement, can be independently evaluated on one side through focusing rules \hyperref[confocus]{\textsc{Low-Focus}} and  \hyperref[absfocus]{\textsc{High-Focus}}.

\paragraph{\bfseries Loops}
In standard Hoare logic, the key to reasoning about while loops is to find a suitable loop invariant, and the same is true in relational Hoare logic.
Typical relational Hoare logics will provide a while rule like \hyperref[relwh]{\textsc{Rel-Wh}},
which is analogous to the while rule in standard Hoare logic.
By treating high-level programs as resources, a loop invariant should specify the high-level program that corresponds to the remaining iterations of the low-level program's loops.
\begin{example}\label{exam:loop}
When additional items need to be recorded in  \bitmaskexample, we often use loops.
In this case, programs iterate through the items in a constant array until index $8$. 
In each iteration, the two programs add the corresponding item in arrays to the collection representation ($\convar{x}$ or $\absvar{s}$).
The goal is to prove the triple $\reltriplenofont{\hspeccolor{\texttt{set\_union\_loop}}}{\texttt{bit\_mask\_loop}}{\convar{x} = \Sigma_{a\in \absvar{s}} 2^a \land \hspeccolor{\cskip}}$ 
to express that after executing the two programs, variable $\convar{x}$ encodes the elements in set $\absvar{s}$ as a sum of powers of 2.

  \begin{tabular}{@{\hspace{1em}}p{0.45\linewidth}|@{\hspace{1em}}p{0.45\linewidth}}
\begin{tabular}{@{}l@{}}
  \textcolor{gray}{\texttt{bit\_mask\_loop}} \\
  \begin{lstlisting}[aboveskip= 0\baselineskip,mathescape=true,language=C, xleftmargin=0.5cm,morekeywords={stack, list}, basicstyle=\normalfont]
$\convar{x}$ = 0;
$\convar{i}$ = 0;
while ($\convar{i}$ < 8){
  $\convar{x}$ = $\convar{x}$ | (1 << $a$[$\convar{i}$]);
  $\convar{i}$ = $\convar{i}$ + 1; }
    \end{lstlisting}
\end{tabular}
  &
\begin{tabular}{@{}l@{}}
  \textcolor{gray}{\texttt{set\_union\_loop}} \\
 \begin{lstlisting}[aboveskip= 0\baselineskip,mathescape=true, language=C, xleftmargin=0.5cm,morekeywords={stack, list}, basicstyle=\normalfont]
$\absvar{s}$ = { };
$\absvar{j}$ = 0;
while ($\absvar{j}$ < 8){
  $\absvar{s}$ = $\absvar{s}$ $\cup$ { $a$[$\absvar{j}$] };
  $\absvar{j}$ = $\absvar{j}$ + 1; }                   
   \end{lstlisting}
\end{tabular} \\
  \end{tabular}
\end{example}

\noindent Obviously, one iteration of the low-level program's loop corresponds to one iteration of the high-level program's loop.
Regardless of which iteration the low-level program is on, the high-level program that corresponds to the remaining iterations is always the while statement. 
Then we can use the following loop invariant to prove this example:
\begin{equation}
  \exists \, l\, n.\  \lasrt{\convar{x} = \sum\nolimits_{a\in l} 2^a \land \convar{i} = n } \land \hasrt{\absvar{s}= l \land \absvar{j} = n} \land \hspeccolor{\cwhile{\absvar{j}<8}{\ldots}}
  \label{eq:toyinv}
\end{equation} 
\noindent The detailed proof  is provided in Appendix \ref{app:example}.

In the previous examples, the low-level programs are structurally aligned with the high-level programs, enabling smooth verification. 
When structural alignment is absent, the programs-as-resources approach is still effective in handling such cases.
For instance, the following two variants of the low-level program in Example \ref{exam:loop} also refine the high-level program of this example. 
Here one variant extracts the first step from the loop, and the other executes two steps per iteration instead of one.
The proofs  can be found in Appendix \ref{app:example}.
\begin{example}\label{exam:3}
   The left low-level program performs initialization steps before executing  the loop, and the right low-level program executes two assignments in each loop iteration. 

    \begin{tabular}{@{\hspace{1em}}p{0.45\linewidth}|@{\hspace{1em}}p{0.45\linewidth}}
\begin{tabular}{@{}l@{}}
    \begin{lstlisting}[aboveskip= 0\baselineskip,mathescape=true, language=C, xleftmargin=0.5cm,morekeywords={stack, list}, basicstyle=\normalfont]
$\convar{x}$ = 0 | (1 << $a$[0]);
$\convar{i}$ = 1;
while ($\convar{i}$ < 8){
  $\convar{x}$ = $\convar{x}$ | (1 << $a$[$\convar{i}$]);
  $\convar{i}$ = $\convar{i}$ + 1; }
      \end{lstlisting}
\end{tabular}
    &
\begin{tabular}{@{}l@{}}
   \begin{lstlisting}[aboveskip= 0\baselineskip,mathescape=true,language=C, morekeywords={stack, list}, basicstyle=\normalfont]
$\convar{x}$ = 0; $\convar{i}$ = 0;
while ($\convar{i}$ < 8){
  $\convar{x}$ = $\convar{x}$ | (1 << $a$[$\convar{i}$]);
  $\convar{x}$ = $\convar{x}$ | (1 << $a$[$\convar{i}$ + 1]);
  $\convar{i}$ = $\convar{i}$ + 2; }             
     \end{lstlisting}
\end{tabular} \\
    \end{tabular}
\end{example}




Building on the idea of treating high-level programs as resources, relational Hoare logic uses relational Hoare triples as its primary judgments.
Then the relational verification processes feature two key components: one is the independent evaluation of singleton statements through focusing rules  \hyperref[confocus]{\textsc{Low-Focus}} and \hyperref[absfocus]{\textsc{High-Focus}}; the other involves applying rules similar to those in standard Hoare logic for  ``while'' and ``if'' statements.
In this paper, we present an encoding theory that allows relational verification incorporating all these features to be conducted in standard Hoare logic.

\section{\label{sec:relax}Reformulating Validity with Configuration Refinement}
In this paper, we assume that the programming languages are equipped with denotational semantics.
\begin{definition}[Denotational Semantics] For program $c \in \text{Prog}$,  $\nrmdeno{c}$ denotes its denotational semantics,
$(\sigma,\sigma') \in \nrmdeno{c}$ if executing $c$ from initial state $\sigma$ can terminate at state $\sigma'$.
\label{def:nrm}
\end{definition}


Recall from  \defref{def:reltriple_small}, the  multi-step transition $\rightarrow^*$ based on small-step semantics is used to capture the update of the high-level configuration.
In this paper, we propose to relax that update with a configuration refinement relation $\hookrightarrow$, which is defined as follows:

\begin{definition}[Configuration Refinement]
    For any programs $c_1$ $c_2$ and any states $\sigma_1$ $\sigma_2 \in \Sigma$,  configuration $(\sigma_2, c_2)$ refines $(\sigma_1, c_1)$, denoted as $(\sigma_1, c_1) \hookrightarrow  (\sigma_2, c_2)$,  if for any state $\sigma_3$, $(\sigma_{2} ,\sigma_{3}) \in \nrmdeno{c_2}$ implies $(\sigma_{1} ,\sigma_{3}) \in \nrmdeno{c_1}$.
    \label{def:cr}
\end{definition}


\noindent
Here, the configuration refinement is defined based on denotational semantics.
This choice ensures we maintain a consistent semantics  throughout the paper.
Nevertheless, if both small-step and denotational semantics characterize the  high-level language, then by \propref{pro:sd}  one can also give an equivalent small-step definition of configuration refinement.
Concretely, the implication that $(\absst_{2} ,\absst_{3}) \in \nrmdeno{\absc_2}$ implies $(\absst_{1} ,\absst_{3}) \in \nrmdeno{\absc_1}$  can be  replaced by that $(\absst_{2} ,\absc_2) \rightarrow^* (\absst_3, \cskip)$ implies $(\absst_{1} ,\absc_1) \rightarrow^* (\absst_3, \cskip)$.
Compared to the multi-step transition $\rightarrow^*$, configuration refinement $\hookrightarrow$ is more relaxed, as evidenced by the following proposition.
\begin{proposition}
    If $(\absst_1, \absc_1)  \rightarrow^* (\absst_2, \absc_2)$, then $(\absst_1, \absc_1)  \hookrightarrow (\absst_2, \absc_2)$. 
\end{proposition}
\noindent It is worth noting that the reverse implication does not necessarily hold.
For instance, we can show that $(\pvar{x} = 0, \cskip) \hookrightarrow (\pvar{x} = 1, \pvar{x} := \pvar{x} - 1)$, whereas there is no multi-step transition from the left configuration to the right one.

Based on configuration refinement, we adopt an alternative definition  for relational Hoare triples.

\begin{definition}[Alternative Definition of Relational Hoare Triples] The relational Hoare triple $\langle \mathcal{P} \rangle \ \conc \ \langle \mathcal{Q} \rangle$ is valid if for any $(\const_1, \absst_1, \absc_1) \models \mathcal{P}$ and any $\const_2$ such that $(\const_1, \const_2) \in  \nrmdeno{\conc}$, {there exist} $\absst_2$ and $\absc_2$ {such that} $(\absst_1, \absc_1) \hookrightarrow  (\absst_2, \absc_2)$ and $(\const_2, \absst_2, \absc_2) \models \mathcal{Q}$.
      \label{def:drht}
\end{definition}

\noindent Using this definition, the triple  $\reltriplenofont{\mathbb{P} \land \hspeccolor{\absc}}{\conc}{\mathbb{Q} \land \hspeccolor{\cskip}}$ still represents program refinement shown in \figref{fig:refinement} since the following proposition holds.
\begin{proposition}
  $(\absst_1, \absc) \hookrightarrow (\absst_2, \cskip)$ iff. $(\absst_1, \absst_2) \in \nrmdeno{\absc}$
\end{proposition}
\noindent Intuitively, when the high-level program in the postcondition is not $\cskip$, it can be interpreted as a continuation, which represents the remaining abstract behavior that must be completed by the low-level context.
This allows triples to capture not just full refinements, but intermediate steps in the control flow.
Besides, useful proof rules (such as rules in \secref{sec:rhtproof}) remain sound. 
A small benefit of employing this new definition is allowing us to adhere to denotational semantics throughout the paper, thus avoiding the complexity of translating between small-step and denotational semantics when presenting theories.
This coherence also benefits our Rocq formalization to rely on  one semantics as well.
The far greater significance of this definition, however, lies in its decomposition property.
That is, we can equivalently break down $(\absst_1, \absc_1) \hookrightarrow (\absst_2, \absc_2)$ into implications on \textit{weakest (liberal) preconditions} \cite{10.5555/550359} for an arbitrary set of final states. 
Formally:

\begin{theorem}[Decomposition]
    The following two propositions are equivalent:
    \begin{enumerate}[label=(\alph*)]
        \item \label{pro:semred} $(\absst_1, \absc_1) \hookrightarrow (\absst_2, \absc_2)$
        \item \label{pro:safeimply}$\forall (X \subseteq \absSigma). \, \absst_1 \models \weakestpre{\absc_1}{X} \Rightarrow  \absst_2  \models \weakestpre{\absc_2}{X}$
    \end{enumerate}   
    \label{theo:decom}
\end{theorem}
\begin{definition}[Weakest Precondition]For any program $c$ and any set $X\subseteq\Sigma$, which also serves as a postcondition over program states, 
    $\weakestpre{c}{X}$ gives the weakest precondition such that all terminal states resulting from the execution of program $c$ satisfy $X$ and is defined as:
    \begin{align*}
        \sigma \models \weakestpre{c}{X} & \quad \text{ iff. } \quad  \forall \sigma_0. \, (\sigma, \sigma_0) \in \nrmdeno{c} \Rightarrow  \sigma_0 \models X 
    \end{align*}
    \label{def:wp}
\end{definition}

\begin{figure}[t]
    \centering
    \includegraphics[width=0.5\textwidth]{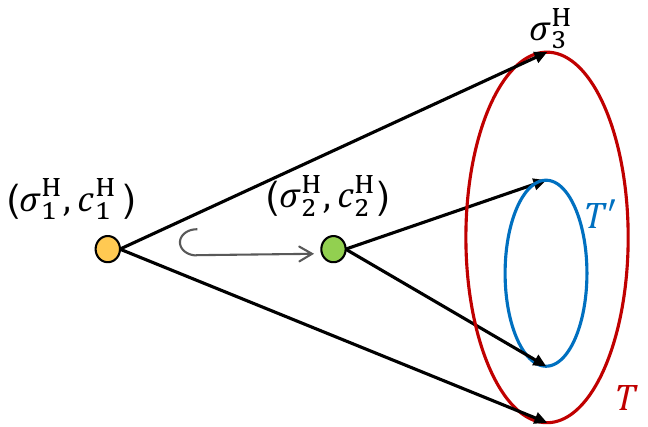}
    \caption{\label{fig:encodeX}State Transition Relation.}
    \end{figure}
\noindent The decomposition theorem (\theoref{theo:decom}) plays a central role in our encoding theory, which we will explain later in \secref{sec:enc}. 
Before proving the theorem, we recall a basic but useful set-theoretic equivalence that will be used in the proof:
\begin{proposition}[A poset fact]\label{prop:9}
 for any set $A$ $B$, $B \subseteq A $ iff. for any set $C$, if $A \subseteq C$ then $B \subseteq C$.
\end{proposition}
\noindent The left-to-right direction of this fact is obvious, while the right-to-left direction can be proved by instantiating $C$ as $A$.
We now prove \theoref{theo:decom}, where
\figref{fig:encodeX} illustrates the intuition.
According to the definition of configuration refinement, \hyperref[pro:semred]{$(a)$} states that any terminal state $\absst$ reachable from high-level programs configuration $(\absst_2, \absc_2)$ must also be reachable from the high-level program configuration $(\absst_1, \absc_1)$.
This is equivalent to the set inclusion $T' \subseteq T$.
Here $T'$ and $T$ represent the terminal state sets for $(\absst_2, \absc_2)$ and $(\absst_1, \absc_1)$ respectively:
\begin{align*}
    &T \defeq  \, \{ \absst_3 \, | \,(\absst_1,\absst_3) \in \nrmdeno{\absc_1} \} \qquad 
    T' \defeq \,  \{ \absst_3 \, |\, (\absst_2,\absst_3) \in \nrmdeno{\absc_2} \}
\end{align*}
According to \propref{prop:9}, the set inclusion $T' \subseteq T$ implies that for any $X \subseteq \absSigma$, if any terminal state in $T$ satisfies $X$, then every terminal state in $T'$ also satisfies $X$, which corresponds to \hyperref[pro:safeimply]{$(b)$}.
To derive \hyperref[pro:safeimply]{$(b)$} $\Rightarrow$ \hyperref[pro:safeimply]{$(a)$}, we employ a similar approach to the right-to-left proof of \propref{prop:9} as follows:
\begin{proposition}\label{prop:10}
    let $X$ be $\{ \absst_3 \, | \,(\absst_1,\absst_3) \in \nrmdeno{\absc_1} \}$, then  proposition \hyperref[pro:safeimply]{$(b)$} implies proposition \hyperref[pro:safeimply]{$(a)$}.
\end{proposition}

\section{\label{sec:enc} Encode Relational Hoare Triples}


In this section, we introduce our encoding theory.
We start by explaining the encoding goal and providing an intuitive overview through a naive attempt (\secref{sec:encgoal}).
Then, we formally present the encoding theory and prove its correctness (\secref{sec:encTH}).



\subsection{\label{sec:encgoal}Intuition Behind the Encoding Theory: a Naive Attempt}
At the core of the encoding theory is to define the assertion encoding $\encasrt{-}$, which transforms a program-as-resource  assertion $\mathcal{P}\subseteq \conSigma \times \absSigma \times \absStmts$ into a unary assertion $\encasrt{\mathcal{P}} \subseteq \conSigma $ and ensures:
\begin{equation*}
    \langle \mathcal{P} \rangle \  \conc \ \langle \mathcal{Q}\rangle \text{ is valid}    \quad \text{iff.} \quad    \ \demovdash \{ \encasrt{\mathcal{P}} \} \ \conc \ \{ \encasrt{ \mathcal{Q} } \}. 
\end{equation*}

\noindent To define the encoding, review \defref{def:drht} which states that
given any initial states $(\const_1, \absst_1, \absc_1)$$\models \mathcal{P}$: 
\begin{itemize}
    \item for any low-level state $\const_2$ such that $(\const_1 ,\const_2) \in \nrmdeno{\conc}$, \textbf{there exist high-level state $\absst_2$ and program $\absc_2$ such that
    $(\absst_1, \absc_1) \hookrightarrow (\absst_2, \absc_2)$ and $ (\const_2 ,\absst_2, \absc_2) \models \mathcal{Q}$}.
\end{itemize}
In standard Hoare logic, the assertions only concern low-level states $\const_1$ and $\const_2$.
If we could remove the angelic part (everything after the exist quantifier) in bold font away from this validity definition, and embed them into the postcondition, then the encoded postcondition will become an assertion over 
low-level states.
Under this transformation, the validity definition corresponds exactly to that of standard Hoare logic.
This leads to a naive encoding for the postcondition:
\begin{align*}
    \encasrt{\relasrt{Q}}(\absst_1, \absc_1) \defeq \lambda \const.\, & \exists \absst_2, \absc_2.\, (\absst_1, \absc_1) \hookrightarrow (\absst_2, \absc_2) \land  (\const, \absst_2, \absc_2) \models \relasrt{Q}.
\end{align*}
Therefore, we have 
    \begin{align*}
        & {\langle \mathcal{P} \rangle \  c \ \langle \mathcal{Q}\rangle}
          \text{ is valid}   \quad \text{iff.}  \quad  \forall \absst \, \absc. \, {\{ \lambda \const. (\const, \absst, \absc) \in \mathcal{P}  \} \ c \ \{ \encasrt{ \mathcal{Q}}(\absst, \absc)  \}} \text{ is valid}
     \end{align*}

\noindent Note that the encoded postcondition $\encasrt{\relasrt{Q}}(\absst, \absc)$ is parameterized by $\absst$ and $\absc$.
The two parameters ensure that the resulting high-level state and program are obtained through a valid configuration refinement from the initial high-level state and program.
However, $\absst$ and $\absc$ depend on the precondition $\mathcal{P}$.
This dependency prevents the naive encoding method from providing a uniform encoding for both preconditions and postconditions.
Without a uniform encoding, we cannot  derive rules from standard Hoare logic to relational Hoare logic.
For example, consider the rule \hyperref[relwh]{\textsc{Rel-Wh}}  for addressing while loops in low-level programs.  
Using the naive encoding, the invariant $\mathcal{P}$ in the precondition and postcondition will be encoded differently.
Then this rule can not be  derived from  the while rule for standard Hoare logic.

Interestingly,  although the naive encoding does not work, it suggests  applying existential quantification to abstract the high-level state and the high-level program from program-as-resource assertions.
Existential quantification provides a straightforward means to encode the $\exists$ component of the $\forall\exists$ pattern in relational Hoare triples into the $\forall$ structure of standard Hoare triples. 
For the $\forall$ component, existential quantification in the precondition exactly corresponds to some universal property, as rule \hyperref[intro:ex]{\textsc{ExIntro}} shows.
This discussion suggests that program-as-resource 
precondition $\mathcal{P}$ and postcondition $\mathcal{Q}$ can be encoded into: 
\begin{itemize}
    \item $ \lambda \const_1. \, \exists \, \absst_1 \absc_1. \,  (\const_1, \absst_1, \absc_1) \models \mathcal{P} \land  \text{some condition}$
    \item $\lambda \const_2. \, \exists \, \absst_2 \absc_2. \,  (\const_2, \absst_2, \absc_2) \models \mathcal{Q} \land  \text{some condition}$
\end{itemize}
\noindent
Here ``some condition'' is independent of low-level states $\const_1$ and $\const_2$.
However, using this encoding form alone leads to an encoded postcondition that becomes disconnected from the encoded precondition. 
As a result, the encoded relational Hoare triples lack key information to ensure configuration refinement $(\absst_1, \absc_1)  \hookrightarrow ( \absst_2, \absc_2) $.
To address this, we need to introduce a necessary placeholder $X$.
This placeholder serves as the connection between the encoded precondition and the encoded postcondition.
Given a placeholder $X$, a program-as-resource assertion $\mathcal{P}$ can be encoded into: 
$$\lambda \const. \, \exists \absst \absc. \,  (\const, \absst, \absc) \models \mathcal{P} \land  \text{some condition}(\absst,\absc,X)$$


\subsection{\label{sec:encTH}The Encoding Theory}

Recall from \theoref{theo:decom}, it suggests that configuration refinement $(\absst_1,\absc_1) \hookrightarrow (\absst_2,\absc_2)$ is equivalent to the requirement that for every subset $X$, if $\absst_1$ satisfies the weakest precondition of $\absc_1$ for $X$, then $\absst_2$ satisfies the weakest precondition of $\absc_2$ for $X$.
Therefore, we can employ the weakest precondition to formalize ``some condition'' and define the assertion encoding.

\begin{definition}[\label{enc:asrtenc}Assertion Encoding] For any program-as-resource assertion $\mathcal{P}$ $\subseteq \conSigma \times \absSigma \times \absStmts$ and subset of high-level states $X \subseteq \absSigma$, define the encoding as
    $$\const \models \encasrt{ \mathcal{P}}_X  \quad \text{ iff. }\quad \exists \absst \absc. \,(\const, \absst, \absc) \models  \mathcal{P} \land  \absst \models \weakestpre{\absc}{X}.$$
\end{definition}

\noindent Subsequently, we can encode relational Hoare triples and  prove that this encoding is correct.

\begin{customthm}{4}[Encoding Relational Triples] For any low-level statement $\conc$, and assertions $\mathcal{P}$, $\mathcal{Q}$ $\subseteq \conSigma \times \absSigma \times \absStmts$:
 \begin{align*}
    \underbrace{\langle \mathcal{P} \rangle \  \conc \ \langle \mathcal{Q}\rangle}_{\mathcal{J}}
     \text{ is valid}   \quad \text{iff.} \quad 
     \underbrace{\forall X. \ \demovdash \{ \encasrt{ \mathcal{P}}_X \} \ \conc \ \{ \encasrt{ \mathcal{Q} }_X \}}_J 
 \end{align*} 
 \label{theo:enc_sec4}
\end{customthm}
\begin{proof}  we prove two directions:
    \begin{itemize}
        \item $\Rightarrow$:  \,  For any $X$ and any $\const_1$ in $\encasrt{\relasrt{P}}_X$, there exist  $\absst_1$ and $\absc_1$ such that $\absst_1 \models \weakestpre{\absc_1}{X}$ and $(\const_1, \absst_1, \absc_1) \models \relasrt{P}$. Then for any terminating state $\const_2$ such that $(\const_1,\const_2) \in \nrmdeno{\conc}$, according to relational triple $\mathcal{J}$, there exist $\absst_2$ and $\absc_2$ such that $(\absst_1, \absc_1) \hookrightarrow (\absst_2, \absc_2)$ and 
        $(\const_2, \absst_2, \absc_2) \models \relasrt{Q}$. Then by \theoref{theo:decom}, we have 
        $\absst_2  \models \weakestpre{\absc_2}{X}$.
        Therefore, according to \defref{enc:asrtenc} we derive 
        $\const_2 \models$ $\encasrt{\mathcal{Q}}_X$.
        \item $\Leftarrow$: Based on \propref{prop:10}, for any $(\const_1, \absst_1, \absc_1) \models \relasrt{P}$,  we instantiate $X$ as 
        the high-level terminal states set
        $\{ \absst_3 \,| \, (\absst_1,\absst_3) \in \nrmdeno{\absc_1}\}$. 
        By the definition of assertion encoding, $\const_1 \models \encasrt{\mathcal{P}}_X$.
        According to standard triple $J$, for any $\const_2 $ such that $(\const_1, \const_2) \in \nrmdeno{\conc}$, we have $\const_2 \models  \encasrt{\mathcal{Q}}_X$.
        By the definition of  assertion encoding, there exist $\absst_2$ and $\absc_2$ such that $\absst_2  \models \weakestpre{\absc_2}{X}$ and $(\const_2,\absst_2, \absc_2) \models \relasrt{Q}$.
        According to the definition of weakest preconditions, for any $\absst_3$ such that $(\absst_2, \absst_3) \in \nrmdeno{\absc_2}$, $\absst_3 \models X$.
        That is for any $\absst_3$ such that $(\absst_2, \absst_3) \in \nrmdeno{\absc_2}$, we have $(\absst_1, \absst_3) \in \nrmdeno{\absc_2}$.
        Therefore, we finally derive $(\absst_1, \absc_1) \hookrightarrow (\absst_2, \absc_2)$ and $(\const_2,\absst_2, \absc_2) \models \relasrt{Q}$.
    \end{itemize}
\end{proof}

\section{Syntactic Encoding of Decomposed Assertions}\label{sec:encdecompose}
Up to this point, we have treated assertions semantically as subsets of program states and configurations, and presented the assertion encoding \(\encasrt{-}_X\) semantically.
This semantic formulation has helped highlight the foundational ideas behind our encoding theory and its soundness guarantees.
In this section, to make our encoding theory more explicit and to apply our encoding theory within an actual verification tool, we present a syntactic encoding that translates decomposed program-as-resource assertions into unary assertions over low-level states.
We begin by establishing equivalent transformations that describe how decomposed assertions behave under the semantic encoding $\encasrt{-}_X$ (\secref{sec:semencode_for_decomposed}).
These equivalences justify the structure of our syntactic encoding.
We then define the syntax for decomposed assertions and present the syntactic encoding (\secref{sec:syntactic}).

Using existential logical variables and pure logical conditions, a program-as-resource assertion can typically be written into components concerning low-level states, high-level states, and the high-level program. 
This decomposition allows for concise focusing rules in real verification tasks.
\[
\exists \vec{a}.\, B(\vec{a}) \wedge \lasrt{\conasrt{P}(\vec{a})} \wedge \hasrt{\absasrt{P}(\vec{a})} \wedge \hspeccolor{\absc}
\]

\subsection{\label{sec:semencode_for_decomposed}Equivalent Transformations for Encoded Decomposed Assertions}
Based on assertion encoding \(\encasrt{-}_X\)  (\defref{enc:asrtenc}), when encoding a decomposed assertion, the existential variables $\vec{a}$, pure logical conditions $B(\vec{a})$, and unary assertion $\conasrt{P}(\vec{a})$ about the low-level states can be lifted out.
This is captured by the following assertion transformation theorem:
\begin{theorem}[\label{the:encodedecomposed}Encoded Assertion Transformation] ~

  \begin{enumerate}[label=(\alph*)]
    \item $ \encasrt{\exists a. \mathcal{P}(a)}_X \iff \exists a. \,\encasrt{\mathcal{P}(a)}_X$.
    \item $ \encasrt{B \land \mathcal{P}}_X \iff  B \land \encasrt{\mathcal{P}}_X$.
    \item $ \encasrt{\lasrt{\conasrt{P}} \land \mathcal{P} }_X \iff \conasrt{P} \land \encasrt{\mathcal{P}}_X $. 
    \item\label{item:enc-disj} $ \encasrt{\mathcal{P}_1 \vee \mathcal{P}_2 }_X \iff \encasrt{\mathcal{P}_1}_X \vee \encasrt{\mathcal{P}_2}_X $. 
\end{enumerate}   
\end{theorem}

\noindent 
Here we include a transformation \hyperref[item:enc-disj]{$(d)$} for disjunction to account for the fact that program-as-resource assertions may include disjunctions of decomposed forms.
The proof of this theorem is straightforward and provided in Appendix~\ref{app:theorem}.
These equivalences show that the encoding of a decomposed assertion can be rewritten as follows:
\[
\encasrt{ \exists \vec{a}.\, B(\vec{a}) \wedge \lasrt{\conasrt{P}} \wedge \hasrt{\absasrt{P}} \wedge \hspeccolor{\absc} }_X
\iff
\exists \vec{a}.\, B(\vec{a}) \wedge \conasrt{P}(\vec{a}) \wedge \encasrt{ \hasrt{\absasrt{P}(\vec{a})} \wedge \hspeccolor{\absc} }_X.
\]
\noindent 
The final component, \(\encasrt{ \hasrt{\absasrt{P}} \wedge \hspeccolor{\absc} }_X\), captures the execution behavior of the high-level program.
Crucially, it is pure: it does not mention or depend on any low-level program state. 
In the following, we use this insight to define a syntactic encoding based on a pure logical predicate, the \textit{execution predicate} \(\needExecX{\absasrt{P}}{\absc}{X}\), which represents this component in the low-level assertion language.

\subsection{Syntactic Encoding for Decomposed Assertions\label{sec:syntactic}}
While the previous subsection defined the encoding semantically, we now make it more concrete by introducing a class of syntactic decomposed assertions and a corresponding syntactic encoding.
We begin with the syntax of basic assertions. 
A pure logical predicate $B$ is a first-order formula over logical variables, built from user-defined predicates $p$ (such as equality or reachability) using standard connectives and quantifiers.
Our encoding is parametric over the assertion languages, so we do not fix the syntax of low-level assertions $\conasrt{P}$ or high-level assertions $\absasrt{P}$.
We assume both include pure predicates $B$, user-defined predicates $p$ over low-level ($\conexp$) or high-level expressions ($\absexp$), conjunction ($\wedge$), and the $\exists$ quantifier to introduce logical variables.
These languages can be extended as needed, e.g., with universal quantification ($\forall$) or separation logic constructs such as the empty heap predicate ($\emp$) and separating conjunction ($*$).

\[
\begin{array}{rcl}
B & ::= & \ptrue \mid \pfalse \mid p(\vec{a}) \mid FO(B) \\
\conasrt{P} & ::= & B \mid p(\vec{e}^{\normalfont L}) \mid \conasrt{P} \wedge \conasrt{P}  \mid \exists a.\, \conasrt{P}(a) \mid \ldots \\
\absasrt{P} & ::= & B \mid p(\vec{e}^{\normalfont H}) \mid \absasrt{P} \wedge \absasrt{P} \mid \exists a.\, \absasrt{P}(a) \mid  \ldots
\end{array}
\]

Building on the syntax of basic assertions, we define the syntax of program-as-resource assertions as finite disjunctions of decomposed forms, where $\hspeccolor{\absc}$ specifies the high-level program to be fulfilled, and \( \lasrt{\conasrt{P}} \), \( \hasrt{\absasrt{P}} \) specify properties over the low-level and high-level states, respectively.
\[
\begin{array}{rcl}
\mathcal{P} & ::= & \bigvee_{i} (\exists \vec{a}.\, B_i(\vec{a}) \wedge \lasrt{\conasrt{P}_i(\vec{a})} \wedge \hasrt{\absasrt{P}_i(\vec{a})} \wedge \hspeccolor{\absc_i})
\end{array}
\]

Based on the syntax of program-as-resource assertions and the semantic equivalences established in \theoref{the:encodedecomposed}, we now define how such assertions are transformed into unary assertions over low-level states. 
To express the high-level behavior within a unary assertion, we introduce the \textit{execution predicate} $\needExecX{\absasrt{P}(\vec{a})}{\absc}{X}$ to represent $\encasrt{ \hasrt{\absasrt{P}(\vec{a})} \wedge \hspeccolor{\absc}}_X$.
\begin{definition}[Syntactic Encoding] Given a high-level postcondition $X$,  the encoding of a syntactic decomposed assertion into a unary assertion over low-level states is defined as follows:
\[
\begin{array}{rcl}
\mathsf{Enc}_X (\exists \vec{a}.\, B(\vec{a}) \wedge \lasrt{\conasrt{P}(\vec{a})} \wedge \hasrt{\absasrt{P}(\vec{a})} \wedge \hspeccolor{\absc})  & \triangleq & \exists \vec{a}.\, B(\vec{a}) \wedge \conasrt{P}(\vec{a}) \wedge \needExecX{\absasrt{P}(\vec{a})}{\absc}{X} \\
\mathsf{Enc}_X ( \mathcal{P}_1\vee \mathcal{P}_2 ) & \triangleq & \mathsf{Enc}_X ( \mathcal{P}_1 )  \vee \mathsf{Enc}_X ( \mathcal{P}_2) 
\end{array}
\]
\end{definition}
\noindent The semantic interpretation of this syntactic encoding aligns with the equivalences in \theoref{the:encodedecomposed}. That is, for any decomposed program-as-resource assertion $\mathcal{P}$ and high-level postcondition $X$:
\[\mathsf{Enc}_X(\mathcal{P}) \Leftrightarrow \encasrt{\mathcal{P}}_X. \]
By transforming program-as-resource assertions into unary assertions over low-level states, this encoding enables relational reasoning to be conducted using standard Hoare logic.
For instance, in the \bitmaskexample, the relational triple \eqref{eq:rel-triple}  can be written in decomposed form as \eqref{eq:standard-triple}, which then is encoded into a standard triple \eqref{eq:standard-triple}  for any high-level postcondition $X$.
\begin{align}
& \reltriplenofont{\hspeccolor{\texttt{set\_union}}}{\texttt{bit\_mask}}{ \convar{x} = \sum\nolimits_{a \in \absvar{s}} 2^a \land \hspeccolor{\cskip} }
\label{eq:rel-triple} \\
& \reltriplenofont{\hspeccolor{\texttt{set\_union}}}{\texttt{bit\_mask}}{ \exists\,l.\, \lasrt{\convar{x} = \sum\nolimits_{a \in l} 2^a} \land \hasrt{\absvar{s} = l} \land \hspeccolor{\cskip} }
\label{eq:decomposed-triple} \\ 
& \left\{ \needExecX{\ptrue}{\texttt{set\_union}}{X} \right\}~\texttt{bit\_mask}~\{ \exists\,l.\, \onlyExec_X{(\absvar{s} = l,~\hspeccolor{\cskip})} \land \convar{x} = \sum\nolimits_{a \in l} 2^a \}
\label{eq:standard-triple}
\end{align}
In the remainder of the paper, we adopt the notation $\encasrt{\mathcal{P}}_X$ uniformly, as it is interchangeable with $\mathsf{Enc}_X(\mathcal{P})$ when $\mathcal{P}$ is a decomposition assertion, which is the typical case in practical verification.

Note that the execution predicate is a pure logical predicate, and we can extend the low-level assertion language to include it:
\[
\conasrtenc ::=  B \mid \needExecX{\absasrt{P}}{\absc}{X} \mid p(\vec{e}^{\normalfont L}) \mid \conasrtenc \wedge \conasrtenc  \mid  \exists a.\, \conasrtenc(a) \mid \conasrtenc  \vee \conasrtenc  \mid \ldots 
\]
\noindent 
$\needExecX{\absasrt{P}}{\absc}{X}$ expresses that there exists an initial state satisfying $\absasrt{P}$ such that any final state after executing $\absc$ satisfies $X$. Its semantic interpretation can be defined as follows:

\begin{definition}[\label{def:exec}Semantic Interpretation of the Execution Predicate]
For any high-level assertion $\absasrt{P}$, postcondition $X$, and program $\absc$, 
\[
\needExecX{\absasrt{P}}{\absc}{X}  \text{ holds \quad iff. \quad there exists } \absst \text{ such that } \absst \models \absasrt{P} \land \textsf{wlp}(\absc, X)
\]
\end{definition}
\noindent
In the next section, we will present proof rules for reasoning about $\needExecX{\absasrt{P}}{\absc}{X}$.

\section{\label{sec:encRule}Encoded Proof Rules and Reasoning with the Execution Predicate}
In this section,  we show how relational reasoning can be carried out within standard Hoare logic, building on our encoding.
We first encode core relational proof rules as derivable standard rules and introduce proof rules for the execution predicate $ \needExecX{\absasrt{P}}{\absc}{X}$ (\secref{subsec:encrules}).
Then we demonstrate examples to show how to leverage our encoding to conduct relational proofs in standard Hoare logic (\secref{subsec:encproofs}).
Together, these results show that standard Hoare logic, extended with the execution predicate and related proof rules, suffices for $\forall\exists$ relational proofs. 
Finally, we show that vertical composition rules can also be encoded to support compositional reasoning (\secref{subsec:encvcrules}).

To support the rules presented in this section, we assume both the low-level and high-level languages are standard imperative languages. Their syntax includes the following constructs:
\[
\begin{array}{rcl}
c & ::= & \cskip \mid \pvar{x} \coloneq e \mid \pvar{x} \coloneq  \textsf{nondet}(e, e)  \mid \ctest{b} \mid  \choice{c}{c}   \mid  \cwhile{b}{c} \mid c;c
\end{array}
\]
\noindent
In programs, expressions $e$ refer only to program variables and constants.
The non-deterministic assignment $\pvar{x} \coloneq  \textsf{nondet}(e_1, e_2)$ assigns to $\pvar{x}$ an arbitrary value between $e_1$ and $e_2$.
We include statement $\ctest{b}$, which behaves like $\cskip$ when $b$ holds and blocks otherwise, and a nondeterministic choice construct $\choice{c_1}{c_2}$, which executes either $c_1$ or $c_2$ nondeterministically.
Then standard $\text{if}\,b\,\text{then}\,c_1\,\text{else}\, c_2 $ can be written as $\choice{\ctest{b};c_1 }{\ctest{\neg b};c_2}$.

\subsection{\label{subsec:encrules}Proof Rules}
\begin{figure}[t]
\centering
\small
  \begin{tabular}{p{0.40\textwidth}p{0.55\textwidth}}
    \begin{minipage}{\linewidth}
    \begin{mathpar}
    \inferrule[\label{rule:lowfocus}\textsc{Low-Focus}]
    { \demovdash \{\conasrt{P}\} \ \conc_1 \ \{\conasrt{R}\} \\
      \reltriplenofont{\lasrt{\conasrt{R}} \land \hasrt{\absasrt{P}} \land \hspeccolor{\absc}}{\conc_2}{\mathcal{Q}} }
    { \reltriplenofont{\lasrt{\conasrt{P}} \land \hasrt{\absasrt{P}} \land \hspeccolor{\absc}}{\conc_1;\conc_2}{\mathcal{Q}} }
    \end{mathpar}
    \end{minipage}
    &
    \begin{minipage}{\linewidth}
    \begin{mathpar}
    \inferrule
    {\demovdash \{\conasrt{P}\} \ \conc_1 \ \{\conasrt{R}\} \\
     \demovdash \{\needExecX{\absasrt{P}}{\absc}{X} \land\conasrt{R}\} \ \conc_2 \ \{ \encasrt{\mathcal{Q}}_X\} }
    {\demovdash \{ \needExecX{\absasrt{P}}{\absc}{X} \land \conasrt{P}\} \ \conc_1;\conc_2 \ \{\encasrt{\mathcal{Q}}_X\} }
    \end{mathpar}
    \end{minipage} \\
        \begin{minipage}{\linewidth}
\begin{mathpar}
\inferrule
  [\label{rule:rel-choice}Choice]{
    \reltriplenofont{\mathcal{P}}{\conc_1}{\mathcal{Q}} \\
     \reltriplenofont{\mathcal{P}}{\conc_2}{\mathcal{Q}} 
    }{ \reltriplenofont{\mathcal{P}}{\choice{\conc_1}{\conc_2}}{\mathcal{P}}}
\end{mathpar}
    \end{minipage}
    &
    \begin{minipage}{\linewidth}
    \begin{mathpar}
    \inferrule
    {
    \demovdash \unarytriple{\encasrt{ \mathcal{P}}_X }{\conc_1}{\encasrt{ \mathcal{Q}}_X} \\
       \demovdash \unarytriple{\encasrt{ \mathcal{P}}_X }{\conc_2}{\encasrt{ \mathcal{Q}}_X}
    }{ \demovdash \unarytriple{\encasrt{ \mathcal{P}}_X}{\choice{\conc_1}{\conc_2}}{\encasrt{ \mathcal{Q}}_X}}
    \end{mathpar}
    \end{minipage}\\
    \begin{minipage}{\linewidth}
\begin{mathpar}
\inferrule
  [\label{rule:relwh}Rel-Wh]{
    \reltriplenofont{\mathcal{P}\land \lasrt{\conb}}{\conc}{\mathcal{P}}
    }{ \reltriplenofont{\mathcal{P}}{\text{\textsf{ while} } \conb \text{\textsf{ do }} \conc}{\mathcal{P} \land \lasrt{\neg \conb}}}
\end{mathpar}
    \end{minipage}
    &
    \begin{minipage}{\linewidth}
    \begin{mathpar}
    \inferrule
    {
    \demovdash \unarytriple{\encasrt{ \mathcal{P}}_X \land {\conb}}{\conc}{\encasrt{ \mathcal{P}}_X}
    }{ \demovdash \unarytriple{\encasrt{ \mathcal{P}}_X}{\text{\textsf{ while} } \conb \text{\textsf{ do }} \conc}{\encasrt{ \mathcal{P}}_X\land {\neg \conb}}}
    \end{mathpar}
    \end{minipage}\\
    \end{tabular}
\caption{Low-level Dependent Relational rules and their corresponding encoded rules}
\label{fig:enc-low-levelrules}
\end{figure}
\paragraph{\bfseries Low-level Rules}
Relational Hoare logic provides proof rules depending on low-level statements, such as sequencing (\hyperref[rule:lowfocus]{\textsc{Low-Focus}}),  loops (\hyperref[rule:relwh]{\textsc{Rel-Wh}}), and nondeterministic choice (\hyperref[rule:rel-choice]{Choice}).
These rules are transformed into standard Hoare rules using our encoding theory, as shown in \figref{fig:enc-low-levelrules}.
For example, 
the rule \hyperref[rule:lowfocus]{\textsc{Low-Focus}}  is encoded into a special case of sequencing rule \hyperref[intro:seq]{\textsc{Seq}}.
Theoretically, our encoding theory explains the similarities in the proof rules of relational Hoare logic and standard Hoare logic.
Practically, our encoding theory also suggests proof rules in standard Hoare logic can be used to prove program refinement in a similar way to relational Hoare logic.

\begin{figure}[t]
\centering
\small
\makebox[\linewidth][l]{\normalfont\textbf{Relational high-focus rule and its corresponding encoded rule}} 
\begin{tabular}{p{0.40\textwidth}p{0.55\textwidth}}
        \begin{minipage}{\linewidth}
    \begin{mathpar}
    \inferrule[\label{rule:absfocus}\textsc{High-Focus}]
    { \angelvdash \unarytriple{\absasrt{P}}{\absc_1}{\absasrt{R}}\\ 
      \reltriplenofont{\lasrt{\conasrt{F}} \land \hasrt{\absasrt{R}} \land \hspeccolor{\absc_2}}{\conc}{\mathcal{Q}} }
    {  \reltriplenofont{\lasrt{\conasrt{F}} \land  \hasrt{\absasrt{P}} \land \hspeccolor{\absc_1;\absc_2}}{\conc}{\mathcal{Q}} }
    \end{mathpar}
    \end{minipage}
    &
    \begin{minipage}{\linewidth}
\begin{mathpar}
  \inferrule
      {\angelvdash \unarytriple{\absasrt{P_1}}{\absc_1}{\absasrt{P_2}} }
      { \needExecX{\absasrt{P_1}}{\absc_1;\absc_2}{X} \Rightarrow \needExecX{\absasrt{P_2}}{\absc_2}{X} }
\end{mathpar}
    \end{minipage} \\ 
    \end{tabular}
\makebox[\linewidth][l]{\normalfont\textbf{Proof rules for updating the execution predicate $\needExecX{\absasrt{P}}{\absc}{X}$ }} 
\begin{mathpar}
\inferrule[\label{execrule:assign}Exec-Assign]
{ }
{ \needExecX{\absasrt{P}[\absexp/\absvar{x}]}{\absvar{x} := \absexp}{X} 
  \Rightarrow  
  \needExecX{\absasrt{P}}{\cskip}{X} } 
  \hfill
\inferrule[\label{execrule:choicel}Exec-ChoiceL]
{ }
{ \needExecX{\absasrt{P}}{\choice{\absc_1}{\absc_2}}{X} 
  \Rightarrow 
  \needExecX{\absasrt{P}}{\absc_1}{X} } \\
  \inferrule[\label{execrule:nondet}Exec-Nondet]
{ \absasrt{P}[v/\absvar{x}] \Rightarrow \absexp_1 \leq v \leq \absexp_2 }
{ \needExecX{\absasrt{P}[v/\absvar{x}]}{\absvar{x} := \textsf{nondet}(\absexp_1,\absexp_2)}{X} 
  \Rightarrow  
  \needExecX{\absasrt{P}}{\cskip}{X} }\hfill
  \inferrule[\label{execrule:assume}Exec-Assume]
{ \absasrt{P} \Rightarrow \absbexp }
{ \needExecX{\absasrt{P}}{\ctest{\absbexp }}{X} 
  \Rightarrow 
  \needExecX{\absasrt{P}}{\cskip}{X} }   \\
    \inferrule[\label{execrule:pure}Exec-Pure]
{ \forall \,X.\, \needExecX{\absasrt{P}}{\absc_1}{X} 
  \Rightarrow 
  \needExecX{\absasrt{P}}{\absc_0}{X} }
{ \forall \,X.\, \needExecX{\absasrt{P}}{\absc_1 ; \absc_2}{X} 
  \Rightarrow 
  \needExecX{\absasrt{P}}{\absc_0 ; \absc_2}{X} }\hfill
\inferrule[\label{execrule:whileend}Exec-While-End]
{ \absasrt{P} \Rightarrow \neg \absbexp  }
{ \needExecX{\absasrt{P}}{\cwhile{\absbexp}{\absc}}{X} 
  \Rightarrow 
  \needExecX{\absasrt{P}}{\cskip}{X} } \\
  \inferrule[\label{execrule:seq}Exec-Seq]
{ \forall \,X.\, \needExecX{\absasrt{P}}{\absc_1}{X} 
  \Rightarrow 
  \needExecX{\absasrt{Q}}{\cskip}{X} }
{ \forall \,X.\, \needExecX{\absasrt{P}}{\absc_1 ; \absc_2}{X} 
  \Rightarrow 
  \needExecX{\absasrt{Q}}{\absc_2}{X} } \hfill
\inferrule[\label{execrule:whileunroll}Exec-While-Unroll]
{ \absasrt{P} \Rightarrow \absbexp }
{ \needExecX{\absasrt{P}}{\cwhile{\absbexp}{\absc}}{X} 
  \Rightarrow 
  \needExecX{\absasrt{P}}{\absc ; \cwhile{\absbexp}{\absc}}{X} } 

\end{mathpar}
\caption{Proof rules for Evaluating High-level Programs}
\label{fig:needexec-rules}
\end{figure}

\paragraph{\bfseries High-level Rules} 
Relational Hoare logic also provides proof rules for independently evaluating high-level programs.
These high-level steps can be encoded into the consequence rule  \hyperref[intro:conseq]{\textsc{Conseq-Pre}} in standard Hoare logic by updating the execution predicate as follows:
\begin{mathpar}
    \inferrule
    {\needExecX{\absasrt{P}}{\absc_1}{X} \Rightarrow \needExecX{\absasrt{R}}{\absc_2}{X} \\
     \demovdash \{ \needExecX{\absasrt{R}}{\absc_2}{X} \land\conasrt{F}\} \ \conc \ \{ \encasrt{\mathcal{Q}}_X\} }
    {\demovdash \{ \needExecX{\absasrt{P}}{\absc_1}{X} \land \conasrt{F}\} \ \conc \ \{ \encasrt{\mathcal{Q}}_X\} }
\end{mathpar}
As shown in \figref{fig:needexec-rules}, rule \hyperref[rule:absfocus]{\textsc{High-Focus}} allows one to angelically execute part of the high-level program in isolation. 
Our encoding reflects this rule by allowing a successful angelic evaluation of $\absc_1$ to update $\needExecX{\absasrt{P_1}}{\absc_1;\absc_2}{X}$ into $\needExecX{\absasrt{P_2}}{\absc_2}{X}$.
The rest of \figref{fig:needexec-rules} presents proof rules for updating the execution predicate, covering the basic imperative constructs.
For example, angelic nondeterminism is captured by two rules: rule \hyperref[execrule:nondet]{\textsc{Exec-Nondet}} allows any value $v$ within the admissible range to be picked, while rule \hyperref[execrule:choicel]{\textsc{Exec-ChoiceL}} permits the left branch of a nondeterministic choice to be selected (a symmetric right-branch rule is omitted here).
Sequential composition is handled by two rules: \hyperref[execrule:pure]{\textsc{Exec-Pure}} allows rewriting the first subcommand without modifying state; \hyperref[execrule:seq]{\textsc{Exec-Seq}} ensures that sequential composition can be evaluated step by step.
For loops, the pair of rules \hyperref[execrule:whileunroll]{\textsc{Exec-While-Unroll}} and \hyperref[execrule:whileend]{\textsc{Exec-While-End}}  express that the execution predicate unrolls the loop as long as the guard $\absbexp$ holds and terminates when the guard fails.

\subsection{\label{subsec:encproofs}Proofs with the Execution Predicate}
To demonstrate how our encoding enables relational proofs within standard Hoare logic, we revisit the examples from \secref{sec:rhtproof}, and prove them using standard Hoare rules and the execution predicate.

\paragraph{\bfseries Nondeterministic Programs.}
Consider the two nondeterministic programs from Example~\ref{exam:nondet}. Our goal is to prove the following standard Hoare triple:
\[
\{\needExecX{\ptrue}{\inlinecodes{$\absvar{y}\coloneq\,\textsf{ nondet}$(0,2)}}{X}\}\,{{\inlinecodes{$\convar{x}\coloneq\,$ $\textsf{ nondet}$(0,1)}}}\, \{\exists\, n.\,  \needExecX{\absvar{y}=n}{\cskip}{X} \wedge \convar{x} = n\}.
\]
We begin by evaluating the low-level assignment $\convar{x} \coloneq \textsf{nondet}(0,1)$ demonically. 
This yields an postcondition: $\exists\, n\in\{0,1\}.\,  \needExecX{\ptrue}{\inlinecodes{$\absvar{y}\coloneq\,\textsf{ nondet}$(0,2)}}{X} \wedge \convar{x} = n$.
Next, we apply rule \hyperref[execrule:nondet]{\textsc{Exec-Nondet}} and angelically pick the same value $n$ to update the execution predicate.
Then we establish the desired postcondition.
This standard Hoare logic proof, along with updating the execution predicate, mirrors the relational reasoning presented in \secref{sec:rhtproof}.

\paragraph{\bfseries Simple Sequential Programs. }
Recall the  \bitmaskexample, where the high-level program builds a set and the low-level program constructs a bitmask. Our goal is to prove:
\[
\unarytriple{\needExecX{\ptrue}{\texttt{set\_union}}{X} }{\texttt{bit\_mask}}{\exists\,l.\, \needExecX{\absvar{s}= l}{\cskip}{X}\land {\convar{x} = \sum\nolimits_{a\in l}2^a}}.
\]
\noindent As shown in \figref{fig:introexample_stdproof}, we begin with the high-level assignment $\absvar{s} := \emptyset$, updating the execution predicate using rules \hyperref[execrule:seq]{\textsc{Exec-Seq}} and \hyperref[execrule:assign]{\textsc{Exec-Assign}}. 
This yields the updated predicate: 
$\onlyExec_X{\color{RoyalBlue}(}{\absvar{s}=\emptyset}{\color{RoyalBlue},}$ ${\color{purple}\absvar{s} \coloneq \absvar{s} \cup a_0;\, \absvar{s} \coloneq \absvar{s} \cup a_1}{\color{RoyalBlue})}$.
Next, we handle the low-level assignment $\convar{x} \coloneq 0$, and apply the sequencing and assignment rules from standard Hoare logic.
Then the goal becomes$\{ \onlyExec_X{\color{RoyalBlue}(} {\absvar{s}=\emptyset}{\color{RoyalBlue},}$ ${\color{purple}\absvar{s} \coloneq \absvar{s} \cup a_0;\, \absvar{s} \coloneq \absvar{s} \cup a_1}{\color{RoyalBlue})}$$\wedge \convar{x} = 0 \}$ $\convar{x} \coloneq \convar{x} | (1 << a_0);\, \convar{x} \coloneq \convar{x} | (1 << a_1)$ $\{\exists\,l.\, \onlyExec_X{\color{RoyalBlue}(} {\absvar{s} = l{\color{RoyalBlue},} }$ ${\color{purple}\cskip}{\color{RoyalBlue})}$ $\wedge{\convar{x} = \sum\nolimits_{a\in l}2^a}\}.$
We proceed similarly for the remaining assignments: updating the execution predicate on the high-level side and applying standard Hoare rules on the low-level side. 
The final postcondition is $\needExecX{\absvar{s} = \{a_0, a_1\}}{\cskip}{X} \wedge \convar{x} = 2^{a_0} + 2^{a_1}$, which implies the desired postcondition.

\paragraph{\bfseries Loops. }
When more items need to be recorded, we use a loop-based version as shown in  Example~\ref{exam:loop}.  
Our goal is to prove the following standard Hoare triple:
\[
\unarytriple{\needExecX{\ptrue}{\texttt{set\_union\_loop}}{X} }{\texttt{bit\_mask\_loop}}{\exists\,l.\, \needExecX{\absvar{s}= l}{\cskip}{X}\land {\convar{x} = \sum\nolimits_{a\in l}2^a}}.
\]
As before, we can evaluate the initial assignments to $\absvar{s}$, $\absvar{j}$, $\convar{x}$, and $\convar{i}$ using standard Hoare rules along with the \hyperref[execrule:seq]{\textsc{Exec-Seq}} and \hyperref[execrule:assign]{\textsc{Exec-Assign}} rules to update the execution predicate. 
This reduces the goal to $\{\needExecX{\absvar{s}=\emptyset \wedge \absvar{j}=0}{\cwhile{\absvar{j} < 8}{\ldots}}{X} \wedge \convar{x}=0 \wedge \convar{i}=0\}$ $\cwhile{\convar{i} < 8}{ \ldots }$ $\{\exists\,l.\, \needExecX{\absvar{s}= l}{\cskip}{X}\land {\convar{x} = \sum\nolimits_{a\in l}2^a}\}.$
We then apply the standard while rule with the following invariant, which corresponds to the relational invariant \eqref{eq:toyinv}:
\[
\exists\, n\,l.\, \needExecX{\absvar{s}=l \wedge \absvar{j}=n}{\cwhile{\absvar{j} < 8}{\ldots}}{X} \wedge \convar{x}=\sum\nolimits_{a\in l}2^a \wedge \convar{i}=n
\]
This invariant expresses that both programs have processed the first $n$ elements of the array. 
The set $\absvar{s}$ contains the first $n$ items, and the integer $\convar{x}$ encodes the same set as a bitmask.
As long as the guard $\convar{i} < 8$ holds, we know $\absvar{j} < 8$ also holds, allowing us to apply \hyperref[execrule:whileunroll]{\textsc{Exec-While-Unroll}} to unroll the high-level while statement.
Each loop iteration then updates both sides independently as before.
Once the loop exits, \hyperref[execrule:whileend]{\textsc{Exec-While-End}} applies on the high-level side, reducing its program to $\cskip$ and yielding the desired postcondition.

\subsection{\label{subsec:encvcrules}Encoded Vertical Composition Rules and Compositional Reasoning}
Relational Hoare logic supports two forms of vertical composition: combining a refinement proof with a standard proof of the high-level program (\hyperref[rule:vc]{\textsc{VC-FC}}), and composing two refinement proofs (\hyperref[rule:vcr]{\textsc{VC-Refine}}). 
Here the linking operators $\odot$ and $\circ$ used in the rules are defined as follows: 
\begin{definition}[Assertion Linking]For any binary assertion $\mathbb{P} \subseteq \conSigma \times \absSigma$ and any unary assertion $\absasrt{P} \subseteq \absSigma$:
  \begin{center}
   $\const \models \mathbb{P} \odot \absasrt{P}$ $\quad \text{iff.} \quad $ $\exists\, \absst. \, (\const, \absst) \models \mathbb{P} \land \absst \models \absasrt{P}$.
  \end{center}
For any binary assertions $\mathbb{P}_1 \subseteq \Sigma_1 \times \Sigma_2$ and $\mathbb{P}_2 \subseteq \Sigma_2 \times \Sigma_3$:
   \begin{center}
    $(\sigma_1,\sigma_3)  \models \mathbb{P}_1 \circ \mathbb{P}_2$ $\quad \text{iff.} \quad $ $\exists\, \sigma_2. \, (\sigma_1,\sigma_2)  \models \mathbb{P}_1 \land (\sigma_2,\sigma_3)  \models \mathbb{P}_2$.
   \end{center}
\end{definition}
\noindent The operator $\odot$ links a binary assertion with a unary assertion by existentially projecting over the high-level state, while $\circ$ composes two binary assertions by joining over their shared intermediate state. Both operators are special cases of the relational join operator~\cite{DBLP:journals/tods/AhoBU79}.

Under our encoding, rules \hyperref[rule:vc]{\textsc{VC-FC}} and \hyperref[rule:vcr]{\textsc{VC-Refine}} are transformed into the following forms:
\begin{mathpar}
  \inferrule
  { \forall X. \demovdash \unarytriple{ \encasrt{ \mathbb{P} \land \hspeccolor{\absc}}_X
  }{\conc}{ \encasrt{ \mathbb{Q} \land \hspeccolor{\cskip}}_X}  \\ 
  \demovdash \{  \absasrt{P} \} \ \absc \ \{\absasrt{Q}\} 
  }
  { \demovdash \{  \mathbb{P} \odot \absasrt{P} \} \  \conc \ \{\mathbb{Q} \odot \absasrt{Q} \} } \\
  \inferrule
  {
    \forall X_1. \demovdash \unarytriple{ \encasrt{ \mathbb{P}_1 \land \hspeccolor{c_2}}_{X_1}
    }{c_1}{ \encasrt{ \mathbb{Q}_1 \land \hspeccolor{\cskip}}_{X_1}}  \\  
  \forall X_2. \demovdash \unarytriple{ \encasrt{ \mathbb{P}_2 \land \hspeccolor{c_3}}_{X_2}
  }{c_2}{ \encasrt{ \mathbb{Q}_2 \land \hspeccolor{\cskip}}_{X_2}}  
  }
  { \forall X. \demovdash \unarytriple{ \encasrt{ \mathbb{P}_1 \circ  \mathbb{P}_2 \land \hspeccolor{c_3}}_X
  }{c_1}{ \encasrt{ \mathbb{Q}_1 \circ  \mathbb{Q}_2  \land \hspeccolor{\cskip}}_X}  }
\end{mathpar}
\noindent The two encoded rules resemble specialized forms of the standard consequence rule. 
In particular, both can be derived using the consequence rule together with the rule \hyperref[intro:ex]{\textsc{ExIntro}} in standard Hoare logic.
For the first rule, the key idea is extracting the initial high-level state $\absst_0$ from the existential quantification in $\mathbb{P} \odot \absasrt{P}$, and instantiating $X$ as the set of high-level states of executing $\absc$ starting from $\absst_0$.
The second rule is then proved by instantiating $X_2$ as $X$ and applying the first rule.
Complete proofs are provided in Appendix~\ref{app:theorem}.

These rules become convenient when used with decomposed assertions. 
For example, a refinement judgment that expresses $\conc$ refines $\absc$ often appears as follows.
\[ \reltriplenofont{\exists \, u. \lasrt{\constorepre (u)} \wedge \hasrt{ \absstorepre(u)} \wedge \hspeccolor{\absc} }{\conc}{\exists \, v. \lasrt{\constorepost(v)} \wedge \hasrt{ \absstorepost(v)} \wedge \hspeccolor{\cskip} } \]
Here,  the logical variables $u$ and $v$  serve to relate the low-level and high-level states and represent the initial and final data, respectively.
For instance, in the \bitmaskexample, the relational triple can be written as  
$\langle {\lasrt{\ptrue} \wedge \hasrt{ \ptrue} \wedge \hspeccolor{\texttt{set\_union}} } \rangle \, {\texttt{bit\_mask}}\, $  $\langle\exists \, v.\, \lasrt{\convar{x} = \sum\nolimits_{a\in v}2^a} \wedge \hasrt{ \absvar{s} = v} \wedge \hspeccolor{\cskip}  \rangle $.
Under encoding, this form enables a convenient vertical composition with a high-level standard triple as follows, where the precondition assumes a constraint $B_1$ on the data stored in the state and the postcondition asserts that constraint $B_2$ will hold after execution. 
\[
    \inferrule*{
       \forall X. \demovdash \left\{\exists \, u.\, 
      {\begin{aligned}& \needExecX{\absstorepre(u)}{\absc}{X}\\ & \land \constorepre(u)\end{aligned}}
  \right\} \, \conc \, 
  \left\{\exists \, v.\, 
      {\begin{aligned}& \needExecX{\absstorepost(v)}{\absc}{X}\\ & \land \constorepost(v)\end{aligned}}
  \right\}
   \\ 
  \demovdash \{\exists\, u.\,  B_1(u) \wedge \absstorepre(u)\} \ \absc \ \{ \forall\, v.\,  \absstorepost(v) \Rightarrow B_2(v) \}
  \\
\forall \,u.\, B_1(u) \Rightarrow \mathsf{inhabitant}( \absstorepre(u))
    }{
       \demovdash \{\exists\, u.\, B_1(u) \wedge \constorepre(u)\} \  \conc \ \{ \exists \, v. \, B_2(v) \wedge \constorepost(v)\} 
    }
\]
The side condition $\forall \,u.\, B_1(u) \Rightarrow \mathsf{inhabitant}( \absstorepre(u))$ ensures that, for any logical variable $u$ such that $B_1(u)$, there exists some high-level state satisfying $ \absstorepre(u)$. 
It reflects the existential quantification of $\mathbb{P} \odot \absasrt{P}$ in the encoded \textsc{VC-FC} rule.
Intuitively, this requirement lies in that to derive properties of the low-level program based on the high-level one, we need to ensure that there exists a valid high-level execution satisfying the properties.
The proof is given in Appendix~\ref{app:theorem}.

\begin{example}
  Assume we have proved that  \texttt{set\_union} ensures $\absvar{s}$ is non-empty after termination, as described by the triple $\unarytriple{\ptrue}{\texttt{set\_union}}{\absvar{s} \neq \emptyset}$. 
  The goal is to prove that  \texttt{bit\_mask} ensures $\convar{x}$  is larger than 0 after termination, as described by the triple $\unarytriple{\ptrue}{\texttt{bit\_mask}}{\convar{x} > 0}$.
\end{example}
\noindent We apply the vertical composition rule with $B_1(u)\defeq \ptrue$ and $B_2(v) \defeq v \neq \emptyset$.
This establishes a low-level triple $\unarytriple{\ptrue }{\texttt{bit\_mask}}{\exists \,v.\, v \neq \emptyset \wedge \convar{x} = \sum\nolimits_{a\in v}2^a}$, whose postcondition implies $\convar{x} > 0$.


\section{\label{sec:casestudy}Case Studies}
This paper shows that standard Hoare logic, extended with the execution predicate $\needExecX{\absasrt{P}}{\absc}{X}$ and its associated proof rules, can express and verify $\forall\exists$ refinement properties.
In \secref{subsec:encproofs}, we have already demonstrated how standard Hoare reasoning applies to several simple examples, including nondeterminism, sequencing, and loops. 
Beyond these, we have formalized a collection of more realistic case studies in Rocq, covering heap mutation, abstract data structures, and function calls:
\begin{itemize}
    \item  \textbf{Mergesort}: A low-level implementation over singly linked lists refines an abstract version over lists. This captures the correspondence between heap-based lists and pure lists.
    \item \textbf{KMP String Matching}: An array-based KMP implementation refines a list-based program. The refinement proof involves reasoning about array index manipulation.
    \item \textbf{Binary Search Trees}: A loop-based insertion program over a mutable tree refines a recursive version over an abstract binary tree model. The two programs differ in control structure but preserve the same search behavior.
    \item \textbf{Depth-First Search}: A recursive DFS program that traverses a graph represented by adjacency lists refines an iterative version over an abstract graph model. 
    The relational specification ensures both programs explore the same set of reachable nodes.
\end{itemize}

In all examples, the low-level programs are written in a heap-based imperative language.
Their specifications are written in a standard separation logic extended with the execution predicate.
The assertion language extends that in \secref{sec:encdecompose} with standard heap assertions as follows, where $\emp$ denotes the empty heap, $\conexp_1 \mapsto \conexp_2$ specifies a singleton heap mapping address $\conexp_1$ to value of $\conexp_2$, and $\conasrtenc_1 * \conasrtenc_2$ asserts that the heap can be split into two disjoint parts satisfying $\conasrtenc_1$ and $\conasrtenc_2$, respectively.
\[
\conasrtenc ::=  B \mid \needExecX{\absasrt{P}}{\absc}{X} \mid \emp \mid \conexp \mapsto \conexp  \mid  p(\vec{e}^{\normalfont L}) \mid \conasrtenc \wedge \conasrtenc  \mid  \exists a.\, \conasrtenc(a) \mid \conasrtenc  \vee \conasrtenc  \mid \conasrtenc * \conasrtenc
\]
Notably, as supported by the syntactic encoding in \secref{sec:syntactic}, the inclusion of separation logic constructs in low-level assertions enables the application of the standard frame rule:
\[
\inferrule{\unarytriple{ \conasrt{P} }{\conc}{\conasrt{Q}}}{\unarytriple{ \conasrt{P} * \conasrt{F}}{\conc}{\conasrt{Q} * \conasrt{F}} }
\]
This is because we reason entirely within standard Hoare logic, using the low-level proof rules together with updating rules for the execution predicate.
In contrast, the high-level programs are written using standard first-order assertions without separation logic. We believe that extending separation logic to high-level programs has limited benefit for sequential verification, though we include a supplementary discussion of such an extension in Appendix~\ref{app:sep}.

In the remainder of this section, we present a detailed proof of the merging algorithm from the mergesort example. 
While we have formalized the full refinement between the low-level and high-level implementations of mergesort in Rocq, we focus here on the merge step, as it is sufficient to show how standard Hoare logic with the execution predicate can be used to prove the refinement between heap-based list manipulation and abstract list concatenation.

\begin{figure}[t]
  \centering
 \begin{tabular}{@{\hspace{0.1em}}p{0.48\linewidth}|@{\hspace{0.5em}}p{0.45\linewidth}}
\begin{tabular}{@{}l@{}}
\begin{lstlisting}[aboveskip=0\baselineskip,mathescape=true, language=C, breaklines=true,
    numbersep=5pt,                  
    showspaces=false,                
    showstringspaces=false,
    showtabs=false,
    tabsize=2,basicstyle=\small,
    basewidth = {.45em}]
struct list{ int data; struct list *next; };
struct list * $\texttt{merge}^{\text{\normalfont L}}$(struct list *$\convar{x}$,  
                      struct list *$\convar{y}$)
{ struct list * $\convar{r}$;
  struct list ** $\convar{t}$ = &$\convar{r}$;
  while (true) {
    if ($\convar{x}$ == NULL) {*$\convar{t}$ = $\convar{y}$;   return $\convar{r}$;}
    else if ($\convar{y}$ == NULL)
      {*$\convar{t}$ = $\convar{x}$; return $\convar{r}$;}
    else if ($\convar{x}$ -> data < $\convar{y}$ -> data)
      { *$\convar{t}$ = $\convar{x}$;
        $\convar{t}$ = &($\convar{x}$->next); $\convar{x}$ = $\convar{x}$->next; }
    else { *$\convar{t}$ = $\convar{y}$;
        $\convar{t}$ = &($\convar{y}$->next); $\convar{y}$ = $\convar{y}$->next; } } }
    \end{lstlisting}
  \end{tabular} & 
    \begin{tabular}{@{}l@{}}
   \begin{lstlisting}[aboveskip=0\baselineskip,mathescape=true, language=C, breaklines=true,
    numbersep=5pt,                  
    showspaces=false,                
    showstringspaces=false,
    showtabs=false,
    tabsize=2,
    basewidth = {.45em}, basicstyle=\small, morekeywords={assume, rec, then,return, list, choice}]
$\texttt{merge}^{\text{\normalfont H}}$($\absvar{p}$, $\absvar{q}$) $\defeq$  
  $\absvar{r} \coloneq \nil$;
  while ($\absvar{p} \neq \nil$ $\vee$ $\absvar{q} \neq \nil$) {
    if ($\absvar{p} = \nil$) then $\absvar{r} \coloneq \absvar{r}   \absvar{q}$
    else if ($\absvar{q} = \nil$) then $\absvar{r} \coloneq \absvar{r}    \absvar{p}$
    else {$\absvar{z}_1\coloneq$ head($\absvar{p}$); 
      $\absvar{z}_2\coloneq$ head($\absvar{q}$);
      choice(assume ($\absvar{z}_1 \leq \absvar{z}_2$));
          $\absvar{p} \coloneq$ tail($\absvar{p}$);
          $\absvar{r} \coloneq$ $\absvar{r}\absvar{z}_1$,
          assume ($\absvar{z}_2 \leq \absvar{z}_1$));
          $\absvar{q} \coloneq$ tail($\absvar{q}$);
          $\absvar{r} \coloneq$ $\absvar{r}\absvar{z}_2$ } };
  return $\absvar{r}$
    \end{lstlisting}
  \end{tabular}\\
  \end{tabular}
  \caption{The merge example: the low-level program (left) merges two singly linked lists using explicit pointer manipulation, while the high-level program (right) merges two abstract lists. }
  \label{fig:mergeprograms} 
\end{figure}

\subsection{Heap-Based Merge Refines list-based Merge}
Merging two lists is naturally specified nondeterministically: when head elements are equal, either can be chosen. In contrast, real-world implementations (e.g., in C language) make this choice deterministically, as shown in \figref{fig:mergeprograms}. 
Our goal is to verify that such a deterministic, heap-based implementation refines a nondeterministic list-based program.

On the left of \figref{fig:mergeprograms} is the low-level implementation \textsf{merge}$^{\text{L}}$, which merges two sorted singly linked lists, $\convar{x}$ and $\convar{y}$, into a new list.
The key idea  is to maintain a pointer-to-pointer variable $\convar{t}$ that always refers to the tail of the result list being constructed. 
On the right is the high-level version \textsf{merge}$^{\text{H}}$, which operates over abstract lists using standard list operations: $l_1l_2$ for concatenation, $l_1z_1$ to append an element, $\textsf{head}(l)$ to access the first element, and $\textsf{tail}(l)$ to get the rest.
Besides differing in data representation, the two implementations also differ in behavior.
First, when both input lists are empty, the high-level program skips the loop, while the low-level version enters.
Second, when both head elements are equal, the high-level program allows either element to be selected, while the low-level program always chooses the element from $\convar{y}$.
Their refinement can be expressed as a relational Hoare triple, where the separation logic predicate $\sll{p}{l}$ relates a linked list segment starting at address $p$ in the heap to an abstract list $l$:
\[
\left\langle\exists \,l_x, l_y.\, 
      {\begin{aligned}
       & \lasrt{ \sll{\convar{x}}{l_x} * \sll{\convar{y}}{l_y}} \wedge \\
      &  \hasrt{\absvar{p} = l_x \wedge \absvar{q} = l_y} \wedge \hspeccolor{\absmerge(\absvar{p},\absvar{q} )}) \end{aligned}}
  \right\rangle \, \conmerge(\convar{x},\convar{y}) \, 
 \left\langle \exists \, l_r.\, 
      {\begin{aligned}
      & \lasrt{ \sll{\convar{r}}{l_r}} \wedge \\
      &  \hasrt{\absvar{r} = l_r} \wedge \hspeccolor{\cskip} \end{aligned}}
  \right\rangle 
\]
This triple states that if the initial heap contains linked lists at $\convar{x}$ and $\convar{y}$ corresponding to high-level variables $\absvar{p}$ and $\absvar{q}$, then the low-level program produces a list $\convar{r}$ that matches $\absvar{r}$.

With the execution predicate,  we can express the refinement as a standard Hoare triple:
\[
\left\{\exists \,l_x, l_y.\, 
      {\begin{aligned}&  \onlyExec_X{\color{RoyalBlue}(} {\absvar{p} = l_x \wedge \absvar{q} = l_y}{\color{RoyalBlue},} {\color{purple}\absmerge(\absvar{p},\absvar{q} )}{\color{RoyalBlue})}\\ 
      & \wedge \sll{\convar{x}}{l_x} * \sll{\convar{y}}{l_y} \end{aligned}}
  \right\} \, \conmerge(\convar{x},\convar{y}) \, 
 \left\{\exists \, l_r.\, 
      {\begin{aligned}&  \onlyExec_X{\color{RoyalBlue}(} {\absvar{r} = l_r}{\color{RoyalBlue},}{\color{purple}\cskip}{\color{RoyalBlue})}\\ 
      & \wedge \sll{\convar{r}}{l_r} \end{aligned}}
  \right\} 
\]
As in \secref{subsec:encproofs}, we begin by applying standard Hoare rules together with \hyperref[execrule:seq]{\textsc{Exec-Seq}} and \hyperref[execrule:assign]{\textsc{Exec-Assign}} to handle the initial setup before entering the loop.
The core of the proof then lies in establishing a loop invariant that relates the current state of the low-level heap to the high-level state.
Intuitively, the invariant should track three components: the remaining portions of the input lists, and the portion of the result list that has already been constructed from pointer $\convar{r}$ to pointer $\convar{t}$.
To express the result list, we introduce a predicate $\sllbseg{p}{q}{l}$ to describe a list segment $l$ that begins at the address stored at pointer $p$ and ends at pointer $q$.
The loop invariant is then stated as:
\[
\left. \exists \,l_x, l_y,l_r, p_t. 
      {\begin{aligned}& \onlyExec_X{\color{RoyalBlue}(} {\absvar{p} = l_x \wedge \absvar{q} = l_y \wedge \absvar{r} = l_r}{\color{RoyalBlue},} {\color{purple}\cwhile{\ldots}{\ldots}}{\color{RoyalBlue})}  \\
      & \wedge \sll{\convar{x}}{l_x} * \sll{\convar{y}}{l_y} * \sllbseg{\&\convar{r}}{\convar{t}}{l_r}* \convar{t} \mapsto p_t
      \end{aligned}}
  \right.
\]

We verify that the loop invariant is preserved across all paths through the loop body.
When either input list is empty, we assume without loss of generality that $\convar{x} = \pnull$ (the case where $\convar{y} = \pnull$ is symmetric).
In the low-level program, $\ast\convar{t}$ is assigned to $\convar{y}$ and $\convar{b}$ is set to 1.
We apply the standard assignment rule to handle these assignments, and use the frame rule to preserve the parts of the heap that are unchanged (e.g., the already constructed result list).
This results in a heap satisfying the invariant with $\convar{t}\mapsto \convar{y}$ and $\convar{b} = 1$.
On the high-level side, the execution predicate remains unchanged, and the loop invariant holds.
When both input lists are non-empty, we assume that $l_x = z_xl_x'$ and $l_y = z_yl_y'$.
For the high-level program, we apply the rule \hyperref[execrule:whileunroll]{\textsc{Exec-While-Unroll}} to unroll one iteration of the loop.
Then, \hyperref[execrule:assign]{\textsc{Exec-Assign}} allows us to update the execution predicate into $\onlyExec_X{\color{RoyalBlue}(} {\absvar{p} = l_x \wedge \absvar{q} = l_y \wedge \absvar{r} = l_r \wedge \absvar{z}_1 = z_x \wedge \absvar{z}_2 = z_y}{\color{RoyalBlue},}{\color{purple}\choice{\ldots}{\ldots};\cwhile{\ldots}{\ldots}}{\color{RoyalBlue})}$.
If $z_x < z_y$, the low-level program appends the node from $\convar{x}$ to the result list.
On the heap, we unfold $\sll{\convar{x}}{l_x}$ and transfer its head node to the list segment described by $\sllbseg{\&\convar{r}}{\convar{t}}{l_r}$, extending it to $l_rz_x$.
The pointer $\convar{t}$ is then updated to point to the next insertion point, which is exactly the new $\convar{x}$.
These steps are handled by standard separation logic reasoning with the assignment and frame rules.
For the high-level program, we apply \hyperref[execrule:pure]{\textsc{Exec-Pure}} and \hyperref[execrule:choicel]{\textsc{Exec-ChoiceL}} to select the left branch.
We then use \hyperref[execrule:assume]{\textsc{Exec-Assume}} to validate the condition, and \hyperref[execrule:assign]{\textsc{Exec-Assign}} to update both $\absvar{p}$ and $\absvar{r}$ appropriately.
This ensures that the execution predicate and heap remain in sync with the invariant.
If $z_x \geq z_y$, the argument is symmetric: the low-level program appends the node from $\convar{y}$, and the high-level program selects the right branch of the nondeterministic choice.

The loop exits when either $l_x = \nil$ or $l_y = \nil$.
Hence, the high-level loop condition ($\absvar{p} \neq \nil \vee \absvar{q} \neq \nil$) must be false, and rule \hyperref[execrule:whileend]{\textsc{Exec-While-End}} can reduce the high-level program in the execution predicate to $\cskip$.
For the low-level side, we consider $l_x = \nil$ (the case where $l_x = \nil$ is symmetric).
Then $\sll{\convar{x}}{l_x}$ simplifies to $\emp$, and combining the remaining heaps gives $\sll{\convar{r}}{l_rl_y}$, which matches the final postcondition in the desired Hoare triple.

\section{\label{sec:discuss}  Discussion}

\paragraph{\bfseries This paper has modified the validity of the relational Hoare triple with configuration refinement. Does this modification change the meaning of the refinement represented in relational triples?}  Indeed, under our modified definition, some triples that were previously invalid become valid. 
For example, the triple $\reltriplenofont{\pvar{x} = \pvar{y} \land \hspeccolor{\cskip}}{\pvar{x} := \pvar{x} + 1}{\pvar{x} = \pvar{y} \land \hspeccolor{\pvar{y} := \pvar{y} - 1}}$ is valid based on configuration refinement.
This is because we no longer require updates of high-level program configuration to precisely track program reductions. 
Instead, we allow rewriting the high-level configuration into a new one that ultimately produces the same final results.
However, for both relational Hoare triples based on multi-step transitions and configuration refinement, when used to express refinement, the high-level program in the postcondition must terminate (e.g., a $\cskip$ statement). 
That is to say, under both definitions, the relational triple $\reltriplenofont{\mathbb{P} \land \hspeccolor{\absc}}{\conc}{\mathbb{Q} \land \hspeccolor{\cskip}}$ always represents that the low-level program $\conc$ refines the high-level program $\absc$, as shown in \figref{fig:refinement}.
Besides, we have formalized proof rules in Rocq, confirming that the proof theory itself is preserved and thus demonstrating that our relaxed definition is sound.

\paragraph{\bfseries Does the final states set $X$ outside the encoded standard Hoare triple introduce any complications?} Simply put: No. Although our theory (as shown in \theoref{theo:enc_sec4}) transforms a relational Hoare triple into a standard Hoare triple universally quantified by a subset $X$, this final-state set $X$ is merely a placeholder. 
In practice, proving refinement reduces to proving a standard Hoare triple where we do not need to reason about what $X$ is in the verification process.
For instance, the examples in \secref{subsec:encproofs} and our case studies show that $X$ is just a placeholder during verification. 
In fact, universally quantified Hoare triples (not universally quantified assertions) are widely used in deduction-based program verification. 
For example, in the C program verification tool VST, a C function's specification has the form: $\forall \, a.\, \{P(x,y,z, \cdots, a)\} \, \pvar{ret} = f(x,y,z, \cdots,z) \, \{Q(\pvar{ret},a)\}$, where $a$ serves as a logical variable.

\paragraph{\bfseries What existing verification infrastructure from standard Hoare logic can be reused, and how?}
Our encoding reduces verifying $\forall\exists$ relational properties to proving standard Hoare triples about low-level programs.
Consequently, any proof strategy or tool compatible with standard Hoare logic and the execution predicate can be reused.
For example, tactics provided by VST-floyd \cite{DBLP:journals/jar/CaoBGDA18}—such as the 
\texttt{forward}  tactic for automatically forwardly stepping through code—remain fully reusable. 
Provers can thus rely on well-tested automation for local reasoning steps (e.g., pointer manipulations in the low-level program). 
In particular, we have integrated the execution predicate into a standard Hoare logic tool \cite{anon} and applied it in case studies. 
In these relational proofs, we reuse the tool’s symbolic executor, verification condition generator, and logic entailment infrastructure.

\paragraph{\bfseries Does reasoning with the execution predicate $\needExecX{\absasrt{P}}{\absc}{X}$ increase complexity relative to existing relational approaches, and is it amenable to automation?}
We argue it does not introduce additional complexity. 
Owing to our encoding theory, the original relational proof rules can be encoded into standard rules that incorporate the execution predicate. 
In prior relational works, updating the high-level configuration involves an angelic triple, as shown in rule  \hyperref[intro:absfocus]{\textsc{High-Focus}}.
In our encoding, the update to the execution predicate is accomplished by a consequence rule, and this update can also be captured by an angelic triple.
So, any strategies that can automate angelic Hoare triples can be adapted to handle  $\needExecX{\absasrt{P}}{\absc}{X}$.
For instance, verifier \textsc{Hypra} \cite{DBLP:journals/pacmpl/DardinierL024} provides a \texttt{hint} annotation to guide the instantiation of existential quantifiers in the SMT solver and thereby facilitates proof automation.
We believe a similar strategy can be applied to automate updating the execution predicate.
Furthermore, we provide update rules, such as \hyperref[execrule:pure]{\textsc{Exec-Pure}} in \figref{fig:needexec-rules}, to evaluate the high-level program within the execution predicate.
Existing $\forall\exists$ relational approaches also require high-level side rules, such as rule \textsc{PURE-R} in ReLoC~\cite{DBLP:journals/corr/abs-2006-13635}.
In particular, our update rules resemble these high-level rules.
Thus, our encoding neither fundamentally changes the way
people reason about program refinement nor adds complexity, and it should be readily automated with existing proof infrastructure.

\paragraph{\bfseries Since the $\forall\exists$ relational Hoare logic can already be encoded into the Hoare logic using ghost states and invariants, why propose a separate encoding based on standard Hoare logic}
In principle, a foundationally formalized verification tool could choose either standard Hoare logic or the Hoare logic with ghost states and invariants as its base framework. 
Our main contribution is theoretical and shows a different way of unifying relational Hoare logic with standard Hoare logic. 
While employing the Hoare logic with ghost states and invariants is common in concurrent program verification, standard Hoare logic sometimes enjoys better meta-theoretical properties in some sense. 
For example, the conjunction rule typically remains valid in standard Hoare logic but may fail for Hoare logics with ghost states.
Moreover, the soundness of standard Hoare logic is also easier to formalize, compared to the Hoare logic with ghost states and invariants.
For example,  as presented in Iris \cite{DBLP:journals/jfp/JungKJBBD18}, the weakest precondition definition involving ghost states and invariants (in its Sec. 7.3) is significantly more complex than the one without them (in its Sec. 6.3).


\section{\label{sec:relatedwork}Related Work}
To the best of our knowledge, this work provides the first theory that encodes the $\forall\exists$ relational Hoare logic into a standard Hoare logic framework. By introducing the execution predicate $\needExecX{\absasrt{P}}{\absc}{X}$  along with its associated proof rules, our encoding allows developers to leverage standard Hoare logic for relational reasoning without modifying the core framework or re-verifying its soundness.
In this section, we discuss related works including approaches to encoding one type of relational Hoare logic into some Hoare logic and frameworks for the $\forall\exists$ relational Hoare logic.

\subsection{Logic Encoding}
Although our primary contribution focuses on encoding $\forall\exists$ relational properties, there are alternative  relational logics. 
For instance, as discussed in the introduction, when dealing with deterministic programs, two-safety properties suffice to establish refinement. Below, we also discuss how such logics can also be encoded within some Hoare logic frameworks.

\paragraph{Encoding Two-Safety}
Barthe \cite{1310735} proposed self-composition, which enables proving $2$-safety properties in standard Hoare logic. 
This method is both straightforward and effective: for two programs, $c_1$ and $c_2$, variables in $c_2$ are renamed to ensure disjoint program states, producing a renamed program $c_2'$. The two programs can then be sequentially composed as $c_1;c_2'$.
Therefore, the 2-safety property for programs $c_1$ and $c_2$ is reduced to some property of all possible executions of program $c_1;c_2'$, allowing reasoning within standard Hoare logic. 
D’Osualdo et al. \cite{10.1145/3563298} employed the weakest liberal precondition assertion \footnote{The weakest precondition assertion \textsf{wp} used in their paper is referred to as weakest liberal precondition \textsf{wlp} in this paper and other works \cite{10.5555/550359}.} (\textsf{wlp}) and introduced a Logic for Hyper-triple Composition (LHC) to provide various composition rules for reasoning about $k$-safety properties—a generalization of $\forall\forall$ relational Hoare logic.
Focusing on $2$-safety, the hyper-triple $\mathbb{P} \Rightarrow \textsf{wlp}({[\conc , \absc]},{\mathbb{Q}})$ used in their work precisely encodes the $\forall\forall$ relational Hoare logic.
In contrast,  this paper focuses on representing the $\forall\exists$ relational Hoare logic using standard Hoare logic.
This task introduces additional complexities, particularly when addressing nondeterministic behavior.

\paragraph{Encoding Data Refinement}
Roever and Engelhardt \cite{Roever_Engelhardt_1998} explored how to prove a low-level program refines a high-level specification  within a generalized version of Hoare logic.
To be precise, their framework extends standard Hoare logic by introducing \textit{specification statements} in the form $P \leadsto Q$.
The statement $P \leadsto Q$ is used to represent the most general program $c$ that satisfies the Hoare triple $\unarytriple{P}{c}{Q}$. 
They demonstrated that the refinement relationship between a low-level program $\conc$ and a high-level specification $\absasrt{P} \leadsto \absasrt{Q}$ can be encoded as a Hoare triple of the low-level program $\conc$.
For example, they proved that given a binary assertion $\mathbb{R}$, the forward simulation  between  a low-level program $\conc$ and a high-level specification $\absasrt{P} \leadsto \absasrt{Q}$, denoted as  
\[ \conc \subseteq_{\mathbb{R}}^{L} \absasrt{P} \leadsto \absasrt{Q} \]
can be encoded as the following Hoare triple \footnote{This encoding requires that $\mathbb{R}$ is total and functional. That is for any low-level state $\const$, there exists a unique high-level state $\absst$ such that $(\const,\absst) \models R$.}:
\begin{equation}
    \unarytriple{\lambda \const.\,  \exists \absst. \, (\const, \absst) \models \mathbb{R} \land \absst \models \absasrt{P} }{\conc}{\lambda \const.\,  \exists \absst. \, (\const, \absst) \models \mathbb{R} \land \absst \models \absasrt{Q} }
    \label{eq:datarefine}
\end{equation}
\noindent 
In their framework,  specification statements play a central role, and the goal is to prove that a low-level program  refines a high-level specification.
Instead, our encoding theory does not require specification statements and focuses on establishing refinement directly between a low-level program and a high-level program. 
Such direct refinement is crucial because a  high-level program can be used to prove various specifications and to enable further refinements.
For instance, a recursive DFS implementation using heaps can be incrementally refined into a recursive DFS without heaps, and later into an iterative DFS without heaps \cite{lammich2015framework}. 
Note that that specifying the high-level  specification as $(\absst = \absst_0) \leadsto (\absst_0, \absst) \in \nrmdeno{\absc}$  implies that the encoding \eqref{eq:datarefine}  captures exactly the refinement of low-level program $\conc$ and high-level program $\absc$:
\[ \unarytriple{\lambda \const.\,  (\const, \absst_0) \models \mathbb{R} }{\conc}{\lambda \const.\,  \exists \absst. \, (\const, \absst) \models \mathbb{R} \land(\absst_0, \absst) \in \nrmdeno{\absc}} \]
This encoding still suffers from the same limitation as the naive approach in \secref{sec:enc}. Precisely, since it treats pre- and postconditions differently, it is unable to support powerful relational proof rules (e.g., \hyperref[intro:relseq]{\textsc{Rel-Seq}} and  \hyperref[relwh]{\textsc{Rel-Wh}} ).
Overcoming this limitation requires a uniform encoding for preconditions and postconditions, which our coding theory achieves.
Besides, the refinement relationship between a low-level program $\conc$ and a high-level program $\absasrt{P} \leadsto \absasrt{Q}$ can be proved using the encoded vertical composition rule.

\paragraph{Encoding $\forall\exists$ Properties into Product Programs}
In the context of product programs, Barthe \cite{DBLP:conf/lfcs/BartheCK13} has incorporated $\forall\exists$ pattern into the program semantics to support refinement reasoning.
They introduced an asymmetric concept of product programs through control-flow graphs.
This concept involves the use of universal quantification over the execution traces of the low-level program and existential quantification over those of the high-level program.
However, in this paper, the  $\forall\exists$ pattern is hidden in the encoded assertions and we employ the denotational semantics to capture observational behaviors. 

\begin{figure}[ht]
    \centering
    \vspace{-25pt}
   \includegraphics[width = 0.55\textwidth]{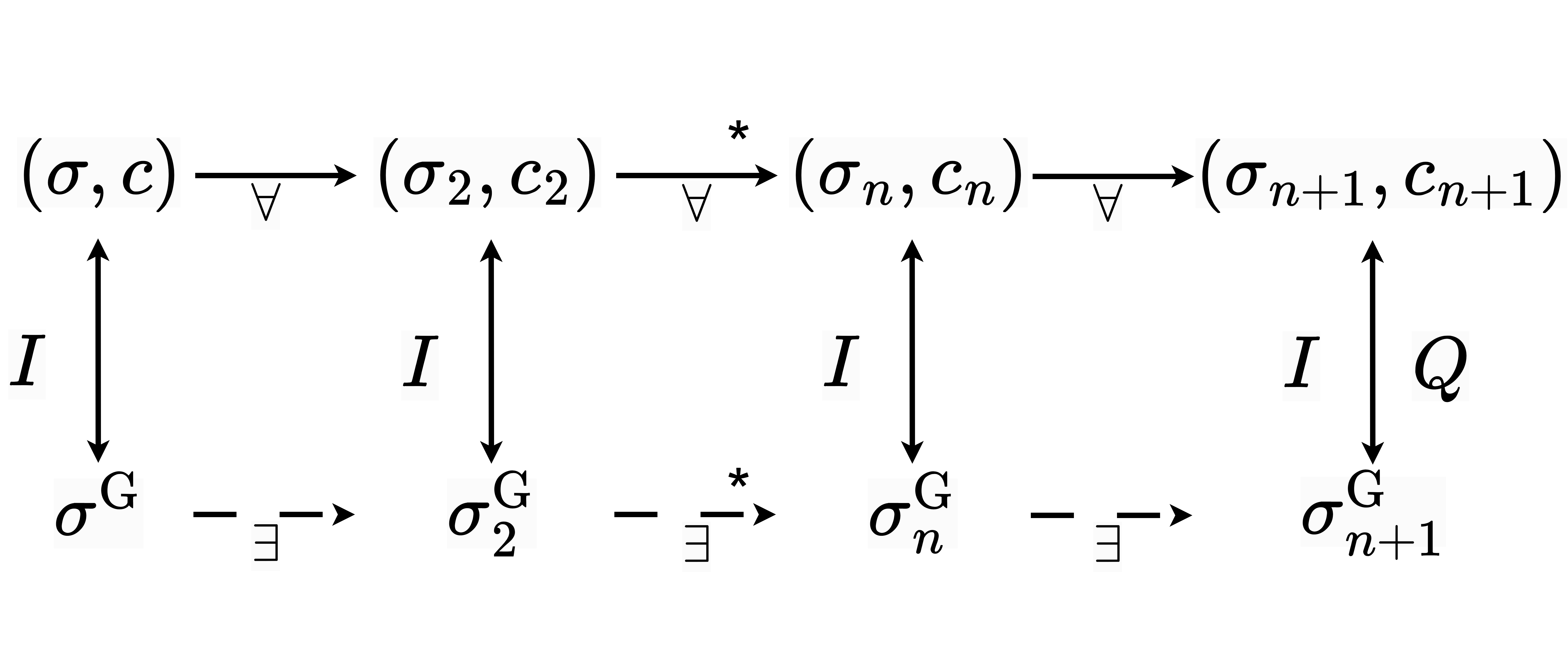}
   \vspace{-15pt}
        \caption{\label{fig:ghost}Relational Hoare Triples}
\end{figure}

\paragraph{Encoding $\forall\exists$ Properties into a Hoare logic with Ghost States and Invariants}
In concurrent program verification, Turon et al. \cite{10.1145/2500365.2500600} proposed treating high-level programs as resources, and  Iris \cite{DBLP:conf/popl/JungSSSTBD15,DBLP:journals/jfp/JungKJBBD18} established a $(\forall\exists)^{\omega}$ separation Hoare logic with ghost states and invariant. 
Iris employs weakest preconditions as primitive and a Hoare triple $\unarytriple{P}{c}{Q}$ can be written as $P \Rightarrow \wpre{c}{Q}$ \footnote{The triple is actually defined as $\square (P \sepimp \wpre{c}{Q})$. We omit the persistence modality $\square$ and replace the separating implication $\sepimp$ with $\Rightarrow$ for simplicity. }.
When ghost states and invariants are involved, we write   $\wpreghost{c}{Q}{I}$ to denote the weakest precondition.
Intuitively, as \figref{fig:ghost} \footnote{The logic in Iris is  a separation logic; however, for simplicity, frames have been omitted in this figure.} shows, $\wpreghost{c}{Q}{I}$ is the largest set of physical/ghost-state  pairs such that:
\begin{itemize}
    \item if program $c$ can reduce, then  we open the invariant $I$ and demonically obtain a physical state $\sigma$ and a matching ghost state $\gst$  such that $(\sigma,\gst) \models I$. 
    For any possible small-step transition $(\sigma,c) \rightarrow (\sigma_2, c_2)$,  we can angelically update  $\gst$ to  some new ghost state $\gst_2$ in a way that preserves the invariant $I$.
    This update process continues for $c_2$.
    \item if program $c$ ends, then  postcondition $Q$ holds.
\end{itemize}
Subsequent frameworks \cite{DBLP:conf/esop/TassarottiJ017,DBLP:journals/corr/abs-2006-13635,Transfinite_Iris} leverage this $(\forall\exists)^{\omega}$ Hoare logic to encode the $\forall\exists$ relational Hoare logic for verifying program refinement.
A common strategy is to instantiate the ghost state as a high-level program configuration and set the invariant  $\mathcal{I}$ as all possible configurations from an initial configuration $ (\absst_0, \absc_0)$ \cite{DBLP:journals/corr/abs-2006-13635}.
\[ \mathcal{I} \defeq \lambda \const,\absst,\absc.\, (\absst_0, \absc_0) \rightarrow^* (\absst, \absc) \]
Since this invariant imposes constraints only on the ghost state rather than the physical state,  the low-level program’s $\forall$ transitions do not depend on the previous $\exists$ updates of the high-level configuration.
Then $\exists$ can be moved behind, $\forall$ can be moved forward, and all $\forall$s and all $\exists$s can be gathered together, reducing the resulting weakest precondition to a $\forall\exists$ pattern.
Then the refinement between programs $\conc$ and $\absc$ is encoded as $\exists \absst_0, \absc_0.\,  \mathcal{I} \Rightarrow \wpreghost{\conc}{\text{StateMatch}}{\mathcal{I}}$.
In contrast, this paper aims to encode the $\forall\exists$ relational Hoare logic into the $\forall$-structured standard Hoare logic.
\subsection{\label{subsec:rhlworks}{Relational Hoare logic frameworks}}

Relational Hoare logic, introduced by Benton \cite{DBLP:conf/popl/Benton04}, extends standard Hoare logic to relate two programs. 
Many relational verification frameworks have been proposed to verify relational properties, including the $\forall\forall$ properties \cite{DBLP:journals/tcs/Yang07,DBLP:conf/pldi/KunduTL09,DBLP:journals/pacmpl/0001DLC18,DBLP:journals/pacmpl/MaillardHRM20,DBLP:journals/toplas/BanerjeeNNN22} and the $\forall\exists$ properties \cite{DBLP:conf/lfcs/BartheCK13,DBLP:journals/pacmpl/ClochardMP20}.
To handle function calls in relational program verification, Song et al. \cite{DBLP:journals/pacmpl/SongCLHSD23} introduced the judgment $S \vdash \conc \sqsubseteq \absc$, which represents that $\conc$ refines $\absc$ when the modules are used according to the specification $S$.
They proposed to insert \textsf{assume} and \textsf{assert} statements in programs to encode the preconditions and postconditions of functions, and demonstrated that these inserted statements can be safely erased. 
In our work, since we reduce relational Hoare logic to standard Hoare logic, function calls can be  straightforwardly handled due to the direct application of the call rules in standard Hoare logic.

Some relational frameworks \cite{DBLP:journals/pacmpl/MaillardHRM20,DBLP:conf/isola/Naumann20, DBLP:journals/toplas/BanerjeeNNN22,DBLP:conf/aplas/DickersonYZD22} employ relational Hoare quadruples as the primary judgment and derive various proof rules for quadruples.
Notably, quadruples representing program refinement, denoted as $\quadruple{P}{\conc}{\absc}{Q}$, also exhibit a $\forall\exists$ pattern:
for any execution of the low-level program $\conc$, there exist an execution of the high-level program $\absc$ to capture it.
Then quadruples can be transformed into equivalent relational Hoare triples, which in turn can be encoded into standard Hoare triples. 
However, these works do not investigate how to encode the $\forall\exists$-structured  relational quadruples directly into $\forall$-structured standard triples.
Besides, some works \cite{DBLP:journals/pacmpl/AntonopoulosKLNNN23, DBLP:conf/lics/NagasamudramN21} based on quadruples  focus on the alignment problem between two programs. 
We believe that approaches for alignment in these frameworks can also be adapted to relational Hoare triples.

Hyper Hoare logic \cite{DBLP:journals/corr/abs-2301-10037} extends standard Hoare logic by employing assertions over sets of states as preconditions and postconditions. This work supports universal and existential quantifiers over these sets and can be used to reason about the $\forall\exists$ properties of a single program. 
However, this work does not address the reduction of $\forall\exists$ relational Hoare logic to standard Hoare logic.

Various verification tools, including Dafny \cite{10.5555/1939141.1939161}, Verus \cite{10.1145/3586037}, VeriFast \cite{10.1007/978-3-642-20398-5_4}, and Why3 \cite{10.1007/978-3-642-37036-6_8}, employ ghost states and ghost code to assist in proving properties of the real (i.e., compiled or executable) code.
These tools ensure that real code with ghost extensions simulates the original real code after the ghost elements are erased. 
In other words, they do not ensure that a low-level (real) program refines a high-level (ghost) program and then prove the correctness of that high-level program. 

IronFleet~\cite{DBLP:conf/sosp/HawblitzelHKLPR15}, built on Dafny, verifies that the pre- and post-states of the implementation match a transition in a high-level (abstract) state machine.
This approach treats the high-level program as a relation over pre- and post-states, similar to the $P \leadsto Q$ specifications of Roever and Engelhardt~\cite{Roever_Engelhardt_1998}.
The high-level programs in both approaches lack structural constructs or intermediate states.
Consequently, when the low-level implementation includes loops, it is difficult to relate intermediate iterations to specific control points in the high-level program.
Refinement proofs in both approaches are indeed standard Hoare proofs showing that the execution of the low-level program preserves a relational specification between its initial and final states.

\subsection{Other Related Work}

Incorrectness logic \cite{DBLP:journals/pacmpl/OHearn20}, which is proposed to reason about the presence of bugs, shows a $\forall\exists$  pattern.
Its judgment $[P] \ c \ [Q]$ describes that for any state $\sigma$ within postcondition $Q$, there exists a valid execution from some state in precondition $P$,  which is exactly the reverse direction of the angelic Hoare triple $\angelvdash \unarytriple{P}{c}{Q}$ in this paper.

\section{\label{sec:concl}Conclusion and Future Work}
In this paper, we introduce the first encoding theory that reduces the $\forall\exists$ relational Hoare logic to the $\forall$-structured standard Hoare logic. 
This encoding theory suggests that standard Hoare logic, despite its unary assertions, can verify program refinement.
Consequently, relational Hoare logic can be directly established based on the existing formalizations of standard Hoare logic and then inherit proof rules embedded in these formalizations.

However, our syntactic encoding targets decomposed assertions, which requires users to explicitly write relational decomposed assertions when employing standard Hoare logic with our execution predicate to verify $\forall\exists$ properties. Additionally, this work does not yet address concurrency or liveness properties, which are essential in many verification scenarios.

In the future, we aim to adapt the encoding theory to support reasoning about refinements that involve more complex observational behaviors. We also plan to extend the approach to additional language features, including concurrent programs.

%
\begin{acks}
We thank David A. Naumann, Zhenjiang Hu, Zhongye Wang, Xinyi Wan, Yiyuan Cao, and Zhiyi Wang for their valuable feedback. 
We are also grateful to the anonymous reviewers for their constructive comments and suggestions.
This work was partially supported by NSF China 62472274.
\end{acks}

\section*{Data-Availability Statement}
An artifact supporting this paper is available on Zenodo \cite{wu2025enc_rel_theory}.
It provides a Rocq-based formalization of our encoding theory, machine-checked proofs, and case studies (BST, DFS, mergesort, KMP) demonstrating relational verification in standard Hoare logic.

\bibliographystyle{ACM-Reference-Format}
\bibliography{main}

\appendix

\newpage
\section{Notations}
The notations used in this paper are summarized in the reference table, as shown in Table \ref{tab:notation}.

\begin{table*}[ht]
\centering
\renewcommand{\arraystretch}{1.4} 
\begin{tabular}{|c|l|}
\hline
\textbf{Notation} & \textbf{Description} \\ \hline
$\conSigma$ & The set of low-level program states \\ \hline
$\absSigma$ & The set of high-level program states \\ \hline
$\conprog$ & The set of low-level programs \\ \hline
$\absprog$ & The set of high-level programs \\ \hline
$\const$($\in \conSigma$) & A low-level program state \\ \hline
$\absst$($\in \absSigma$) & An high-level program state \\ \hline
$\conc$($\in \conprog$) & A low-level program \\ \hline
$\absc$($\in \absprog$) & An high-level program \\ \hline
$\nrmdeno{c}$($\subseteq \Sigma \times \Sigma$) & Normal termination case of denotational semantics of program $c$ \\ \hline
$\errdeno{c}$($\subseteq \Sigma$) & Error case of denotational semantics of program $c$ \\ \hline
$\conasrt{P}$($\subseteq \conSigma$) & Unary assertion about low-level program states \\ \hline
$\absasrt{P}$($\subseteq \absSigma$)  & Unary assertion about high-level program states \\ \hline
$\mathbb{P}$($\subseteq \conSigma \times \absSigma$)  & Binary assertion about states of both programs \\ \hline
$\mathcal{P}$($\subseteq \conSigma \times \absSigma \times \absprog$)  & Program-as-resource assertion \\ \hline
$\lasrt{\conasrt{P}}$ & Lifted program-as-resource assertion for $\conasrt{P}$ \\ \hline
$\hasrt{\absasrt{P}}$ & Lifted program-as-resource assertion for $\absasrt{P}$ \\ \hline
$\hspeccolor{\absc}$ & Lifted program-as-resource assertion for $\absc$ \\ \hline
$\encasrt{\mathcal{P}}_X$ & Encoding of a program-as-resource assertion based on state set $X$ \\ \hline
$\mathsf{Enc}_X{\mathcal{P}}$ & Syntactic Encoding  based on postcondition $X$ \\ \hline
$\conasrtenc$ & Syntactic low-level assertion extended with the execution predicate \\ \hline
$\reltriple{P}{\conc}{Q}$ & Relational Hoare triple \\ \hline
$\demovdash \unarytriple{P}{c}{Q}$ & Standard Hoare triple \\ \hline
$\angelvdash \unarytriple{P}{c}{Q}$ & Angelic Hoare triple \\ \hline
$\textsf{wlp}(c, Q)$ & Weakest liberal precondition \\ \hline
$\needExecX{\absasrt{P}}{\absc}{X}$ & The execution predicate  about high-level program execution \\ \hline
\end{tabular}
\vspace{2mm}
\caption{Notation table for key symbols and definitions.}
\label{tab:notation}
\end{table*}

\section{\label{app:theorem}Theorems}
Here we present the conclusions in this paper and provide detailed proofs for them.

\begin{customthm}{4}[Encoding Relational Triples] For any low-level statement $\conc$, and assertions $\mathcal{P}$, $\mathcal{Q}$ $\subseteq \conSigma \times \absSigma \times \absStmts$:
 \begin{align*}
    \underbrace{\langle \mathcal{P} \rangle \  \conc \ \langle \mathcal{Q}\rangle}_{\mathcal{J}}
     \text{ is valid}   \quad \text{iff.} \quad 
     \underbrace{\forall X. \ \demovdash \{ \encasrt{ \mathcal{P}}_X \} \ \conc \ \{ \encasrt{ \mathcal{Q} }_X \}}_J 
 \end{align*} 
\end{customthm}
\begin{proof}  we prove two directions:
    \begin{itemize}
        \item $\Rightarrow$:  \,  For any $X$ and any $\const_1$ in $\encasrt{\relasrt{P}}_X$, there exist  $\absst_1$ and $\absc_1$ such that $\absst_1 \models \weakestpre{\absc_1}{X}$ and $(\const_1, \absst_1, \absc_1) \models \relasrt{P}$. Then for any terminating state $\const_2$ such that $(\const_1,\const_2) \in \nrmdeno{\conc}$, according to relational triple $\mathcal{J}$, there exist $\absst_2$ and $\absc_2$ such that $(\absst_1, \absc_1) \hookrightarrow (\absst_2, \absc_2)$ and 
        $(\const_2, \absst_2, \absc_2) \models \relasrt{Q}$. Then by \theoref{theo:decom}, we have 
        $\absst_2  \models \weakestpre{\absc_2}{X}$.
        Therefore, according to \defref{enc:asrtenc} we derive 
        $\const_2 \models$ $\encasrt{\mathcal{Q}}_X$.
        \item $\Leftarrow$: Based on \propref{prop:10}, for any $(\const_1, \absst_1, \absc_1) \models \relasrt{P}$,  we instantiate $X$ as 
        the high-level terminal states set
        $\{ \absst_3 \,| \, (\absst_1,\absst_3) \in \nrmdeno{\absc_1}\}$. 
        By the definition of assertion encoding, $\const_1 \models \encasrt{\mathcal{P}}_X$.
        According to standard triple $J$, for any $\const_2 $ such that $(\const_1, \const_2) \in \nrmdeno{\conc}$, we have $\const_2 \models  \encasrt{\mathcal{Q}}_X$.
        By unfolding the assertion encoding, there exist $\absst_2$ and $\absc_2$ such that $\absst_2  \models \weakestpre{\absc_2}{X}$ and $(\const_2,\absst_2, \absc_2) \models \relasrt{Q}$.
        According to the definition of weakest preconditions, for any $\absst_3$ such that $(\absst_2, \absst_3) \in \nrmdeno{\absc_2}$, $\absst_3 \models X$.
        That is for any $\absst_3$ such that $(\absst_2, \absst_3) \in \nrmdeno{\absc_2}$, we have $(\absst_1, \absst_3) \in \nrmdeno{\absc_2}$.
        Therefore, we finally derive $(\absst_1, \absc_1) \hookrightarrow (\absst_2, \absc_2)$ and $(\const_2,\absst_2, \absc_2) \models \relasrt{Q}$.
    \end{itemize}
\end{proof}

\begin{customthm}{16}[Encoded Assertion Transformation] ~

 \begin{enumerate}[label=(\alph*)]
    \item $ \encasrt{\exists a. \mathcal{P}(a)}_X \iff \exists a. \,\encasrt{\mathcal{P}(a)}_X$.
    \item $ \encasrt{B \land \mathcal{P}}_X \iff  B \land \encasrt{\mathcal{P}}_X$.
    \item $ \encasrt{\lasrt{\conasrt{P}} \land \mathcal{P} }_X \iff \conasrt{P} \land \encasrt{\mathcal{P}}_X $. 
    \item $ \encasrt{\mathcal{P}_1 \vee \mathcal{P}_2 }_X \iff \encasrt{\mathcal{P}_1}_X \vee \encasrt{\mathcal{P}_2}_X $. 
\end{enumerate} 
\end{customthm}
\begin{proof}  ~
    \begin{enumerate}[label=(\alph*)]
\item for any $\const \models \encasrt{\exists a. \mathcal{P}(a)}_X$, we have $ \exists a. \, \exists \absst \absc. \,$ $(\const, \absst, \absc) \models  \mathcal{P}  \land  \absst \models \weakestpre{\absc}{X}$, which is equivalent to that 
 there exists $a$ such that 
 $\const \models \encasrt{\mathcal{P}(a)}$.
 \item for any $ \const \models  \encasrt{B \land \mathcal{P}}_X$, we have $ \exists \absst \absc. \,(\const, \absst, \absc) \models  \mathcal{B \land \mathcal{P}} \land  \absst \models \weakestpre{\absc}{X}$.
 Since $(\const, \absst, \absc) \models  {B \land \mathcal{P}}$ is equivalent to $B \land  (\const, \absst, \absc) \models  {\mathcal{P}}$, then the $ \const \models  \encasrt{B \land \mathcal{P}}_X$
  is equivalent to that 
 $ B \land \exists \absst \absc. \,  (\const, \absst, \absc) \models  \mathcal{\mathcal{P}} \land  \absst \models \weakestpre{\absc}{X}$. Then
 $\const \models B \land \encasrt{\mathcal{P}}_X$.
 \item for any $ \const \models  \encasrt{\lasrt{\conasrt{P}} \wedge \mathcal{P} }_X$, we have $ \exists \absst \absc_0. \, $ $(\const, \absst, \absc_0) \models  {\lasrt{\conasrt{P}} \wedge \mathcal{P} } \land  \absst \models \weakestpre{\absc}{X}$.
 Here $(\const, \absst, \absc_0) \models  {\lasrt{\conasrt{P}} }$ is equivalent to $\const \models \conasrt{P} $.
 Then we know $ \const \models  \conasrt{P} \land \encasrt{\mathcal{P}}_X$.
 \item for any $\const \models \encasrt{\mathcal{P}_1 \vee \mathcal{P}_2}_X$, we have $ \exists a. \, \exists \absst \absc. \,$ $(\const, \absst, \absc) \models  \mathcal{P}_1 \vee \mathcal{P}_2  \land  \absst \models \weakestpre{\absc}{X}$, which is equivalent to that 
 $(\const, \absst, \absc) \models  \mathcal{P}_1 \land  \absst \models \weakestpre{\absc}{X}$ or $(\const, \absst, \absc) \models  \mathcal{P}_2 \land  \absst \models \weakestpre{\absc}{X}$.
 \end{enumerate}
\end{proof}

\begin{corollary}
$\encasrt{ \exists \vec{a}.\, B(\vec{a}) \wedge \lasrt{\conasrt{P}} \wedge \hasrt{\absasrt{P}} \wedge \hspeccolor{\absc} }_X
\iff
\exists \vec{a}.\, B(\vec{a}) \wedge \conasrt{P}(\vec{a}) \wedge \needExecX{\absasrt{P}}{\absc}{X}$
\end{corollary}
\begin{proof}
    for any $ \const \models  \encasrt{\lasrt{\conasrt{P}} \land \hasrt{\absasrt{P}} \land \hspeccolor{\absc} }_X$, we have $ \exists \absst \absc_0. \, $ $(\const, \absst, \absc_0) \models  \lasrt{\conasrt{P}} \land \hasrt{\absasrt{P}} \land  \hspeccolor{\absc} $$ \land  \absst \models \weakestpre{\absc_0}{X}$.
 Here $(\const, \absst, \absc_0) \models  {\lasrt{\conasrt{P}} \land \hasrt{\absasrt{P}} \land \hspeccolor{\absc} }$ is equivalent to $\const \models \conasrt{P} \land \absst \models \absasrt{P} \land \absc_0 = \absc $.
 Then we know $ \const \models  \encasrt{\lasrt{\conasrt{P}} \land \hasrt{\absasrt{P}} \land \hspeccolor{\absc} }_X$ is equivalent to $\const \models \conasrt{P} \land \exists \absst.\, \absst \models \absasrt{P} \land \absst \models \weakestpre{\absc}{X} $.
\end{proof}

\begin{corollary}~
    Rule \hyperref[rule:vc]{\textsc{VC-FC}} and 
    rule \hyperref[rule:vcr]{\textsc{VC-Refine}} in relational Hoare logic can be derived using the consequence rule and rule \hyperref[intro:ex]{\textsc{ExIntro}} in standard Hoare logic.
  \end{corollary}
  \begin{proof} 
    Based on encoding theorem, the two relational rules are transformed into the following two  rules:
    \begin{mathpar}
        \inferrule[\label{rule:tradvc}\scshape Enc-VC-FC]
        { \forall X. \demovdash \unarytriple{ \encasrt{ \mathbb{P} \land \hspeccolor{\absc}}_X
        }{\conc}{ \encasrt{ \mathbb{Q} \land \hspeccolor{\cskip}}_X}  \\ 
        \demovdash \{  \absasrt{P} \} \ \absc \ \{\absasrt{Q}\} 
        }
        { \demovdash \{  \mathbb{P} \odot \absasrt{P} \} \  \conc \ \{\mathbb{Q} \odot \absasrt{Q} \} } \\
        \inferrule[\label{rule:tradvcr}\scshape Enc-VC-Refine]
        {
          \forall X_1. \demovdash \unarytriple{ \encasrt{ \mathbb{P}_1 \land \hspeccolor{c_2}}_{X_1}
          }{c_1}{ \encasrt{ \mathbb{Q}_1 \land \hspeccolor{\cskip}}_{X_1}}  \\  
        \forall X_2. \demovdash \unarytriple{ \encasrt{ \mathbb{P}_2 \land \hspeccolor{c_3}}_{X_2}
        }{c_2}{ \encasrt{ \mathbb{Q}_2 \land \hspeccolor{\cskip}}_{X_2}}  
        }
        { \forall X. \demovdash \unarytriple{ \encasrt{ \mathbb{P}_1 \circ  \mathbb{P}_2 \land \hspeccolor{c_3}}_X
        }{c_1}{ \encasrt{ \mathbb{Q}_1 \circ  \mathbb{Q}_2  \land \hspeccolor{\cskip}}_X}  }
      \end{mathpar}
Then we prove the soundness of the two rules:

    \begin{itemize}
      \item rule \hyperref[rule:tradvc]{\textsc{Enc-VC-FC}}: to prove the triple $ \{ \mathbb{P} \odot \absasrt{P}  \} \  \conc \ \{\mathbb{Q} \odot  \absasrt{Q} \}$, according to the definition of assertion linking,
      we can first apply  rule \hyperref[intro:ex]{\textsc{ExIntro}}. 
      The goal then becomes proving that for any $\absst_1$, the triple  $ \{ \lambda \const_1.\,  (\const_1, \absst_1) \models \mathbb{P} \land \absst_1 \models \absasrt{P}  \} \  \conc \ \{\mathbb{Q} \odot  \absasrt{Q} \}$ is $\forall$-valid.
      Next, we instantiate $X$ as  $\{\absst \, | \, (\absst_1, \absst) \in \nrmdeno{\absc}\}$, and 
      then the precondition $(\lambda \const_1.\,  (\const_1, \absst_1) \models \mathbb{P} \land \absst_1 \models \absasrt{P})$ implies the encoded precondition $\encasrt{\mathbb{P} \land \hspeccolor{\absc}}_X$.
      Focus on the encoded postcondition $\encasrt{\mathbb{Q} \land \hspeccolor{\cskip}}_X$, it is equivalent to $(\lambda \const_2.\, \exists \absst_2.\,  (\const_2, \absst_2) \models \mathbb{Q} \land \absst_2 \models \weakestpre{\cskip}{X})$.
      Here by definition,  $\absst_2 \models \weakestpre{\cskip}{X}$ is equivalent to $(\absst_1,\absst_2) \in \nrmdeno{\absc} $.
      Then, by the triple $\{ \absasrt{P} \} \ \absc \ \{\absasrt{Q}\}$, we know that $\absst_2 \models \absasrt{Q}$.
      Thus, we  derive  $\mathbb{Q} \odot  \absasrt{Q}$.
      \item rule \hyperref[rule:tradvcr]{\textsc{Enc-VC-Refine}}:
      for any $X$, we need to prove the triple $ \{ \encasrt{ \mathbb{P}_1 \circ  \mathbb{P}_2 \land \hspeccolor{c_3}}_X
        \} \, {c_1} \, \{ \llparenthesis \mathbb{Q}_1 \circ  \mathbb{Q}_2  \land \hspeccolor{\cskip} \rrparenthesis_X\} $. 
    Next, we instantiate $X_2$ as $X$ and unfold the assertion encoding in  triple $ \unarytriple{ \encasrt{ \mathbb{P}_2 \land \hspeccolor{c_3}}_{X_2}
        }{c_2}{ \encasrt{ \mathbb{Q}_2 \land \hspeccolor{\cskip}}_{X_2}} $, thus resulting  the following triple:
            \begin{equation*}
                \begin{aligned}
                  & \left\{ \begin{array}{l}
                    { \lambda \sigma_2. \, \exists \sigma_3. \, (\sigma_2, \sigma_3) \models \mathbb{P}_2 \land \sigma_3 \models \weakestpre{c_3}{X}
                    }
                  \end{array}  \right\} \\ 
                  & \quad\,\,\,\, c_2  \\
                  & \left\{ \begin{array}{l}
                    \lambda \sigma_2'. \, \exists \sigma_3'. \, (\sigma_2', \sigma_3') \models \mathbb{Q}_2 \land \sigma_3 \models \weakestpre{\cskip}{X}
                  \end{array}\right\}
                \end{aligned}
              \end{equation*}
    Then we can apply the  rule \hyperref[rule:tradvc]{\textsc{Enc-VC-FC}}, and the resulted precondition and postcondition are:
    \begin{footnotesize}
    \begin{align*}
        & \lambda \sigma_1. \, \exists \sigma_2.\,  ( \sigma_1 ,\sigma_2) \models \mathbb{P}_1  \land \exists \sigma_3. \, (\sigma_2, \sigma_3) \models \mathbb{P}_2 \land \sigma_3 \models \weakestpre{c_3}{X}
          \\ 
        & \lambda \sigma_1'. \, \exists \sigma_2'.\,  ( \sigma_1' ,\sigma_2') \models \mathbb{Q}_1 \land \exists \sigma_3'. \, (\sigma_2', \sigma_3') \models \mathbb{Q}_2 \land \sigma_3 \models \weakestpre{\cskip}{X}
    \end{align*}
\end{footnotesize}

    \noindent 
    In fact, the resulting precondition and postcondition correspond to the precondition and postcondition of the goal, which are as follows:
    \begin{footnotesize}
    \begin{align*}
        & \lambda \sigma_1. \, \exists \sigma_3 \sigma_2. ( \sigma_1 ,\sigma_2) \models \mathbb{P}_1 \land  ( \sigma_2 ,\sigma_3) \models \mathbb{P}_2 \land \sigma_3 \models \weakestpre{c_3}{X} \\ 
        & \lambda \sigma_1'. \, \exists \sigma_3' \sigma_2'. ( \sigma_1' ,\sigma_2') \models \mathbb{Q}_1 \land  ( \sigma_2' ,\sigma_3') \models \mathbb{Q}_2 \land \sigma_3' \models \weakestpre{\cskip}{X}
    \end{align*}
\end{footnotesize}
    \end{itemize}

    \end{proof}

\begin{corollary}
    \[
    \inferrule*{
       \forall X. \demovdash \left\{\exists \, u.\, 
      {\begin{aligned}& \needExecX{\absstorepre(u)}{\absc}{X}\\ & \land \constorepre(u)\end{aligned}}
  \right\} \, \conc \, 
  \left\{\exists \, v.\, 
      {\begin{aligned}& \needExecX{\absstorepost(v)}{\absc}{X}\\ & \land \constorepost(v)\end{aligned}}
  \right\}
   \\ 
  \demovdash \{\exists\, u.\,  B_1(u) \wedge \absstorepre(u)\} \ \absc \ \{ \forall\, v.\,  \absstorepost(v) \Rightarrow B_2(v) \}
  \\
\forall \,u.\, B_1(u) \Rightarrow \mathsf{inhabitant}( \absstorepre(u))
    }{
       \demovdash \{\exists\, u.\, B_1(u) \wedge \constorepre(u)\} \  \conc \ \{ \exists \, v. \, B_2(v) \wedge \constorepost(v)\} 
    }
\]
\end{corollary}
\begin{proof}
To prove the triple $\{\exists\, u.\, B_1(u) \wedge \constorepre(u)\} \  \conc \ \{ \exists \, v. \, B_2(v) \wedge \constorepost(v)\} $,  we can first apply Hoare rules to introduce the logical variable $u$ such that $B1(u)$.
According to  $\forall \,u.\, B_1(u) \Rightarrow \mathsf{inhabitant}( \absstorepre(u))$, we know that there exists some  $\absst_1 \models \absstorepre(u)$. 
Next, we instantiate $X$ as  $\{\absst \, | \, (\absst_1, \absst) \in \nrmdeno{\absc}\}$, and 
then the predicate $\needExecX{\absstorepre(u)}{\absc}{X}$ holds.
Thus, the encoded precondition $\needExecX{\absstorepre(u)}{\absc}{X} \wedge \constorepre(u)$ holds.
Then for any $\const_2$ such that $(\const_1,\const_2) \in \nrmdeno{\absc}$, there exists $v$ such that $\const_2 \models \needExecX{\absstorepost(v)}{\absc}{X}\land \constorepost(v)$.
Then there exists $\absst_2$ such that $\absst_2 \models \weakestpre{\cskip}{X})$ and $\absst_2 \models \absstorepost(v)$.
Here by definition,  $\absst_2 \models \weakestpre{\cskip}{X}$ is equivalent to $(\absst_1,\absst_2) \in \nrmdeno{\absc} $.
Then, by the triple $\{\exists\, u.\,  B_1(u) \wedge \absstorepre(u)\} \ \absc \ \{ \forall\, v.\,  \absstorepost(v) \Rightarrow B_2(v) \}$, we know that $B_2(v) $ holds.
Thus, we  derive  $\const_2 \models B_2(v) \wedge \constorepost(v)$.
\end{proof}

\section{\label{app:ext}Extension of the Encoding Theory} 

\subsection{\label{sec:pcall}Interaction with Function Calls}
In this section, we focus on  programs that include function calls.
Building on the idea of standard contextual Hoare triples, we propose relational contextual Hoare triples, which incorporate proof contexts for low-level programs while remaining context-free for high-level programs.
Then we extend the encoding theory and transform relational contextual Hoare triples into standard contextual Hoare triples.

\subsubsection{Preliminary: Denotational Semantics and Hoare Logic for Function Calls}
For program languages instrumented with function calls, the denotational semantics should be parameterized by the environment $\chi$, which specifies the behavior of the callees.
\begin{definition}[Denotational Semantics with Call] Let $\fid$ denote the set of function names, and suppose $\chi : \fid \rightarrow \mathscr{P}(\Sigma \times \Sigma)$ is an environment that defines callee functions' behavior. For any program $c$, $\nrmfcdeno{c}$ denotes its denotational semantics. In particular, the denotational semantics for function calls is defined as follows:
  \begin{equation*}
      \begin{aligned}
          & \nrmfcdeno{\ccall{f}} \defeq \{(\sigma_1, \sigma_2) |  (\sigma_1, \sigma_2)\in \chi(f) \} 
      \end{aligned}
  \end{equation*}
  \end{definition}

\begin{definition}[Configuration Refinement with Call]
  Given the environment  $\chi : \fid \rightarrow \mathscr{P}(\Sigma \times \Sigma)$,  $(\sigma_1, c_1)$ can transition to  $(\sigma_2, c_2)$, denoted as $(\sigma_1, c_1) \hookrightarrow^{\chi}  (\sigma_2, c_2)$,  if for any state $\sigma_3$, $(\sigma_{2} ,\sigma_{3}) \in \nrmfcdeno{c_2}$ implies $(\sigma_{1} ,\sigma_{3}) \in \nrmfcdeno{c_1}$.
\end{definition}

\noindent Moreover, for a closed program $\pro$ composed of finite pairs of function names $f_i$ and corresponding function bodies $c_i$, the program semantics
\(\nrmdeno{\pro}\)
is given by the least fixed point
$\chi$ satisfying
$\forall (f_i : c_i) \in \pro . \ \nrmfcdeno{c_i} = \chi(f_i)$, which can be computed by Kleene fixed-point theorem.
\begin{definition}[Program Semantics] For any program \(\pro\), its semantics \(\nrmdeno{\pro}\) is the least fixed point of \(\Phi : (\fid \to \mathscr{P}(\Sigma \times \Sigma)) \to (\fid \to \mathscr{P}(\Sigma \times \Sigma))\), defined as follows:  

  \begin{enumerate}  
      \item the bottom \(\chi_0 : \fid \to \mathscr{P}(\Sigma \times \Sigma)\): \(\chi_0(f) = \emptyset\) for all \(f \in \fid\).  
      
      \item a monotonic function \(\Phi : (\fid \to \mathscr{P}(\Sigma \times \Sigma)) \to (\fid \to \mathscr{P}(\Sigma \times \Sigma))\) such that, for each \((f_i,c_i) \in \rho\),  
      \[
      \Phi(\chi)(f_i) = \nrmdeno{c_i}^{\chi},
      \]
      
      \item  the least fixed point  is given as:  
      \[
        \nrmdeno{\pro} = \bigcup_{n=0}^\infty \Phi^n(\chi_0),
      \]
      
  \end{enumerate} 
\end{definition}



Contextual Hoare triples simplify proofs by introducing a proof context $\Delta$, which consists of the specifications of all functions.
\begin{definition}[Function Specification and Proof Context] A function specification is parameterized by variables of any type and has the form $(f : \Pi (\overrightarrow{a}: A). \{P(\overrightarrow{a})\} \,\{Q(\overrightarrow{a})\})$, where $\overrightarrow{a}$ is a list of the logical variables,  the precondition $P(\overrightarrow{a}) \subseteq \Sigma$ and the postcondition $Q(\overrightarrow{a}) \subseteq \Sigma$. A proof context $\Delta$ for a program $\rho$ is a set of function specifications $(f_i : \Pi (\overrightarrow{a}: A_i). \{P_i(\overrightarrow{a})\} \,\{Q_i(\overrightarrow{a})\})$.
\end{definition}

\noindent Then the proof context $\Delta$ and the environment $\chi$ both characterize the behavior of functions and we use $\text{\normalfont Valid}(\chi, \Delta)$ to denote their consistency.
\begin{definition}[Valid Environment]
For a proof context $\Delta$, the environment $\chi$ is valid with respect to $\Delta$, denoted as $\text{\normalfont Valid}(\chi, \Delta)$, if 
for any function specification $(f : \Pi (\overrightarrow{a}: A). \{P(\overrightarrow{a})\} \,\{Q(\overrightarrow{a})\})$ in $\Delta$ and  any logical variable $\overrightarrow{a} \in A$, the following  holds:
if an initial state $\sigma_1 \models P(\overrightarrow{a})$, then for any terminal state $\sigma_2$ such that $(\sigma_1, \sigma_2) \in \chi(f)$,  we have $\sigma_2\models Q(\overrightarrow{a})$.
\end{definition}

\noindent Given a proof context $\Delta$, we can use any valid environment $\chi$ to account for the behavior of invoked functions in a statement.
Then, the validity of contextual Hoare triples can be defined as follows:
\begin{definition}[Contextual Hoare Triples] For a proof context $\Delta$, a program $c$, and assertions $P$, $Q \subseteq \Sigma$, we have
$\Delta \demovdash \unarytriple{P}{c}{Q}$ (is $\forall$-valid with respect to $\Delta$) if, given any environment $\chi$ such that $\text{\normalfont Valid}(\chi, \Delta)$, and for any initial state $\sigma_1 \models P$ and any final state $\sigma_2$ such that $(\sigma_1, \sigma_2) \in \nrmfcdeno{c}$, we have $\sigma_2 \models Q$.
\label{def:cht}
\end{definition}

\noindent 
The correctness of a  program is constructed from the correctness of all its functions. 
Contextual Hoare logic captures this by enabling users to verify contextual Hoare triples first and then deriving context-free triples through induction over the program’s fixed-point semantics. 
This can be formalized as the following theorem:
\begin{theorem} [Correctness of Contextual Hoare Triples] Given any program $\rho$ and its proof context $\Delta$, if for any function specification $(f_i : \Pi (\overrightarrow{a}: A_i). \{P_i(\overrightarrow{a})\} \{Q_i(\overrightarrow{a})\}) \in \Delta$, the function body $c_i$ can be proved to satisfy this specification  under the context $\Delta$, i.e., $\forall \overrightarrow{a} \in A_i. \, \Delta \demovdash \{P_i(\overrightarrow{a})\} \ c_i \ \{Q_i(\overrightarrow{a})\}$, then the program semantic $\nrmdeno{\rho}$ is valid with respect to $\Delta$, i.e., $\text{\normalfont Valid}(\nrmdeno{\rho}, \Delta)$.
    
\end{theorem}


\subsubsection{Relational Contextual  Hoare Triples}

For programs without function calls, the encoding theory, as described in \theoref{lem:enc}, demonstrates that relational Hoare triples can be encoded to a standard Hoare triple for low-level programs. 
Building on this encoding, a natural extension is to augment relational Hoare triples with proof contexts specific to low-level programs. 
In contrast, there is no need to introduce proof contexts for high-level programs. 
High-level programs are treated as resources within assertions, allowing them to undergo equivalent transformations during verification. 
As a result, function calls in high-level programs can be replaced with their corresponding function bodies.

This paper introduces two distinct proof contexts $\Delta$ and $\reldelta$ to relational Hoare triples. 
The proof context $\Delta$  contains only standard specifications and  $\reldelta$ is a relational proof context 
that consists of the relational specifications of low-level functions in the form $(f : \Pi (\overrightarrow{a} : A). \langle \mathcal{P}(\overrightarrow{a}) \rangle \langle  \mathcal{Q}(\overrightarrow{a})\rangle)$.
Then, we introduce how a low-level environment $\chi $ is valid concerning a relational proof context $\reldelta$ and present the definition for relational contextual Hoare triples.

\begin{definition}[Relational Valid Environment]For a relational proof context $\reldelta$, 
  the environment $\chi$ is valid with respect to $\reldelta$, denoted as $\text{\normalfont RelValid}(\chi, \reldelta)$,
if
for any  specification $(f : \Pi (\overrightarrow{a}: A). \langle \mathcal{P}(\overrightarrow{a})\rangle \langle \mathcal{Q}(\overrightarrow{a})\rangle)$ in $\reldelta$ and given any logical variable $\overrightarrow{a} \in A$,  the following holds:
for any $(\const_1,\absst_1,\absc_1) \models \mathcal{P}(\overrightarrow{a})$ and any low-level terminal state $\const_2$ such that $(\const_1, \const_2) \in \chi(f)$, there must exist high-level state $\absst_2$ and program $\absc_2$ such that  $(\absst_1, \absc_1) \hookrightarrow (\absst_2, \absc_2) $ and $ (\const_2 ,\absst_2, \absc_2) \models \mathcal{Q}(\overrightarrow{a})$.
\end{definition}

\begin{definition}[Relational Contextual  Hoare Triples] 
For a standard proof context $\Delta$,  a relational proof context $\reldelta$, a low-level program $\conc$, and assertions $\mathcal{P}$, $\mathcal{Q}$ $\subseteq \conSigma \times \absSigma \times \absStmts$, we have
$\Delta, \reldelta \vdash \reltriple{P}{\conc}{Q}$ if, given any environment $\chi$ such that $\text{\normalfont Valid}(\chi, \Delta)$ and $\text{\normalfont RelValid}(\chi, \reldelta)$, for any $(\const_1, \absst_1, \absc_1) \models \mathcal{P}$ and any $\const_2$ such that $(\const_1, \const_2) \in  \nrmfcdeno{\conc}$, {there exist} $\absst_2$ and $\absc_2$ {such that} $(\absst_1, \absc_1) \hookrightarrow  (\absst_2, \absc_2)$ and $(\const_2, \absst_2, \absc_2) \models \mathcal{Q}$.

\label{def:hqpc}
\end{definition}

\subsubsection{The Encoding Theory and Call Rules}

Since the relational proof context $\reldelta$ is absent in standard contextual Hoare triples, we should encode it into a standard proof context.
Notice that a function specification is parameterized by a type $A$, serving as logical variables that establish a connection between the precondition and postcondition for function $f$.
We can add a placeholder $X$  to these logical variables.
Then for each function specification in $\reldelta$,
we use the assertion encoding defined in \defref{enc:asrtenc} to encode its preconditions and postconditions into unary assertions.
Precisely, a relational function specification $(f : \Pi (\overrightarrow{a}: A).$ 
$ \langle \mathcal{P}(\overrightarrow{a})\rangle \langle \mathcal{Q}(\overrightarrow{a})\rangle)$ can be encoded into a standard
specification $(f : \lambda X.  \Pi (\overrightarrow{a} : A). \{ \encasrt{\mathcal{P}(\overrightarrow{a})}_X\} \{\encasrt{\mathcal{Q}(\overrightarrow{a})}_X\})  $.
\begin{definition}[Relational Proof Context Encoding]A relational proof context $\reldelta$ is encoded into a standard proof context $\encasrt{\reldelta}$ as follows: 
for any relational function specification $(f : \Pi (\overrightarrow{a}: A).$ 
$ \langle \mathcal{P}(\overrightarrow{a})\rangle \langle \mathcal{Q}(\overrightarrow{a})\rangle)$ in $\reldelta$, it is encoded into specification $(f : \lambda X.  \Pi (\overrightarrow{a} : A). \{ \encasrt{\mathcal{P}(\overrightarrow{a})}_X\} \{\encasrt{\mathcal{Q}(\overrightarrow{a})}_X\})  $, where $\encasrt{-}_X$ remains as defined in \defref{enc:asrtenc}.
\end{definition}

\noindent
Then we can encode relational contextual  Hoare triples into standard contextual Hoare triples.
We claim this encoding is correct through the following theorem:


\begin{lemma} For any low-level environment $\chi : \fid \rightarrow \conSigma \times \conSigma$ and relational proof context $\reldelta$, $\text{\normalfont RelValid}(\chi, \reldelta)$  holds if and only if  $\text{\normalfont Valid}(\chi, \encasrt{\reldelta})$ holds.
\end{lemma}
\begin{proof} ~
\begin{itemize}
    \item $\Rightarrow$: we need to prove that for any relational function specification $(f : \Pi (\overrightarrow{a}: A). \langle \mathcal{P}(\overrightarrow{a})\rangle$ $\langle \mathcal{Q}(\overrightarrow{a})\rangle)$ in $\reldelta$, its encoded specification is valid with respect to $\chi$.
    Then given any $X\subseteq \absSigma$ and  $\overrightarrow{a} \in A$, for any initial state $\const_1 \models \encasrt{\mathcal{P}(\overrightarrow{a})}_X$, we know there exist $\absst_1$ and $\absc_1$ such that $(\const_1,\absst_1,\absc_1) \models \mathcal{P}(\overrightarrow{a})$ and $\absst_1 \models \weakestpre{\absc_1}{X}$.
    Since  $\text{RelValid}(\chi, \reldelta)$  holds, we can derive that for any low-level terminal state $\const_2$ such that $(\const_1, \const_2) \in \chi(f)$, there exist $\absst_2$ and $\absc_2$ {such that} $(\absst_1, \absc_1) \hookrightarrow  (\absst_2, \absc_2)$ and $(\const_2, \absst_2, \absc_2) \models \mathcal{Q}(\overrightarrow{a})$.
     Then by \theoref{theo:decom}, we have 
    $\absst_2  \models \weakestpre{\absc_2}{X}$.
    Therefore, we derive 
    $\const_2 \models$ $\encasrt{\mathcal{Q}(\overrightarrow{a})}_X$.
    \item $\Leftarrow$: we need to prove that for any relational function specification $(f : \Pi (\overrightarrow{a}: A). \langle \mathcal{P}(\overrightarrow{a})\rangle $ $\langle \mathcal{Q}(\overrightarrow{a})\rangle)$ in $\reldelta$, it is valid with respect to $\chi$.
    Then given any  $\overrightarrow{a} \in A$, for any $(\const_1, \absst_1, \absc_1) \models \relasrt{P}$,  we instantiate $X$ of the encoded specification as 
    the high-level terminal states set
        $\{ \absst \,| \, (\absst_1,\absst_3) \in \nrmdeno{\absc_1}\}$ and then $\const_1 \models \encasrt{\mathcal{P}}_X$.
        According to  $\text{Valid}(\chi, \encasrt{\reldelta})$, for any $\const_2 $ such that $(\const_1, \const_2) \in \chi(f)$, we have $\const_2 \models  \encasrt{\mathcal{Q}(\overrightarrow{a})}_X$.
        By unfolding the assertion encoding, there exist $\absst_2$ and $\absc_2$ such that $\absst_2  \models \weakestpre{\absc_2}{X}$ and $(\const_2,\absst_2, \absc_2) \models \relasrt{Q}(\overrightarrow{a})$.
        According to the definition of weakest precondition, we know any $\absst_3$ such that $(\absst_2, \absst_3) \in \nrmdeno{\absc_2}$, $\absst_3 \models X$.
        That is for any $\absst_3$ such that $(\absst_2, \absst_3) \in \nrmdeno{\absc_2}$, we have $(\absst_1, \absst_3) \in \nrmdeno{\absc_2}$.
        Therefore, we finally derive $(\absst_1, \absc_1) \hookrightarrow (\absst_2, \absc_2)$ and $(\const_2,\absst_2, \absc_2) \models \relasrt{Q}(\overrightarrow{a})$.
    
\end{itemize}
\end{proof}
\begin{theorem}[Encoding Relational Contextual Triples] For any proof context $\Delta$, $\reldelta$, low-level statement $\conc$, and assertions $\mathcal{P}$, $\mathcal{Q}$ $\subseteq \conSigma \times \absSigma \times \absprog$:
 \begin{align*}
    \underbrace{ \Delta, \reldelta \vdash  \langle \mathcal{P} \rangle \  \conc \ \langle \mathcal{Q}\rangle}_{\mathcal{J}}
     \quad \text{iff.} \quad 
     \underbrace{\forall X. \, \Delta, \encasrt{\reldelta}  \demovdash \{ \encasrt{ \mathcal{P} }_X \} \ \conc \ \{ \encasrt{ \mathcal{Q}}_X  \}}_J
 \end{align*} 
\end{theorem}
\begin{proof}~
\begin{itemize}
    \item $\Rightarrow$: 
    Given any $\chi$ such that $\text{\normalfont Valid}(\chi, \Delta)$ holds and  $\text{\normalfont Valid}(\chi, \encasrt{\reldelta})$  holds, we know that   $\text{\normalfont Valid}(\chi, \Delta)$ holds and $\text{\normalfont RelValid}(\chi, \reldelta)$  holds.
    For any $X$ and any $\const_1$ in $\encasrt{\relasrt{P}}_X$, there exist  $\absst_1$ and $\absc_1$ such that $\absst_1 \models \weakestpre{\absc_1}{X}$ and $(\const_1, \absst_1, \absc_1) \models \relasrt{P}$. Then for any terminating state $\const_2$ such that $(\const_1,\const_2) \in \nrmfcdeno{\conc}$, according to relational triple $\mathcal{J}$, there exist $\absst_2$ and $\absc_2$ such that $(\absst_1, \absc_1) \hookrightarrow (\absst_2, \absc_2)$ and 
        $(\const_2, \absst_2, \absc_2) \models \relasrt{Q}$. Then by \theoref{theo:decom}, we have 
        $\absst_2  \models \weakestpre{\absc_2}{X}$.
        Therefore, according to \defref{enc:asrtenc} we derive 
        $\const_2 \models$ $\encasrt{\mathcal{Q}}_X$.
    \item $\Leftarrow$: Given any $\chi$ such that $\text{\normalfont Valid}(\chi, \Delta)$ holds and $\text{\normalfont RelValid}(\chi, \reldelta)$  holds, we know   $\text{\normalfont Valid}(\chi, \Delta)$ holds and  $\text{\normalfont Valid}(\chi, \encasrt{\reldelta})$  holds.
    For any $(\const_1, \absst_1, \absc_1) \models \relasrt{P}$,  we instantiate $X$ as 
        the high-level terminal states set
        $\{ \absst \,| \, (\absst_1,\absst_3) \in \nrmdeno{\absc_1}\}$ and then $\const_1 \models \encasrt{\mathcal{P}}_X$.
        According to standard triple $J$, for any $\const_2 $ such that $(\const_1, \const_2) \in \nrmfcdeno{\conc}$, we have $\const_2 \models  \encasrt{\mathcal{Q}}_X$.
        By unfolding the assertion encoding, there exist $\absst_2$ and $\absc_2$ such that $\absst_2  \models \weakestpre{\absc_2}{X}$ and $(\const_2,\absst_2, \absc_2) \models \relasrt{Q}$.
        According to the definition of weakest precondition, we know any $\absst_3$ such that $(\absst_2, \absst_3) \in \nrmdeno{\absc_2}$, $\absst_3 \models X$.
        That is for any $\absst_3$ such that $(\absst_2, \absst_3) \in \nrmdeno{\absc_2}$, we have $(\absst_1, \absst_3) \in \nrmdeno{\absc_2}$.
        Therefore, we finally derive $(\absst_1, \absc_1) \hookrightarrow (\absst_2, \absc_2)$ and $(\const_2,\absst_2, \absc_2) \models \relasrt{Q}$.
\end{itemize}
\end{proof}

Subsequently,
relational Hoare logic inherits the call rule  \hyperref[rule:hoare-call]{\textsc{Hoare-Call}} in standard Hoare logic.
\begin{mathpar}
  \textsc{\label{rule:hoare-call}\scshape Hoare-Call}\inferrule
   {\conf: \Pi (\overrightarrow{a}: A). \{ \conasrt{P}(\overrightarrow{a})\} \{ \conasrt{Q}(\overrightarrow{a}) \} \in \Delta } 
  { \Delta \demovdash \{  \conasrt{P}(\overrightarrow{a}) \} \ \text{call } \conf \ \{  \conasrt{Q}(\overrightarrow{a}) \} }
\end{mathpar}
This rule enables provers to evaluate the call individually with a standard specification in the low-level program. 
Furthermore, to evaluate both sides simultaneously, the \hyperref[rule:hoare-call]{\textsc{Hoare-Call}} rule can be specified as follows:
\begin{mathpar}
  \inferrule
  {\conf: \lambda X. \Pi (\overrightarrow{a}: A). \{ \encasrt{ \mathcal{P}(\overrightarrow{a})}_X\} \{ \encasrt{ \mathcal{Q}(\overrightarrow{a}) }_X \} \in \Delta } 
 {\Delta \demovdash \{ \encasrt{ \mathcal{P}(\overrightarrow{a}) }_X\} \ \text{call } \conf \ \{\encasrt{ \mathcal{Q}(\overrightarrow{a})  }_X \} }
\end{mathpar}

\subsection{\label{sec:err}Addressing Errors in Programs}
Real-world programs may result in errors, such as array out-of-bounds violations. 
For example, the low-level program depicted in \figref{fig:while_err} can lead to errors if the program variable $\convar{i}$ exceeds the bounds of the array stored in variable $\convar{A}$.

\begin{figure}[ht]
  \centering
    \begin{tabular}{@{\hspace{1em}}p{0.45\linewidth}|@{\hspace{1em}}p{0.45\linewidth}}
\begin{tabular}{@{}l@{}}
    \begin{lstlisting}[aboveskip= 0\baselineskip,mathescape=true, language=C, xleftmargin=0.5cm,morekeywords={stack, list}, basicstyle=\normalfont]
$\convar{x}$ = 0;
$\convar{i}$ = 0;
while ($\convar{i}$ < 8){

  $\convar{x}$ = $\convar{x}$ | (1 << $\convar{A}$[$\convar{i}$]);
  $\convar{i}$ = $\convar{i}$ + 1;
}
      \end{lstlisting}
\end{tabular}
    &
\begin{tabular}{@{}l@{}}
   \begin{lstlisting}[aboveskip= 0\baselineskip,mathescape=true,language=C, morekeywords={stack, list, assert}, basicstyle=\normalfont]
$\absvar{s}$ = { };
$\pvar{j}$ = 0;
while ($\pvar{j}$ < 8){
  assert($\len{\absvar{A}} \geq \pvar{j}$);
  $\absvar{s}$ = $\absvar{s}$ $\cup$ { $\absvar{A}$[$\pvar{j}$] };
  $\pvar{j}$ = $\pvar{j}$ + 1;
}             
     \end{lstlisting}
\end{tabular} \\
    \end{tabular}
  \caption{Implementations of recording a collection with arrays} 
  \label{fig:while_err}
\end{figure} 
However, if we know that the corresponding variable $\pvar{j}$ in the high-level program stays within the bounds of the corresponding array $\absvar{A}$, then $\convar{i}$ will also remain within bounds.
This follows from the principle that program refinement must ensure: if the high-level program results in no error, then the low-level program must result in no error.
In this section, we extend our encoding theorem to support error refinement and demonstrate how to use high-level program properties in refinement verification.

\subsubsection{Preliminary: Denotational Semantics and Hoare Logic}
To capture program errors, in addition to the normal termination case,  denotational  semantics of a statement $c$ should include an error case:

\begin{definition}[Denotational Semantics] For any program $c \in \text{Prog}$,  the denotational semantics
  $\llbracket c \rrbracket \in Rel(\Sigma)$ comprises two components:
  \begin{itemize}
    \item \text{\normalfont termination case}:  $(\sigma,\sigma') \in \nrmdeno{c}$ if the execution of $c$ from the initial state $\sigma$ may terminate at state $\sigma'$;
    \item \text{\normalfont error case}:  $\sigma\in \errdeno{c}$ if the execution of $c$ from the initial state $\sigma$ may result in an error.
  \end{itemize}

  \label{def:nrmerr}
  \end{definition}

\begin{definition}[Configuration Refinement]
  For any programs $c_1$ $c_2$ and any states $\sigma_1$ $\sigma_2 \in \Sigma$,   $(\sigma_1, c_1) \hookrightarrow (\sigma_2, c_2) $ holds if:
  \begin{itemize}
    \item \text{\normalfont termination case}: for any $\sigma_{3}$ such that $(\sigma_{2} ,\sigma_{3}) \in \nrmdeno{c_2}$, we have $ (\sigma_{1} ,\sigma_{3}) \in \nrmdeno{c_1}$;
    \item \text{\normalfont error case}: if $\sigma_2 \in \errdeno{c_2} $, then $\sigma_1 \in \errdeno{c_1}$.
  \end{itemize}
\end{definition}

\noindent For example, the semantics for an assertion statement $\cassert{P}$
is defined as follows:
\begin{align*}
    (\sigma_1, \sigma_2) \in \nrmdeno{\cassert{{P}}} &\text{\quad iff \quad} \sigma_1 = \sigma_2 \text{ and } \sigma_1 \models {P} \\
    \sigma \in \errdeno{\cassert{{P}}} &\text{\quad iff \quad} \sigma \not\models {P}  
\end{align*}

The definitions of validity for standard Hoare triples and relational Hoare triples are extended to account for errors, with the differing parts highlighted in bold font:
\begin{definition}[Standard Hoare Triples] For any statement $c$, and assertion $P$, $Q$ $\subseteq \Sigma$, the  Hoare triple $\{P\}\  c\ \{Q\} $ is $\forall$-valid   if  for any $\sigma_1 \models P :$ 

\begin{tabular}{l}
    \tabitem \text{\normalfont\bfseries no error:}  $\sigma_1 \notin \errdeno{c}$  \\
    \tabitem \text{\normalfont correct result:} $\forall \sigma_2. \, (\sigma_1,\sigma_2) \in \nrmdeno{c}  \Rightarrow  \sigma_2 \models Q $
\end{tabular}
\label{def:hte}
\end{definition}



\begin{definition}[Relational Hoare Triples] For low-level program $\conc$, and program-as-resource assertions $\mathcal{P}$, $\mathcal{Q}$ $\subseteq \conSigma \times \absSigma \times \absStmts$, the relational Hoare triple $\langle \mathcal{P} \rangle \  \conc \ \langle \mathcal{Q} \rangle$ is valid if for any $(\const_1, \absst_1, \absc_1) \models \mathcal{P} :$ 
  \begin{itemize}
    \item  {\normalfont \bfseries error simulation}: if $\const_1 \in \errdeno{\conc}$ then $\absst_1 \in \errdeno{\absc_1}$;
    \item {\normalfont result simulation}: for any $\const_2$ such that $(\const_1, \const_2) \in  \nrmdeno{\conc}$, \textbf{either} $\absst_1 \in \errdeno{\absc}$, \textbf{or} there exist
     $\absst_2$ and $\absc_2$ such that $(\absst_1, \absc_1) \hookrightarrow (\absst_2, \absc_2) $ and $ (\const_2 ,\absst_2, \absc_2) \models \mathcal{Q}$ 
  \end{itemize}
 \label{def:rhte}
\end{definition}
\subsubsection{\label{subsubsec:encerr}The Encoding Theory and Derived Rules}
Due to the additional error cases in the definitions of standard and relational Hoare triples, the encoding correctness as stated in \theoref{lem:enc} becomes invalid and the assertion encoding $\encasrt{-}_X$ needs to be adapted to ensure that
\begin{equation*}
    \langle \mathcal{P} \rangle \  \conc \ \langle \mathcal{Q}\rangle \text{ is valid}    \quad \text{iff.} \quad  \forall X.  \ \demovdash \{ \encasrt{\mathcal{P}}_X \} \ \conc \ \{ \encasrt{ \mathcal{Q} }_X \} 
\end{equation*}
In the following, we focus on the error cases of the two triples and assume that $(\const_1, \absst_1, \absc_1) \models \mathcal{P} $ and $\const_1 \models \encasrt{ \mathcal{P} }_X$.
The relational Hoare triple on the left side requires that
$$ \text{if } \const_1 \in \errdeno{\conc} \text{ then } \absst_1 \in \errdeno{\absc_1} $$
and the standard Hoare triple on the right side requires:
$$ \const_1 \notin \errdeno{\conc} $$
Intuitively, the requirement of the former can be restated as that 
$ \text{if } \absst_1 \notin \errdeno{\absc_1}, \text{ then }  \const_1 \notin \errdeno{\conc} $.
Thus, we can guess that the encoded assertion $\encasrt{\mathcal{P}}_X$ contains a condition $\absst_1 \notin \errdeno{\absc_1}$, which coincides with the weakest precondition for programs with errors.

\begin{definition}[Weakest Precondition]For any program $c$ and any set $X\subseteq\Sigma$, which also serves as an assertion over program states, 
    $\weakestpre{c}{X}$ gives the weakest precondition such that all terminal states resulting from the execution of program $c$ satisfy $X$ and is defined as:
  \begin{align*}
      \sigma \models \weakestpre{c}{X} \quad \text{ iff. } \quad
      & \sigma \notin \errdeno{c} \land  \forall \sigma_0. \, (\sigma, \sigma_0) \in \nrmdeno{c} \Rightarrow  \sigma_0 \models X 
  \end{align*}
\end{definition}

\noindent Based on the modified definition of the weakest precondition, the assertion encoding $\encasrt{\mathcal{P}}_X$ remains as defined in  \defref{enc:asrtenc}, and the execution predicate $\needExecX{\absasrt{P}}{\absc}{X}$ remains as interpreted in \defref{def:exec}.
For convenience, we restate these two definitions.
\begin{customdef}{15}[Assertion Encoding] For any program-as-resource assertion $\mathcal{P}$ $\subseteq \conSigma \times \absSigma \times \absStmts$ and subset of high-level states $X \subseteq \absSigma$, define the encoding as
    $$\const \models \encasrt{ \mathcal{P}}_X  \quad \text{ iff. } \quad \exists \absst \absc. \,(\const, \absst, \absc) \models  \mathcal{P} \land  \absst \models \weakestpre{\absc}{X}.$$
\end{customdef}
\begin{customdef}{17}[Semantic Interpretation of the Execution Predicate]For any high-level assertion $\absasrt{P}$, postcondition $X$, and program $\absc$,  $\needExecX{\absasrt{P}}{\absc}{X}$ expresses that there exists an initial state satisfying $\absasrt{P}$ such that the program $\absc$ terminates only in a state satisfying $X$:
\[
\needExecX{\absasrt{P}}{\absc}{X}  \text{ holds \quad iff. \quad there exists } \absst \text{ such that } \absst \models \absasrt{P} \land \textsf{wlp}(\absc, X)
\]
  \end{customdef}

\noindent The correctness of the encoding is preserved, as established in \theoref{lem:enc}. 

\begin{customthm}{D.1}[Encoding Relational Triples with Errors] For any low-level statement $\conc$, and assertions $\mathcal{P}$, $\mathcal{Q}$ $\subseteq \conSigma \times \absSigma \times \absStmts$:
 \begin{align*}
    \underbrace{\langle \mathcal{P} \rangle \  \conc \ \langle \mathcal{Q}\rangle}_{\mathcal{J}}
     \text{ is valid}   \quad \text{iff.} \quad 
     \underbrace{\forall X. \ \demovdash \{ \encasrt{ \mathcal{P}}_X \} \ \conc \ \{ \encasrt{ \mathcal{Q} }_X \}}_J 
 \end{align*} 
\end{customthm}
\begin{proof}  we prove two directions:
    \begin{itemize}
        \item $\Rightarrow$:  \,  For any $X$ and any $\const_1$ in $\encasrt{\relasrt{P}}_X$, there exist  $\absst_1$ and $\absc_1$ such that $\absst_1 \models \weakestpre{\absc_1}{X}$ and $(\const_1, \absst_1, \absc_1) \models \relasrt{P}$. 
        \begin{itemize}
            \item no error: according to the error simulation of relational triple $\mathcal{J}$,
        we know if $\const_1 \in \errdeno{\conc}$, then $\absst_1 \in \errdeno{\absc_1}$.
        However  $\absst_1 \models \weakestpre{\absc_1}{X}$ implies $\absst_1 \notin \errdeno{\absc_1}$, and thus $\const_1 \notin \errdeno{\conc}$.
        \item correct result:  
        for any terminating state $\const_2$ such that $(\const_1,\const_2) \in \nrmdeno{\conc}$, according to relational triple $\mathcal{J}$,
        there exist $\absst_2$ and $\absc_2$ such that $(\absst_1, \absc_1) \hookrightarrow (\absst_2, \absc_2)$ and 
        $(\const_2, \absst_2, \absc_2) \models \relasrt{Q}$. Then by \theoref{theo:decom}, we have 
        $\absst_2  \models \weakestpre{\absc_2}{X}$.
        Therefore, according to \defref{enc:asrtenc} we derive 
        $\const_2 \models$ $\encasrt{\mathcal{Q}}_X$.
        \end{itemize}
        \item $\Leftarrow$:
    For any $(\const_1, \absst_1, \absc_1)$ in $\relasrt{P}$, we instantiate $X$ as $\{\absst_3 \ | \ (\absst_1, \absst_3) \in \nrmdeno{\absc_1}\}$, and observe that if $\absst_1 \notin \errdeno{\absc_1}$ then $\absc_1 \in \encasrt{\mathcal{P}}_X$. Then, we prove two cases:
    \begin{itemize}
        \item error simulation: when $\const_1 \in \errdeno{\conc}$, then $\absst_1 \in \errdeno{\absc_1}$, then this case proved. Otherwise, $\absst_1 \notin \errdeno{\absc_1}$. 
        Then $\const_1 \in \encasrt{\mathcal{P}}_X$ and from the  triple $J$, we know $\const_1 \notin \errdeno{c}$, which is a contradiction.
        \item result simulation: for any $\const_2$ such that $(\const_1, \const_2) \in \nrmdeno{\conc}$, similarly we only need to consider when $\absst_1 \notin \errdeno{\absc_1}$ and we can derive that $\const_2 \in \encasrt{\mathcal{Q}}_X$. By unfolding the encoding, there exists $\absst_2$ $\absc_2$ such that safe$(\absst_2,\absc_2,X)$ and $(\const_2,\absst_2, \absc_2) \in \relasrt{Q}$.
        According to the definition of safe, we know any $\absst_3$ such that $(\absst_2, \absst_3) \in \nrmdeno{\absc_2}$, $\absst_3 \in \{\absst_0 \ | \ (\absst_1, \absst_0) \in \nrmdeno{\absc_2}\}$, and $\absst_2 \notin \errdeno{\absc_2}$.
        Therefore, we get $(\absst_1, \absc_1) \hookrightarrow (\absst_2, \absc_2)$.
    \end{itemize}

    \end{itemize}
\end{proof}

Then for the statement $\cassert{\absasrt{P}}$ in high-level programs, the rule \hyperref[absassertseq]{\textsc{High-AssertSeq}} can be derived, which is specifically designed to extract the assertions made by the high-level program.
Furthermore, if the precondition for the high-level program contains a pure fact $B$, we can extract it from the execution of the high-level program.
\begin{mathpar}
  \textsc{\label{absassertseq}High-Assert}\inferrule{ }
    {\needExecX{\absasrt{P}_1}{\cassert{\absasrt{P}_2} }{X} \Rightarrow \needExecX{\absasrt{P}_1 \land \absasrt{P}_2}{\cskip}{X} }
\end{mathpar}

\subsubsection{Leveraging Properties of High-level Programs}
This section demonstrates how to leverage  properties of the high-level program in a  refinement proof using the two programs in \figref{fig:while_err}.

The goal is to prove the  relational  triple
$\langle {\convar{A} = \absvar{A} \land \hspeccolor{\absc}} \rangle $ ${\conc}\, \langle {\convar{x} = \Sigma_{a\in \absvar{s}} 2^a \land \hspeccolor{\cskip}} \rangle $.
This is equivalent to the following triple, where the assertions are written in decomposed form.
Here we assume that the values of the variables $\convar{A}$ and $\absvar{A}$ in the two programs are represented by the same array $l$.
\begin{equation*}
  \begin{aligned}
  & \left\langle 
    \lasrt{ {\convar{A}} = {l}} \land \hasrt{\absvar{A} = l} \land \hspeccolor{\absc} 
   \right\rangle \, \conc \,
    \left\langle 
      \exists l_0. \lasrt{\convar{x} = \Sigma_{a\in l_0} 2^a} \land \hasrt{\absvar{s} = l_0} \land  \hspeccolor{\cskip}
   \right\rangle
  \end{aligned}
\end{equation*}
Based on the encoding theory, we can transform this relational triple into the following standard Hoare triple with a placeholder $X$.
\begin{equation*}
  \begin{aligned}
    & \left\{ 
    \needExecX{\absvar{A} = l}{\absc}{X} \land {\convar{A}}={l}
    \right\}  \, \conc \, \left\{ 
    \exists l_0. \,  \needExecX{ \absvar{s} = l_0 }{\cskip}{X} \land  {\convar{x} = \Sigma_{a\in l_0}} 2^a 
    \right\}
  \end{aligned}
\end{equation*}
During verification, we use the following loop invariant, similar to the one presented in \secref{sec:rhtproof}. 
Here we also use $\abswhile$ to refer to the loop in the high-level program.
\begin{equation*}
  \begin{aligned}
    \exists l_0\, n.\ & \needExecX{{\absvar{s}= l_0 \land \absvar{j} = n \land \absvar{A} = l}}{\abswhile}{X} \land {\convar{x} = \Sigma_{a\in l_0} 2^a \land \convar{i} = n \land {\convar{A}} = {l}}
  \end{aligned}
  \label{inv:2}
\end{equation*}
Then, we present the proof for the encoded standard triple in \figref{fig:whileHL}.
\begin{figure}[t]
  \centering

    \begin{lstlisting}[aboveskip= 0\baselineskip,mathescape=true,language=C,basicstyle=\footnotesize, escapechar=\&]
// $\{ \needExecX{\absvar{A} = l}{\absvar{s} = \{\,\}; \absvar{j} = 0;  \abswhile}{X} \land {\convar{A}}={l} \}$ 
// $\color{gray}\text{consequence rules}$
// $\{ \needExecX{\absvar{s} = \emptyset \land \absvar{j} = 0\land \absvar{A} = l}{  \abswhile}{X} \land {\convar{A}}={l} \}$ 
$\convar{x} = 0; $ $\convar{i} = 0; $
// $\color{gray}\text{sequencing rules}$
// $\{ \exists l_0 \, n.\  \needExecX{{\absvar{s}= l_0 \land \absvar{j} = n \land \absvar{A} = l}}{\abswhile}{X}  \land  {\convar{x} = \Sigma_{a\in l_0} 2^a \land \convar{i} = n \land {\convar{A}}={l}} \} $  
while ($\convar{i}$ < 8){
  // $\{\needExecX{{\absvar{s}= l_0 \land \absvar{j} = n \land \absvar{A} = l}}{\abswhile}{X} \land  {\convar{x} = \Sigma_{a\in l_0} 2^a \land \convar{i} = n \land {\convar{A}}={l}}   \} $ 
  // $\color{gray}\text{consequence rule: loop unrolling}$
  // $\{\needExecX{\absvar{s}= l_0 \land \absvar{j} = n \land \absvar{A} = l}{\cassert{\len{\absvar{A}} \geq \absvar{j}};\cdots}{X}  \land  {\convar{x} = \Sigma_{a\in l_0} 2^a \land \convar{i} = n \land {\convar{A}}={l} } \} $ 
  // $\color{gray}\text{consequence rule and extract pure fact}$
  // $\{  \needExecX{\absvar{s}= l_0 \land \absvar{j} = n \land \absvar{A} = l}{\absvar{s}= \absvar{s} \cup \{\absvar{A}[\absvar{j}]\};\cdots}{X} \land  {\color{blue} \text{\textsf length}(l) \geq n}  \land {\convar{x} = \Sigma_{a\in l_0} 2^a \land \convar{i} = n \land {\convar{A}}={l} }  \} $ 
  $\convar{x}$ = $\convar{x}$ | (1 << $\convar{A}$[$\convar{i}$]);
  // $\color{gray}\text{consequence and sequencing rules}$
  // $\{ \exists l_0. \, \needExecX{\absvar{s}= l_0 \land \absvar{j} = n \land \absvar{A} = l}{\absvar{j} = \absvar{j} + 1; \abswhile}{X}   \land  {\convar{x} = \Sigma_{a\in l_0} 2^a \land \convar{i} = n \land {\convar{A}}={l} }\} $ 
  $\convar{i}$ = $\convar{i}$ + 1; 
  // $\color{gray}\text{consequence and sequencing rules}$
  // $\{ \exists l_0. \, \needExecX{\absvar{s}= l_0 \land \absvar{j} = n + 1 \land \absvar{A} = l}{ \abswhile}{X}   \land  {\convar{x} = \Sigma_{a\in l_0} 2^a \land \convar{i} = n + 1\land  }\} $ 
}
  \end{lstlisting}
  \caption{Refinement Proof Based on Standard Hoare triples}
  \label{fig:whileHL}
\end{figure}
\noindent
In the loop body, we first  unroll the  loop in the high-level program, encountering a statement $\cassert{\len{A'} \geq \absvar{j}}$. 
Next, we apply the rule \hyperref[absassertseq]{\textsc{High-Assert}} to derive that the length of the array $\absvar{A}$ is greater than or equal to the value of $\absvar{j}$. 
We then extract the pure fact ${\color{blue} \text{\textsf{length}}(l) \geq n}$, highlighted in blue in the proof. 
Therefore, we can safely evaluate the array read expression $\convar{A}[\convar{i}]$ in the low-level program.


\subsection{\label{app:sep}Support for High-level Programs with Separation Logic}
Here we adapt our encoding theory to support the situations 
When high-level programs support dynamically allocated memory and thus employ separation logic \cite{DBLP:conf/lics/Reynolds02}.

\subsubsection{Preliminary}

Suppose $\absst_1 \cupdot \absst_2$ means the disjoint union of state $\absst_1$ and $\absst_2$.
Then we present the separation conjunction in high-level assertions.
\begin{definition}[Separation Conjunction] For any high-level assertions $\absasrt{P}_1$ $\absasrt{P}_2 \subseteq \absSigma$, 
the separation conjunction $\absasrt{P}_1 * \absasrt{P}_2$ is defined as follows:
\begin{align*}
    \absst \models \absasrt{P}_1 * \absasrt{P}_2 \text{ iff. } &\exists \,\absst_1 \, \absst_2.\, \absst_1 \cupdot \absst_2 = \absst \land \\ 
   & \absst_1 \models \absasrt{P}_1 \land \absst_2 \models \absasrt{P}_2
\end{align*}
\end{definition}

\noindent Then we present the adapted validity definition  
for relational Hoare triples  to account for an environment frame in the high-level program:
\begin{definition}[Relational Hoare Triples] The relational Hoare triple $\langle \mathcal{P} \rangle \ \conc \ \langle \mathcal{Q} \rangle$ is valid if given any high-level environment frame $\absst_f$, for any
$(\const_1, \absst_1, \absc_1) \models \mathcal{P}$ and any $\const_2$ such that $(\const_1, \const_2) \in  \nrmdeno{\conc}$, {there exist} $\absst_2$ and $\absc_2$ {such that} $(\absst_1 \cupdot \absst_f, \absc_1) \hookrightarrow  ( \absst_2 \cupdot \absst_f,\absc_2)$ and $(\const_2, \absst_2, \absc_2) \models \mathcal{Q}$.
\end{definition}

\subsubsection{The Encoding Theory}
According to separation logic, the environment untouched by the high-level program should be preserved.
Consequently, we can augment the placeholder $X \subseteq \absSigma$ with a framing assertion $\absasrt{F} \subseteq \absSigma$,  representing  unchanged states during the execution of the high-level program.

\begin{definition}[Assertion Encoding] For any program-as-resource assertion $\mathcal{P}$ $\subseteq \conSigma \times \absSigma \times \absStmts$ and a pair of subsets of high-level states $(X,F) \in \mathfrak{P}(\absSigma) \times \mathfrak{P}(\absSigma) $, define the encoding as
\begin{align*}
    \const \models \encasrt{ \mathcal{P}}_{(X,F)}  \text{ iff. } & \exists \absst  \absst_f  \absc. \,(\const, \absst, \absc) \models  \mathcal{P} \land  \\ 
    & \absst_f \models F \land
    (\absst \cupdot \absst_f) \models \weakestpre{\absc}{X}. 
\end{align*}

\begin{theorem}[Encoding Relational Triples] For any low-level statement $\conc$, and assertions $\mathcal{P}$, $\mathcal{Q}$ $\subseteq \conSigma \times \absSigma \times \absStmts$:
 \begin{align*}
    \underbrace{\langle \mathcal{P} \rangle   \conc  \langle \mathcal{Q}\rangle}_{\mathcal{J}}
     \text{ is valid}   \, \text{iff.} \, 
     \underbrace{\forall (X,F). \ \demovdash \{ \encasrt{ \mathcal{P}}_{(X,F)} \}  \conc  \{ \encasrt{ \mathcal{Q} }_{(X,F)} \}}_J 
 \end{align*} 
\end{theorem}
\begin{proof}  we prove two directions:
    \begin{itemize}
        \item $\Rightarrow$:  \,  For any $X$, $F$ and any $\const_1$ in $\encasrt{\relasrt{P}}_X$, there exist  $\absst_1$, $\absst_f$ and $\absc_1$ such that $\absst_1 \cupdot \absst_f \models \weakestpre{\absc_1}{X}$, $\absst_f \models F$ and $(\const_1, \absst_1, \absc_1) \models \relasrt{P}$. Then for any terminating state $\const_2$ such that $(\const_1,\const_2) \in \nrmdeno{\conc}$, according to relational triple $\mathcal{J}$, there exist $\absst_2$ and $\absc_2$ such that $(\absst_1 \cupdot \absst_f, \absc_1) \hookrightarrow (\absst_2 \cupdot \absst_f, \absc_2)$ and 
        $(\const_2, \absst_2, \absc_2) \models \relasrt{Q}$. Then by \theoref{theo:decom}, we have 
        $\absst_2 \cupdot \absst_f  \models \weakestpre{\absc_2}{X}$.
        Therefore, according to \defref{enc:asrtenc} we derive 
        $\const_2 \models$ $\encasrt{\mathcal{Q}}_X$.
        \item $\Leftarrow$: Based on \propref{prop:10}, for any environment $\absst_f$, any $(\const_1, \absst_1, \absc_1) \models \relasrt{P}$,  we instantiate $X$ as 
        the high-level terminal states set
        $\{ \absst_3 \,| \, (\absst_1 \cupdot \absst_f,\absst_3) \in \nrmdeno{\absc_1}\}$ and instantiate $F$ as the singleton set $\{\absst_f\}$. 
        By the definition of assertion encoding, $\const_1 \models \encasrt{\mathcal{P}}_X$.
        According to standard triple $J$, for any $\const_2 $ such that $(\const_1, \const_2) \in \nrmdeno{\conc}$, we have $\const_2 \models  \encasrt{\mathcal{Q}}_X$.
        By unfolding the assertion encoding, there exist $\absst_2$  and $\absc_2$ such that $\absst_2 \cupdot \absst_f  \models \weakestpre{\absc_2}{X}$ and $(\const_2,\absst_2, \absc_2) \models \relasrt{Q}$.
        According to the definition of weakest preconditions, for any $\absst_3$ such that $(\absst_2 \cupdot \absst_f, \absst_3) \in \nrmdeno{\absc_2}$, $\absst_3 \models X$.
        That is for any $\absst_3$ such that $(\absst_2 \cupdot \absst_f) \in \nrmdeno{\absc_2}$, we have $(\absst_1 \cupdot \absst_f, \absst_3) \in \nrmdeno{\absc_2}$.
        Therefore, we finally derive $(\absst_1 \cupdot \absst_f, \absc_1) \hookrightarrow (\absst_2 \cupdot \absst_f, \absc_2)$ and $(\const_2,\absst_2, \absc_2) \models \relasrt{Q}$.
    \end{itemize}
\end{proof}

\end{definition}

\section{\label{app:example}Examples}
Here we list the examples used in this paper and provide both relational and standard proofs for them.

\begin{appexample}
The low-level program (on the left) assigns the variable $\mathsf{x}$  to either $0$ or $1$. In contrast, the high-level program (on the right) allows $\mathsf{y}$ to nondeterministically take on any value of $0$, $1$, or $2$.
We aim to prove the relational Hoare triple $\reltriplenofont{\hspeccolor{\absc}}{\conc}{ \convar{x} = \absvar{x} \land \hspeccolor{\cskip}}$.
\end{appexample}
\begin{tabular}{@{\hspace{1em}}p{0.45\linewidth}|@{\hspace{1em}}p{0.45\linewidth}}
\begin{tabular}{@{}l@{}}
        \begin{lstlisting}[aboveskip=0\baselineskip,mathescape=true, language=C, 
        xleftmargin=0.5cm,morekeywords={stack, list}]
  $\convar{x}$ $\coloneq$ $\textsf{nondet}$(0,1); 
    \end{lstlisting}
  \end{tabular} & 
    \begin{tabular}{@{}l@{}}
   \begin{lstlisting}[aboveskip=0\baselineskip,mathescape=true,language=C, 
    xleftmargin=0.5cm,morekeywords={stack, list}]
  $\absvar{y}$ $\coloneq$ $\textsf{nondet}$(0,2); 
    \end{lstlisting}
  \end{tabular}\\
  \end{tabular}

~

\begin{relproof} ~

 \begin{lstlisting}[mathescape=true,language=C,basicstyle=\footnotesize, escapechar=\&]
// $\langle \hspeccolor{\absvar{x} = \textsf{nondet}(0,2)} \rangle$ 
$\convar{x}$ = $\textsf{nondet}$(0,1);
// $\color{gray}\text{low-level step}$
// $\langle  \exists n. n \in \{0, 1\} \land \lasrt{\convar{x} = n} \land  \hspeccolor{\absvar{x} = \textsf{nondet}(0,2)} \rangle$ 
// $\color{gray}\text{high-level step}$
// $\langle n \in \{0, 1\} \land \lasrt{\convar{x} = n} \land  \hasrt{\absvar{x} = n} \land \hspeccolor{\cskip} \rangle$ 
  \end{lstlisting}    
\end{relproof}

\begin{traproof}
 \begin{lstlisting}[mathescape=true,language=C,basicstyle=\footnotesize, escapechar=\&]
// $\{  \needExecX{\ptrue}{\absvar{x} = \textsf{nondet}(0,2)}{X} \}$ 
$\convar{x}$ = $\textsf{nondet}$(0,1);
// $\color{gray}\text{sequencing rule}$
// $\{  \exists n. n \in \{0, 1\} \land \needExecX{\ptrue}{\absvar{x} = \textsf{nondet}(0,2)}{X} \land \lasrt{\convar{x} = n} \}$ 
// $\color{gray}\text{consequence rule}$
// $\{ n \in \{0, 1\} \land \needExecX{\absvar{x} = n}{\cskip}{X} \land \lasrt{\convar{x} = n} \}$ 
  \end{lstlisting}
  
\end{traproof}

\begin{appexample}
The low-level program uses bitwise OR operations to set specific bits in program variable $\convar{x}$ while the high-level program directly builds a set $\absvar{s}$.
The goal is to prove the triple $\reltriplenofont{\hspeccolor{\texttt{set\_union\_loop}}}{\texttt{bit\_mask\_loop}}{\convar{x} = \Sigma_{a\in \absvar{s}} 2^a \land \hspeccolor{\cskip}}$ 
to express that after executing the two programs, variable $\convar{x}$ encodes the elements in set $\absvar{s}$ as a sum of powers of 2.
\end{appexample}
\begin{tabular}{@{\hspace{1em}}p{0.45\linewidth}|@{\hspace{1em}}p{0.45\linewidth}}
\begin{tabular}{@{}l@{}}
        \begin{lstlisting}[aboveskip=0\baselineskip,mathescape=true, language=C, 
        xleftmargin=0.5cm,morekeywords={stack, list}]
$\convar{x}$ = 0;
$\convar{x}$ = $\convar{x}$ | (1 << $a_0$);
$\convar{x}$ = $\convar{x}$ | (1 << $a_1$);
    \end{lstlisting}
  \end{tabular} & 
    \begin{tabular}{@{}l@{}}
   \begin{lstlisting}[aboveskip=0\baselineskip,mathescape=true,language=C, 
    xleftmargin=0.5cm,morekeywords={stack, list}]
$\absvar{s}$ = { };
$\absvar{s}$ = $\absvar{s}$ $\cup$ {$a_0$};
$\absvar{s}$ = $\absvar{s}$ $\cup$ {$a_1$};
    \end{lstlisting}
  \end{tabular}\\
  \end{tabular}

~ 

\begin{relproof} ~

 \begin{lstlisting}[mathescape=true,language=C,basicstyle=\footnotesize, escapechar=\&]
// $\langle \hspeccolor{\absvar{s} = \{\,\}; \absvar{s} = \absvar{s} \cup \{a_0\};  ; \absvar{s} = \absvar{s} \cup \{a_1\}} \rangle$ 
// $\color{gray}\text{high-level step}$
// $\langle \hasrt{\absvar{s}= \emptyset} \land \hspeccolor{ \absvar{s} = \absvar{s} \cup \{a_0\}; \absvar{s} = \absvar{s} \cup \{a_1\}} \rangle$  
$\convar{x}$ = 0;
// $\color{gray}\text{low-level step}$
// $\langle  \lasrt{\convar{x} = 0} \land \hasrt{\absvar{s}= \emptyset} \land \hspeccolor{ \absvar{s} = \absvar{s} \cup \{a_0\};\absvar{s} = \absvar{s} \cup \{a_1\}}  \rangle$ 
$\convar{x}$ = $\convar{x}$ | (1 << $a_0$);
// $\color{gray}\text{high-level and low-level steps}$
// $\langle  \lasrt{\convar{x} = 2^{a_0}} \land \hasrt{\absvar{s}= \{a_0\}} \land \hspeccolor{ \absvar{s} = \absvar{s} \cup \{a_1\}} \rangle$ 
$\convar{x}$ = $\convar{x}$ | (1 << $a_1$);
// $\color{gray}\text{high-level and low-level steps}$
// $\langle a_0 \neq a_1 \wedge \lasrt{\convar{x} =  2^{a_0} + 2^{a_1}} \land \hasrt{\absvar{s}= \{a_0, a_1\}} \land \hspeccolor{\cskip} \lor a_0 = a_1 \wedge \lasrt{\convar{x} =  2^{a_0}} \land \hasrt{\absvar{s}= \{a_0\}} \land \hspeccolor{\cskip}\rangle$
// $\color{gray}\text{consequence rule}$
// $\langle \exists \, l.\,  \lasrt{\convar{x} = \Sigma_{a\in l} 2^a} \land \hasrt{\absvar{s}= l} \land \hspeccolor{\cskip}\rangle$ 
  \end{lstlisting}    
\end{relproof}

\begin{traproof}
 \begin{lstlisting}[mathescape=true,language=C,basicstyle=\footnotesize, escapechar=\&]
// $\{ \needExecX{\textsf{True}}{\absvar{s} = \{\,\}; \absvar{s} = \absvar{s} \cup \{a_0\};  \absvar{s} = \absvar{s} \cup \{a_1\}}{X} \}$ 
// $\color{gray}\text{consequence rule}$
// $\{ \needExecX{\absvar{s}= \emptyset}{ \absvar{s} = \absvar{s} \cup \{a_0\}; \absvar{s} = \absvar{s} \cup \{a_1\}}{X} \}$ 
$\convar{x}$ = 0;
// $\color{gray}\text{sequencing rule}$
// $\{\needExecX{\absvar{s}= \emptyset}{ \absvar{s} = \absvar{s} \cup \{a_0\}; \absvar{s} = \absvar{s} \cup \{a_1\}}{X} \land {\convar{x} = 0} \}$ 
$\convar{x}$ = $\convar{x}$ | (1 << $a_0$);
// $\color{gray}\text{consequence and sequencing rules}$
// $\{ \needExecX{\absvar{s}= \{a_0\}}{ \absvar{s} = \absvar{s} \cup \{a_1\}}{X} \land  {\convar{x} = 2^{a_0}} \}$ 
$\convar{x}$ = $\convar{x}$ | (1 << $a_1$);
// $\color{gray}\text{consequence and sequencing rules}$
// $\{ a_0 \neq a_1 \wedge \needExecX{\absvar{s}= \{a_0,a_1\}}{\cskip}{X} \land  {\convar{x} =  2^{a_0} + 2^{a_1}} \lor a_0 = a_1 \wedge \needExecX{\absvar{s}= \{a_0\}}{\cskip}{X} \land  {\convar{x} =  2^{a_0} } \}$
// $\color{gray}\text{consequence rule}$
// $\{\exists \, l.\, \needExecX{\absvar{s}= l}{\cskip}{X} \land  {\convar{x} = \Sigma_{a\in l} 2^a} \}$
  \end{lstlisting}
  
\end{traproof}

\begin{appexample}
Both programs iterate through the items in a constant array $a$ until index $8$. 
In each iteration, the two programs add the corresponding item in arrays to the collection representation ($\convar{x}$ or $\absvar{s}$).
Here we also need to prove the triple $\reltriplenofont{\hspeccolor{\absc}}{\conc}{ \exists l.\  \lasrt{\convar{x} = \Sigma_{a\in l} 2^a} \land \hasrt{\absvar{s}= l} \land \hspeccolor{\cskip}}$.

\begin{tabular}{@{\hspace{1em}}p{0.45\linewidth}|@{\hspace{1em}}p{0.45\linewidth}}
\begin{tabular}{@{}l@{}}
  \textcolor{gray}{\texttt{bit\_mask\_loop}} \\
  \begin{lstlisting}[aboveskip= 0\baselineskip,mathescape=true,language=C, xleftmargin=0.5cm,morekeywords={stack, list}, basicstyle=\normalfont]
$\convar{x}$ = 0;
$\convar{i}$ = 0;
while ($\convar{i}$ < 8){
  $\convar{x}$ = $\convar{x}$ | (1 << $a$[$\convar{i}$]);
  $\convar{i}$ = $\convar{i}$ + 1; }
    \end{lstlisting}
\end{tabular}
  &
\begin{tabular}{@{}l@{}}
  \textcolor{gray}{\texttt{set\_union\_loop}} \\
 \begin{lstlisting}[aboveskip= 0\baselineskip,mathescape=true, language=C, xleftmargin=0.5cm,morekeywords={stack, list}, basicstyle=\normalfont]
$\absvar{s}$ = { };
$\absvar{j}$ = 0;
while ($\absvar{j}$ < 8){
  $\absvar{s}$ = $\absvar{s}$ $\cup$ { $a$[$\absvar{j}$] };
  $\absvar{j}$ = $\absvar{j}$ + 1; }                   
   \end{lstlisting}
\end{tabular} \\
  \end{tabular}
\end{appexample}

\begin{relproof} ~

 \begin{lstlisting}[mathescape=true,language=C,basicstyle=\footnotesize, escapechar=\&]
// $\langle \hspeccolor{\absvar{s} = \{\,\}; \absvar{j} = 0;  \abswhile} \rangle$ 
// $\color{gray}\text{high-level steps}$
// $\langle  \hasrt{\absvar{s}= \emptyset \land \absvar{j} = 0  } \land \hspeccolor{ \abswhile} \rangle$ 
$\convar{x} = 0; $ $\convar{i} = 0; $
// $\langle \lasrt{\convar{x} = 0 \land \convar{i} = 0 } \land \hasrt{\absvar{s}= \emptyset \land \absvar{j} = 0  } \land \hspeccolor{ \abswhile} \rangle$ 
// $\langle \exists l \, n.\  \lasrt{\convar{x} = \Sigma_{a\in l} 2^a \land \convar{i} = n } \land   \hasrt{\absvar{s}= l \land \absvar{j} = n } \land \hspeccolor{\abswhile} \rangle $  
while ($\convar{i}$ < 8){
  // $\langle n < 8 \land \lasrt{\convar{x} = \Sigma_{a\in l} 2^a \land \convar{i} = n } \land \hasrt{\absvar{s}= l \land \absvar{j} = n } \land \hspeccolor{\abswhile} \rangle $ 
  // $\color{gray}\text{high-level step: loop unrolling}$
  // $\langle n < 8 \land \lasrt{\convar{x} = \Sigma_{a\in l} 2^a \land \convar{i} = n } \land $ 
  // $\quad \quad \, \hasrt{\absvar{s}= l \land \absvar{j} = n} \land \hspeccolor{\absvar{s}= \absvar{s} \cup \{a[\absvar{j}]\}; \absvar{j} = \absvar{j} + 1; \abswhile} \rangle $ 
  $\convar{x}$ = $\convar{x}$ | (1 << $a$[$\convar{i}$]);
  // $\color{gray}\text{high-level and low-level steps}$
  // $\langle \exists l. \, n < 8 \land \lasrt{\convar{x} = \Sigma_{a\in l} 2^a \land \convar{i} = n  } \land $ 
  // $\quad \quad \,  \hasrt{\absvar{s}= l \land \absvar{j} = n } \land \hspeccolor{\absvar{j} = \absvar{j} + 1; \abswhile} \rangle $ 
  $\convar{i}$ = $\convar{i}$ + 1; 
  // $\color{gray}\text{high-level and low-level steps}$
  // $\langle \exists l. \, n < 8 \land \lasrt{\convar{x} = \Sigma_{a\in l} 2^a \land \convar{i} = n +1  } \land $ 
  // $\quad \quad \,  \hasrt{\absvar{s}= l \land \absvar{j} = n +1 } \land \hspeccolor{ \abswhile} \rangle $ 
}
\end{lstlisting}    
\end{relproof}

\begin{traproof}
 \begin{lstlisting}[mathescape=true,language=C,basicstyle=\footnotesize, escapechar=\&]
// $\{  \needExecX{\ptrue}{\absvar{s} = \{\,\}; \absvar{j} = 0;  \abswhile}{X} \}$ 
// $\color{gray}\text{consequence rules}$
// $\{  \needExecX{\absvar{s}= \emptyset \land \absvar{j} = 0  }{ \abswhile}{X} \}$ 
$\convar{x} = 0; $ $\convar{i} = 0; $
// $\{ {\convar{x} = 0 \land \convar{i} = 0 } \land \needExecX{\absvar{s}= \emptyset \land \absvar{j} = 0  }{ \abswhile}{X} \}$ 
// $\{ \exists l \, n.\  {\convar{x} = \Sigma_{a\in l} 2^a \land \convar{i} = n } \land   \needExecX{\absvar{s}= l \land \absvar{j} = n }{\abswhile}{X} \} $  
while ($\convar{i}$ < 8){
  // $\{ n < 8 \land {\convar{x} = \Sigma_{a\in l} 2^a \land \convar{i} = n } \land \needExecX{\absvar{s}= l \land \absvar{j} = n }{\abswhile}{X} \} $ 
  // $\color{gray}\text{consequence rule: loop unrolling}$
  // $\langle n < 8 \land {\convar{x} = \Sigma_{a\in l} 2^a \land \convar{i} = n } \land $ 
  // $\quad \, \needExecX{\absvar{s}= l \land \absvar{j} = n}{\absvar{s}= \absvar{s} \cup \{a[\absvar{j}]\}; \absvar{j} = \absvar{j} + 1; \abswhile}{X} \} $ 
  $\convar{x}$ = $\convar{x}$ | (1 << $a$[$\convar{i}$]);
  // $\color{gray}\text{consequence and sequencing rules}$
  // $\{ \exists l. \, n < 8 \land {\convar{x} = \Sigma_{a\in l} 2^a \land \convar{i} = n  } \land $ 
  // $\quad \,  \needExecX{\absvar{s}= l \land \absvar{j} = n }{\absvar{j} = \absvar{j} + 1; \abswhile}{X} \} $ 
  $\convar{i}$ = $\convar{i}$ + 1; 
  // $\color{gray}\text{consequence and sequencing rules}$
  // $\{ \exists l. \, n < 8 \land {\convar{x} = \Sigma_{a\in l} 2^a \land \convar{i} = n +1  } \land $ 
  // $\quad\,  \needExecX{\absvar{s}= l \land \absvar{j} = n +1 }{ \abswhile}{X} \} $ 
}
  \end{lstlisting}
  
\end{traproof}

\begin{appexample}
The left low-level program performs initialization steps before executing the loop.
The goal is also to prove the triple $\reltriplenofont{\hspeccolor{\absc}}{\conc}{ \exists l.\  \lasrt{\convar{x} = \Sigma_{a\in l} 2^a} \land \hasrt{\absvar{s}= l} \land \hspeccolor{\cskip}}$ 
to express that after executing the two programs, variable $\convar{x}$ encodes the elements in set $\absvar{s}$ as a sum of powers of 2.

\begin{tabular}{@{\hspace{1em}}p{0.45\linewidth}|@{\hspace{1em}}p{0.45\linewidth}}
\begin{tabular}{@{}l@{}}
  \begin{lstlisting}[aboveskip= 0\baselineskip,mathescape=true,language=C, xleftmargin=0.5cm,morekeywords={stack, list}, basicstyle=\normalfont]
$\convar{x}$ = 0 | (1 << $a$[0]);
$\convar{i}$ = 1;
while ($\convar{i}$ < 8){
  $\convar{x}$ = $\convar{x}$ | (1 << $a$[$\convar{i}$]);
  $\convar{i}$ = $\convar{i}$ + 1; }
    \end{lstlisting}
\end{tabular}
  &
\begin{tabular}{@{}l@{}}
 \begin{lstlisting}[aboveskip= 0\baselineskip,mathescape=true, language=C, xleftmargin=0.5cm,morekeywords={stack, list}, basicstyle=\normalfont]
$\absvar{s}$ = { };
$\absvar{j}$ = 0;
while ($\absvar{j}$ < 8){
  $\absvar{s}$ = $\absvar{s}$ $\cup$ { $a$[$\absvar{j}$] };
  $\absvar{j}$ = $\absvar{j}$ + 1; }                   
   \end{lstlisting}
\end{tabular} \\
  \end{tabular}
\end{appexample}

\begin{relproof} ~

 \begin{lstlisting}[mathescape=true,language=C,basicstyle=\footnotesize, escapechar=\&]
// $\langle \hspeccolor{\absvar{s} = \{\,\}; \absvar{j} = 0;  \abswhile} \rangle$ 
// $\color{gray}\text{high-level steps}$
// $\langle  \hasrt{\absvar{s}= \emptyset \land \absvar{j} = 0  } \land \hspeccolor{ \abswhile} \rangle$ 
$\convar{x}$ = 0 | (1 << $a$[0]); $\convar{i} = 1; $
// $\langle \lasrt{\convar{x} = 2^{a_0} \land \convar{i} = 1 } \land \hasrt{\absvar{s}= \emptyset \land \absvar{j} = 0  } \land \hspeccolor{ \abswhile} \rangle$ 
// $\color{gray}\text{high-level step: loop unrolling}$
// $\langle \lasrt{\convar{x} = 2^{a_0} \land \convar{i} = 1 } \land \hasrt{\absvar{s}= \emptyset \land \absvar{j} = 0  } \land$
$\quad\quad \, \hspeccolor{\absvar{s}= \absvar{s} \cup \{a[\absvar{j}]\}; \absvar{j} = \absvar{j} + 1; \abswhile} \rangle$ 
// $\color{gray}\text{high-level steps}$
// $\langle \lasrt{\convar{x} = 2^{a_0} \land \convar{i} = 1 } \land \hasrt{\absvar{s}= \{a_0\} \land \absvar{j} = 1  } \land \hspeccolor{ \abswhile} \rangle$ 
// $\langle \exists l \, n.\  \lasrt{\convar{x} = \Sigma_{a\in l} 2^a \land \convar{i} = n } \land   \hasrt{\absvar{s}= l \land \absvar{j} = n } \land \hspeccolor{\abswhile} \rangle $  
while ($\convar{i}$ < 8){
  // $\langle n < 8 \land \lasrt{\convar{x} = \Sigma_{a\in l} 2^a \land \convar{i} = n } \land \hasrt{\absvar{s}= l \land \absvar{j} = n } \land \hspeccolor{\abswhile} \rangle $ 
  // $\color{gray}\text{high-level step: loop unrolling}$
  // $\langle n < 8 \land \lasrt{\convar{x} = \Sigma_{a\in l} 2^a \land \convar{i} = n } \land $ 
  // $\quad \quad \, \hasrt{\absvar{s}= l \land \absvar{j} = n} \land \hspeccolor{\absvar{s}= \absvar{s} \cup \{a[\absvar{j}]\}; \absvar{j} = \absvar{j} + 1; \abswhile} \rangle $ 
  $\convar{x}$ = $\convar{x}$ | (1 << $a$[$\convar{i}$]);
  // $\color{gray}\text{high-level and low-level steps}$
  // $\langle \exists l. \, n < 8 \land \lasrt{\convar{x} = \Sigma_{a\in l} 2^a \land \convar{i} = n  } \land $ 
  // $\quad \quad \,  \hasrt{\absvar{s}= l \land \absvar{j} = n } \land \hspeccolor{\absvar{j} = \absvar{j} + 1; \abswhile} \rangle $ 
  $\convar{i}$ = $\convar{i}$ + 1; 
  // $\color{gray}\text{high-level and low-level steps}$
  // $\langle \exists l. \, n < 8 \land \lasrt{\convar{x} = \Sigma_{a\in l} 2^a \land \convar{i} = n +1  } \land $ 
  // $\quad \quad \,  \hasrt{\absvar{s}= l \land \absvar{j} = n +1 } \land \hspeccolor{ \abswhile} \rangle $ 
}
\end{lstlisting}    
\end{relproof}

\begin{traproof}
 \begin{lstlisting}[mathescape=true,language=C,basicstyle=\footnotesize, escapechar=\&]
// $\{  \needExecX{\ptrue}{\absvar{s} = \{\,\}; \absvar{j} = 0;  \abswhile}{X} \}$ 
// $\color{gray}\text{consequence rules}$
// $\{  \needExecX{\absvar{s}= \emptyset \land \absvar{j} = 0  }{ \abswhile}{X} \}$ 
$\convar{x}$ = 0 | (1 << $a$[0]); $\convar{i} = 1; $
// $\{ {\convar{x} = 2^{a_0} \land \convar{i} = 1 } \land \needExecX{\absvar{s}= \emptyset \land \absvar{j} = 0  }{ \abswhile}{X} \}$ 
// $\color{gray}\text{consequence rule: loop unrolling}$
// $\{ {\convar{x} = 2^{a_0} \land \convar{i} = 1 } \land $
$\quad\quad \,  
\needExecX{\absvar{s}= \emptyset \land \absvar{j} = 0  } {\absvar{s}= \absvar{s} \cup \{a[\absvar{j}]\}; \absvar{j} = \absvar{j} + 1; \abswhile}{X} \}$ 
// $\color{gray}\text{consequence rules}$
// $\{ {\convar{x} = 2^{a_0} \land \convar{i} = 1 } \land \needExecX{\absvar{s}= \{a_0\} \land \absvar{j} = 1  }{ \abswhile}{X} \}$ 
// $\{ \exists l \, n.\  {\convar{x} = \Sigma_{a\in l} 2^a \land \convar{i} = n } \land   \needExecX{\absvar{s}= l \land \absvar{j} = n }{\abswhile}{X} \} $  
while ($\convar{i}$ < 8){
  // $\{ n < 8 \land {\convar{x} = \Sigma_{a\in l} 2^a \land \convar{i} = n } \land \needExecX{\absvar{s}= l \land \absvar{j} = n }{\abswhile}{X} \} $ 
  // $\color{gray}\text{consequence rule: loop unrolling}$
  // $\langle n < 8 \land {\convar{x} = \Sigma_{a\in l} 2^a \land \convar{i} = n } \land $ 
  // $\quad \, \needExecX{\absvar{s}= l \land \absvar{j} = n}{\absvar{s}= \absvar{s} \cup \{a[\absvar{j}]\}; \absvar{j} = \absvar{j} + 1; \abswhile}{X} \} $ 
  $\convar{x}$ = $\convar{x}$ | (1 << $a$[$\convar{i}$]);
  // $\color{gray}\text{consequence and sequencing rules}$
  // $\{ \exists l. \, n < 8 \land {\convar{x} = \Sigma_{a\in l} 2^a \land \convar{i} = n  } \land $ 
  // $\quad \,  \needExecX{\absvar{s}= l \land \absvar{j} = n }{\absvar{j} = \absvar{j} + 1; \abswhile}{X} \} $ 
  $\convar{i}$ = $\convar{i}$ + 1; 
  // $\color{gray}\text{consequence and sequencing rules}$
  // $\{ \exists l. \, n < 8 \land {\convar{x} = \Sigma_{a\in l} 2^a \land \convar{i} = n +1  } \land $ 
  // $\quad\,  \needExecX{\absvar{s}= l \land \absvar{j} = n +1 }{ \abswhile}{X} \} $ 
}
  \end{lstlisting}
  
\end{traproof}

\begin{appexample}
The left low-level program performs two steps each iteration.
Here the goal is also to prove the triple $\reltriplenofont{\hspeccolor{\absc}}{\conc}{ \exists l.\  \lasrt{\convar{x} = \Sigma_{a\in l} 2^a} \land \hasrt{\absvar{s}= l} \land \hspeccolor{\cskip}}$ 
to express that after executing the two programs, variable $\convar{x}$ encodes the elements in set $\absvar{s}$ as a sum of powers of 2.

\begin{tabular}{@{\hspace{1em}}p{0.45\linewidth}|@{\hspace{1em}}p{0.45\linewidth}}
\begin{tabular}{@{}l@{}}
  \begin{lstlisting}[aboveskip= 0\baselineskip,mathescape=true,language=C, xleftmargin=0.5cm,morekeywords={stack, list}, basicstyle=\normalfont]
$\convar{x}$ = 0;
$\convar{i}$ = 0;
while ($\convar{i}$ < 8){
  $\convar{x}$ = $\convar{x}$ | (1 << $a$[$\convar{i}$]);
  $\convar{x}$ = $\convar{x}$ | (1 << $a$[$\convar{i}$ + 1]);
  $\convar{i}$ = $\convar{i}$ + 2; }  
    \end{lstlisting}
\end{tabular}
  &
\begin{tabular}{@{}l@{}}
 \begin{lstlisting}[aboveskip= 0\baselineskip,mathescape=true, language=C, xleftmargin=0.5cm,morekeywords={stack, list}, basicstyle=\normalfont]
$\absvar{s}$ = { };
$\absvar{j}$ = 0;
while ($\absvar{j}$ < 8){
  $\absvar{s}$ = $\absvar{s}$ $\cup$ { $a$[$\absvar{j}$] };
  
  $\absvar{j}$ = $\absvar{j}$ + 1; }                      
   \end{lstlisting}
\end{tabular} \\
  \end{tabular}
\end{appexample}

\begin{relproof} ~

 \begin{lstlisting}[mathescape=true,language=C,basicstyle=\footnotesize, escapechar=\&]
// $\langle \hspeccolor{\absvar{s} = \{\,\}; \absvar{j} = 0;  \abswhile} \rangle$ 
// $\color{gray}\text{high-level steps}$
// $\langle  \hasrt{\absvar{s}= \emptyset \land \absvar{j} = 0  } \land \hspeccolor{ \abswhile} \rangle$ 
$\convar{x} = 0; $ $\convar{i} = 0; $
// $\langle \lasrt{\convar{x} = 0 \land \convar{i} = 0 } \land \hasrt{\absvar{s}= \emptyset \land \absvar{j} = 0  } \land \hspeccolor{ \abswhile} \rangle$ 
// $\langle \exists l \, n.\  \lasrt{\convar{x} = \Sigma_{a\in l} 2^a \land \convar{i} = 2*n } \land   \hasrt{\absvar{s}= l \land \absvar{j} = 2*n } \land \hspeccolor{\abswhile} \rangle $  
while ($\convar{i}$ < 8){
  // $\langle n < 4 \land \lasrt{\convar{x} = \Sigma_{a\in l} 2^a \land \convar{i} = 2* n } \land \hasrt{\absvar{s}= l \land \absvar{j} = 2* n } \land \hspeccolor{\abswhile} \rangle $ 
  // $\color{gray}\text{high-level step: loop unrolling}$
  // $\langle n < 4 \land \lasrt{\convar{x} = \Sigma_{a\in l} 2^a \land \convar{i} = 2* n } \land $ 
  // $\quad \quad \, \hasrt{\absvar{s}= l \land \absvar{j} = 2* n} \land \hspeccolor{\absvar{s}= \absvar{s} \cup \{a[\absvar{j}]\}; \absvar{j} = \absvar{j} + 1; \abswhile} \rangle $ 
  $\convar{x}$ = $\convar{x}$ | (1 << $a$[$\convar{i}$]);
  // $\color{gray}\text{high-level and low-level steps}$
  // $\langle \exists l. \, n < 4 \land \lasrt{\convar{x} = \Sigma_{a\in l} 2^a \land \convar{i} = 2* n  } \land $ 
  // $\quad \quad \,  \hasrt{\absvar{s}= l \land \absvar{j} = 2* n } \land \hspeccolor{\absvar{j} = \absvar{j} + 1; \abswhile} \rangle $ 
  // $\color{gray}\text{high-level steps}$
  // $\langle \exists l. \, n < 4 \land \lasrt{\convar{x} = \Sigma_{a\in l} 2^a \land \convar{i} = 2* n   } \land $ 
  // $\quad \quad \,  \hasrt{\absvar{s}= l \land \absvar{j} = 2* n +1 } \land \hspeccolor{ \abswhile} \rangle $ 
  // $\color{gray}\text{high-level step: loop unrolling}$
  // $\langle n < 4 \land \lasrt{\convar{x} = \Sigma_{a\in l} 2^a \land \convar{i} = 2* n +1 } \land $ 
  // $\quad \quad \, \hasrt{\absvar{s}= l \land \absvar{j} = 2* n + 1} \land \hspeccolor{\absvar{s}= \absvar{s} \cup \{a[\absvar{j}]\}; \absvar{j} = \absvar{j} + 1; \abswhile} \rangle $ 
  $\convar{x}$ = $\convar{x}$ | (1 << $a$[$\convar{i}$ +1]);
  // $\color{gray}\text{high-level and low-level steps}$
  // $\langle \exists l. \, n < 4 \land \lasrt{\convar{x} = \Sigma_{a\in l} 2^a \land \convar{i} = 2* n } \land $ 
  // $\quad \quad \,  \hasrt{\absvar{s}= l \land \absvar{j} = 2* n + 1} \land \hspeccolor{\absvar{j} = \absvar{j} + 1; \abswhile} \rangle $ 
  $\convar{i}$ = $\convar{i}$ + 2; 
  // $\color{gray}\text{high-level and low-level steps}$
  // $\langle \exists l. \, n < 4 \land \lasrt{\convar{x} = \Sigma_{a\in l} 2^a \land \convar{i} = 2* (n + 1)  } \land $ 
  // $\quad \quad \,  \hasrt{\absvar{s}= l \land \absvar{j} = 2* (n +1) } \land \hspeccolor{ \abswhile} \rangle $ 
}
\end{lstlisting}    
\end{relproof}

\begin{traproof}
 \begin{lstlisting}[mathescape=true,language=C,basicstyle=\footnotesize, escapechar=\&]
// $\{  \needExecX{\ptrue}{\absvar{s} = \{\,\}; \absvar{j} = 0;  \abswhile}{X} \}$ 
// $\color{gray}\text{consequence rules}$
// $\{  \needExecX{\absvar{s}= \emptyset \land \absvar{j} = 0  }{ \abswhile}{X} \}$ 
$\convar{x} = 0; $ $\convar{i} = 0; $
// $\{ {\convar{x} = 0 \land \convar{i} = 0 } \land \needExecX{\absvar{s}= \emptyset \land \absvar{j} = 0  }{ \abswhile}{X} \}$ 
// $\{ \exists l \, n.\  {\convar{x} = \Sigma_{a\in l} 2^a \land \convar{i} = 2*n } \land   \needExecX{\absvar{s}= l \land \absvar{j} = 2*n }{\abswhile}{X} \} $  
while ($\convar{i}$ < 8){
  // $\{ n < 4 \land {\convar{x} = \Sigma_{a\in l} 2^a \land \convar{i} = 2*n } \land$
  $ \quad \, \needExecX{\absvar{s}= l \land \absvar{j} = 2*n }{\abswhile}{X} \} $ 
  // $\color{gray}\text{consequence rule: loop unrolling}$
  // $\langle n < 4 \land {\convar{x} = \Sigma_{a\in l} 2^a \land \convar{i} = 2*n } \land $ 
  // $\quad \, \needExecX{\absvar{s}= l \land \absvar{j} = 2*n}{\absvar{s}= \absvar{s} \cup \{a[\absvar{j}]\}; \absvar{j} = \absvar{j} + 1; \abswhile}{X} \} $ 
  $\convar{x}$ = $\convar{x}$ | (1 << $a$[$\convar{i}$]);
  // $\color{gray}\text{consequence and sequencing rules}$
  // $\{ \exists l. \, n < 4 \land {\convar{x} = \Sigma_{a\in l} 2^a \land \convar{i} = 2*n  } \land $ 
  // $\quad \,  \needExecX{\absvar{s}= l \land \absvar{j} = 2*n }{\absvar{j} = \absvar{j} + 1; \abswhile}{X} \} $ 
  // $\color{gray}\text{consequence rule}$
  // $\{ \exists l. \, n < 4 \land {\convar{x} = \Sigma_{a\in l} 2^a \land \convar{i} = 2*n } \land $ 
  // $\quad\,  \needExecX{\absvar{s}= l \land \absvar{j} = 2*n +1 }{ \abswhile}{X} \} $ 
  // $\color{gray}\text{consequence rule: loop unrolling}$
  // $\{ n < 4 \land {\convar{x} = \Sigma_{a\in l} 2^a \land \convar{i} = 2* n  } \land $ 
  // $\quad \, \needExecX{\absvar{s}= l \land \absvar{j} = 2* n + 1}{\absvar{s}= \absvar{s} \cup \{a[\absvar{j}]\}; \absvar{j} = \absvar{j} + 1; \abswhile}{X} \} $ 
  $\convar{x}$ = $\convar{x}$ | (1 << $a$[$\convar{i}$+1]);
  // $\color{gray}\text{consequence and sequencing rules}$
  // $\{ \exists l. \, n < 4 \land {\convar{x} = \Sigma_{a\in l} 2^a \land \convar{i} = 2* n  } \land $ 
  // $\quad  \,  \needExecX{\absvar{s}= l \land \absvar{j} = 2* n + 1}{\absvar{j} = \absvar{j} + 1; \abswhile}{X} \} $ 
  $\convar{i}$ = $\convar{i}$ + 2; 
  // $\color{gray}\text{consequence and sequencing rules}$
  // $\{ \exists l. \, n < 4 \land {\convar{x} = \Sigma_{a\in l} 2^a \land \convar{i} = 2* (n + 1)  } \land $ 
  // $\quad  \,  \needExecX{\absvar{s}= l \land \absvar{j} = 2* (n +1) }{ \abswhile}{X} \} $ 
}
  \end{lstlisting}
  
\end{traproof}

\begin{appexample}
Both programs iterate through the items in an array  until index $8$. 
In each iteration, the two programs add the corresponding item in arrays ($\convar{A}$ or $\absvar{A}$) to the collection representation ($\convar{x}$ or $\absvar{s}$).
The low-level program  can lead to errors if the program variable $\convar{i}$ exceeds the bounds of the array stored in variable $\convar{A}$ while the high-level program asserts
that the program variable $\absvar{j}$ stays within the range.
Here the goal is also to prove the triple $\reltriplenofont{\hspeccolor{\absc}}{\conc}{ \exists l.\  \lasrt{\convar{x} = \Sigma_{a\in l} 2^a} \land \hasrt{\absvar{s}= l} \land \hspeccolor{\cskip}}$ 
to express that after executing the two programs, variable $\convar{x}$ encodes the elements in set $\absvar{s}$ as a sum of powers of 2.

    \begin{tabular}{@{\hspace{1em}}p{0.45\linewidth}|@{\hspace{1em}}p{0.45\linewidth}}
\begin{tabular}{@{}l@{}}
    \begin{lstlisting}[aboveskip= 0\baselineskip,mathescape=true, language=C, xleftmargin=0.5cm,morekeywords={stack, list}, basicstyle=\normalfont]
$\convar{x}$ = 0;
$\convar{i}$ = 0;
while ($\convar{i}$ < 8){

  $\convar{x}$ = $\convar{x}$ | (1 << $\convar{A}$[$\convar{i}$]);
  $\convar{i}$ = $\convar{i}$ + 1;
}
      \end{lstlisting}
\end{tabular}
    &
\begin{tabular}{@{}l@{}}
   \begin{lstlisting}[aboveskip= 0\baselineskip,mathescape=true,language=C, morekeywords={stack, list, assert}, basicstyle=\normalfont]
$\absvar{s}$ = { };
$\pvar{j}$ = 0;
while ($\pvar{j}$ < 8){
  assert($\len{\absvar{A}} \geq \pvar{j}$);
  $\absvar{s}$ = $\absvar{s}$ $\cup$ { $\absvar{A}$[$\pvar{j}$] };
  $\pvar{j}$ = $\pvar{j}$ + 1;
}             
     \end{lstlisting}
\end{tabular} \\
    \end{tabular}
\end{appexample}

\begin{relproof}

  \begin{lstlisting}[aboveskip= 0\baselineskip,mathescape=true,language=C,basicstyle=\footnotesize, escapechar=\&]
// $\color{gray}\text{high-level steps}$
// $\langle  \lasrt{\convar{A}=l} \land \hasrt{\absvar{s} = \emptyset \land \absvar{j} = 0\land \absvar{A} = l} \land \hspeccolor{  \abswhile}  \rangle$ 
$\convar{x} = 0; $ $\convar{i} = 0; $
// $\color{gray}\text{low-level steps}$
// $\langle \exists l_0 \, n.\  \lasrt{\convar{x} = \Sigma_{a\in l_0} 2^a \land \convar{i} = n \land \convar{A}=l} \land $ 
// $\quad  \hasrt{{\absvar{s}= l_0 \land \absvar{j} = n \land \absvar{A} = l}} \land \hspeccolor{\abswhile}  \rangle $  
while ($\convar{i}$ < 7){
  // $\langle \lasrt{\convar{x} = \Sigma_{a\in l_0} 2^a \land \convar{i} = n \land \convar{A}=l} \land $ 
  // $\,\, \hasrt{{\absvar{s}= l_0 \land \absvar{j} = n \land \absvar{A} = l}} \land \hspeccolor{\abswhile} \rangle $ 
  // $\color{gray}\text{high-level step: loop unrolling}$
  // $\langle  \lasrt{\convar{x} = \Sigma_{a\in l_0} 2^a \land \convar{i} = n \land \convar{A}=l }  \land $ 
  // $\,\,  \hasrt{\absvar{s}= l_0 \land \absvar{j} = n \land \absvar{A} = l} \land \hspeccolor{\cassert{\len{\absvar{A}} \geq \absvar{j}};\cdots} \rangle $ 
  // $\color{gray}\text{high-level step and extract pure fact}$
  // $\langle {\color{blue} \text{\textsf length}(l) \geq n}  \land \lasrt{\convar{x} = \Sigma_{a\in l_0} 2^a \land \convar{i} = n \land \convar{A}=l } \land   $ 
  // $\,\, \hasrt{\absvar{s}= l_0 \land \absvar{j} = n \land \absvar{A} = l} \land \hspeccolor{\absvar{s}= \absvar{s} \cup \{\absvar{A}[\absvar{j}]\};\cdots} \rangle $ 
  $\convar{x}$ = $\convar{x}$ | (1 << $\convar{A}$[$\convar{i}$]);
  // $\color{gray}\text{high-level and low-level steps}$
  // $\langle \exists l_0. \, \lasrt{\convar{x} = \Sigma_{a\in l_0} 2^a \land \convar{i} = n \land \convar{A}=l }  \land $ 
  // $\, \, \hasrt{\absvar{s}= l_0 \land \absvar{j} = n \land \absvar{A} = l} \land \hspeccolor{\absvar{j} = \absvar{j} + 1; \abswhile}  \rangle $ 
  $\convar{i}$ = $\convar{i}$ + 1; 
  // $\color{gray}\text{high-level and low-level steps}$
  // $\langle \exists l_0. \, \lasrt{\convar{x} = \Sigma_{a\in l_0} 2^a \land \convar{i} = n + 1\land \convar{A}=l }   \land $ 
  // $\, \, \hasrt{\absvar{s}= l_0 \land \absvar{j} = n + 1 \land \absvar{A} = l} \land \hspeccolor{ \abswhile}  \rangle $ 
}
  \end{lstlisting}
    
\end{relproof}

\begin{traproof}

  \begin{lstlisting}[aboveskip= 0\baselineskip,mathescape=true,language=C,basicstyle=\footnotesize, escapechar=\&]
// $\{ \needExecX{\absvar{A} = l}{\absvar{s} = \{\,\}; \absvar{j} = 0;  \abswhile}{X} \land {\convar{A}}={l} \}$ 
// $\color{gray}\text{consequence rules}$
// $\{ \needExecX{\absvar{s} = \emptyset \land \absvar{j} = 0\land \absvar{A} = l}{  \abswhile}{X} \land {\convar{A}}={l} \}$ 
$\convar{x} = 0; $ $\convar{i} = 0; $
// $\color{gray}\text{sequencing rules}$
// $\{ \exists l_0 \, n.\  \needExecX{{\absvar{s}= l_0 \land \absvar{j} = n \land \absvar{A} = l}}{\abswhile}{X}  \land  $ 
// $\quad {\convar{x} = \Sigma_{a\in l_0} 2^a \land \convar{i} = n \land {\convar{A}}={l}} \} $  
while ($\convar{i}$ < 8){
  // $\{\needExecX{{\absvar{s}= l_0 \land \absvar{j} = n \land \absvar{A} = l}}{\abswhile}{X} \land $ 
  // $\,\, {\convar{x} = \Sigma_{a\in l_0} 2^a \land \convar{i} = n \land {\convar{A}}={l}}   \} $ 
  // $\color{gray}\text{consequence rule: loop unrolling}$
  // $\{\needExecX{\absvar{s}= l_0 \land \absvar{j} = n \land \absvar{A} = l}{\cassert{\len{\absvar{A}} \geq \absvar{j}};\cdots}{X}  \land $ 
  // $\,\, {\convar{x} = \Sigma_{a\in l_0} 2^a \land \convar{i} = n \land {\convar{A}}={l} } \} $ 
  // $\color{gray}\text{consequence rule and extract pure fact}$
  // $\{  \needExecX{\absvar{s}= l_0 \land \absvar{j} = n \land \absvar{A} = l}{\absvar{s}= \absvar{s} \cup \{\absvar{A}[\absvar{j}]\};\cdots}{X} \land   $ 
  // $\,\,{\color{blue} \text{\textsf length}(l) \geq n}  \land {\convar{x} = \Sigma_{a\in l_0} 2^a \land \convar{i} = n \land {\convar{A}}={l} }  \} $ 
  $\convar{x}$ = $\convar{x}$ | (1 << $\convar{A}$[$\convar{i}$]);
  // $\color{gray}\text{consequence and sequencing rules}$
  // $\{ \exists l_0. \, \needExecX{\absvar{s}= l_0 \land \absvar{j} = n \land \absvar{A} = l}{\absvar{j} = \absvar{j} + 1; \abswhile}{X}   \land $ 
  // $\, \, {\convar{x} = \Sigma_{a\in l_0} 2^a \land \convar{i} = n \land {\convar{A}}={l} }\} $ 
  $\convar{i}$ = $\convar{i}$ + 1; 
  // $\color{gray}\text{consequence and sequencing rules}$
  // $\{ \exists l_0. \, \needExecX{\absvar{s}= l_0 \land \absvar{j} = n + 1 \land \absvar{A} = l}{ \abswhile}{X}   \land $ 
  // $\, \, {\convar{x} = \Sigma_{a\in l_0} 2^a \land \convar{i} = n + 1\land  }\} $ 
}
  \end{lstlisting}
    
\end{traproof}

\end{document}